\documentclass[superscriptaddress,prd,aps,amsmath,amssymb,showpacs,showkeys,onecolumn,10pt]{revtex4-2}
\usepackage[dvips]{graphicx,color}          
\usepackage{colortbl}                       
\usepackage{booktabs}                       
\usepackage{amsmath,amssymb}                
\usepackage{subcaption}                     
\usepackage{enumitem}                       
\usepackage{times}                          
\usepackage{float}
\usepackage{adjustbox}
\usepackage{xcolor}
\usepackage[colorlinks=true,urlcolor=blue,linkcolor=red,citecolor=blue]{hyperref}
\usepackage{orcidlink}
\begin{document}

\date{\today}

\title{Modified Black Hole Physics in Quantum Gravity with Quintessence and Topological Defects
}

\author{Faizuddin Ahmed\orcidlink{0000-0003-2196-9622}}
\email[Email: ]{faizuddinahmed15@gmail.com}
\affiliation{Department of Physics, The Assam Royal Global University, Guwahati, 781035, Assam, India}

\author{Ahmad Al-Badawi\orcidlink{0000-0002-3127-3453}}
\email[Email: ]{ahmadbadawi@ahu.edu.jo}
\affiliation{Department of Physics, Al-Hussein Bin Talal University 71111, Ma’an, Jordan}

\author{ \.{I}zzet Sakall{\i}\orcidlink{0000-0001-7827-9476}}
\email[Email: ]{izzet.sakalli@emu.edu.tr (Corresponding author)}
\affiliation{Physics Department, Eastern Mediterranean University, Famagusta 99628, North Cyprus via Mersin 10, Turkey}

\begin{abstract}
We study a static, spherically symmetric black hole (BH) of the quantum Oppenheimer-Snyder (QOS) type surrounded by a quintessence field (QF) and a cloud of strings (CS). Because the line element has the mass-function form, the Einstein tensor is linear in the mass function, and the three contributions enter the effective field equations additively with no cross terms. For the state parameter $w=-2/3$ adopted throughout, the geometry is not asymptotically flat. The horizon condition is a quintic in $r$ whose real positive roots are an inner horizon, the BH horizon, and an outer quintessence horizon, and the static region is bounded by the last two. Working inside that region, we obtain the photon sphere, the critical impact parameter, and the shadow radius seen by a static observer at finite radius, together with null and timelike geodesics, the Lyapunov exponent of circular null orbits, the orbital speed, and the geodetic precession frequency. For scalar, electromagnetic (EM), and Dirac test fields we derive the master equations and establish mode stability, using an S-deformation for the scalar sector and the supersymmetric factorization of the Dirac pair. The associated quasinormal mode (QNM) frequencies are computed at third-order WKB order and benchmarked against Schwarzschild. Damping weakens as the CS parameter and the QF normalization grow, whereas the loop quantum gravity (LQG) deformation acts only weakly. On the thermodynamic side we fix the surface-gravity prescription appropriate to a spacetime with two Killing horizons. The Hawking temperature stays non-negative over the whole admissible branch and vanishes at the degenerate configuration in which the BH horizon merges with the quintessence horizon, so no negative-temperature phase arises. Specific heat and Gibbs free energy are re-examined on that branch.
\end{abstract}


\maketitle

\section{Introduction}\label{sec:intro}

The past century has witnessed a remarkable transformation of BHs from purely theoretical constructs in Einstein's general relativity (GR) to observationally confirmed astrophysical phenomena. This evolution reached two turning points, the detection of gravitational waves by LIGO and Virgo collaborations \cite{isz01,isz02} and the historic direct imaging of BH shadows by the EHT collaboration \cite{isz03,isz03a,isz03b,isz03c,isz03d}. These achievements have inaugurated an unprecedented era in BH physics, offering extraordinary opportunities to explore the strong-field regime of gravity and test theoretical predictions with remarkable precision.

While the classical description of BHs has proven remarkably successful, it is anticipated to break down at the Planck scale where quantum gravitational effects become dominant. To address these fundamental limitations, effective quantum gravity approaches have emerged as promising frameworks that incorporate quantum corrections into Einstein's equations without requiring a complete theory of quantum gravity \cite{EQG1,EQG2}. Among these approaches, LQG has garnered considerable attention due to its non-perturbative and background-independent formulation \cite{LQG1,LQG2}. LQG predicts discrete spacetime geometry at the Planck scale, leading to modifications of classical BH solutions through quantum deformation parameters. These modifications typically manifest as additional terms in the metric function that become significant near the BH core, potentially resolving classical singularities while sharpening the discussion of BH thermodynamics and stability \cite{isz09,EQG2}.

The QOS model is one application of effective quantum gravity to BH physics \cite{JL}. This model incorporates quantum corrections through the Barbero-Immirzi parameter $\gamma$ and the Planck length $\ell_P$, resulting in a quantum deformation parameter $\lambda = 16\sqrt{3}\pi\gamma^3\ell_P^2$ that modifies the classical Schwarzschild BH metric. The QOS solution provides a regularized description of gravitational collapse that avoids singularity formation while preserving essential BH spacetime features. Recent investigations have demonstrated that these quantum corrections significantly influence various aspects of BH physics, including geodesic motion, perturbation dynamics, and thermodynamic properties \cite{FA11,FA12}.

Parallel to developments in quantum gravity, the discovery of cosmic acceleration has sparked renewed interest in dark energy and its potential manifestations in BH environments. QFs, characterized by equation of state parameters $w < -1/3$, have emerged as leading dark energy candidates, providing a dynamic alternative to the cosmological constant \cite{SN1,SN2,DE1}. The interaction between QFs and BHs has attracted theoretical attention, since these fields modify the geometry around BHs and change their observable properties \cite{DE2,iszdn2}. QFs are typically characterized by their energy density $\rho_q$ and pressure $p_q = w_q\,\rho_q$, where the state parameter $w_q$ determines the field's gravitational behavior. For phantom dark energy, $w_q < -1$, while for standard quintessence, $-1 < w_q < -1/3$ \cite{iszdn2,DE2}.

Another aspect of modified BH physics involves topological defects, particularly cosmic strings and their associated gravitational effects. CS configurations represent a specific class of matter distributions that can surround BHs, arising from the interaction of fundamental strings or cosmic strings with the gravitational field \cite{PSL,isz13}. These configurations are characterized by a radially dependent energy density $\rho_c = \alpha/r^2$, where $\alpha$ is the cloud parameter quantifying string density. The gravitational effects of string clouds can be substantial, leading to modifications of metric structure, geodesic motion, and observable BH signatures \cite{DVS6,isz20}. From an astrophysical perspective, cosmic strings are predicted by various beyond-standard-model theories and could have formed during phase transitions in the early universe, making their study relevant for both fundamental physics and cosmology \cite{isz13,isz14,isz15}. The combination of quantum gravitational effects, QFs, and CS in BH spacetimes has been little explored. Each component introduces distinct modifications to the classical Schwarzschild BH geometry, and their interplay can lead to novel phenomena absent in simpler models.

Recent studies have reported various BH solutions surrounded by CS in combination with additional physical fields. A BH solution in the presence of CS and perfect fluid dark matter was presented in \cite{DVS1}, while the effects of Ayón-Beato-García nonlinear electrodynamics coupled with CS were explored in \cite{DVS2}. Other works have investigated BHs coupled to nonlinear electrodynamics in CS backgrounds \cite{DVS3,DVS4}, as well as Bardeen BHs minimally coupled to nonlinear electrodynamics and surrounded by CS \cite{DVS5}. The combined influence of cloud and fluid of strings on BH solutions was studied in \cite{DVS6}, and gravitational lensing by a Bardeen BH in the presence of CS was analyzed in \cite{DVS7}. A Schwarzschild BH embedded in CS and surrounded by perfect fluid dark matter was examined in \cite{DVS8}. Additionally, a BH solution within effective quantum gravity including both CS and QF was presented in \cite{FA11}, a deformed Schwarzschild BH configuration with surrounding CS was analyzed in \cite{FA12}, and an Ayón-Beato-García BH coupled with CS \cite{FA4}.

Understanding geodesic behavior in modified BH spacetimes is fundamental to predicting observable signatures and testing theoretical models against experimental data. Null geodesics determine photon paths and are directly related to gravitational lensing, BH shadows, and other EM phenomena \cite{isz28,isz29}. Timelike geodesics govern massive particle motion and matter for accretion processes, orbital dynamics, and gravitational wave emission from binary systems \cite{isz30,isz31}. The stability of circular orbits, characterized by effective potentials and Lyapunov exponents, tells us about the dynamical behavior of matter near the BH and can serve as probes of modified gravity theories \cite{isz32,isz33}. Recent investigations into null and timelike geodesic motion in various BH solutions, involving different matter field configurations, can be found in \cite{FA1,FA2,FA3,FA4,FA5,FA6,FA7,FA8,FA9,FA11,FA12,FA13} and references therein.

Perturbation analysis represents another powerful tool for investigating BH stability and extracting information about QNMs, which encode characteristic frequencies of BH oscillations \cite{isz34,isz35}. Different perturbation types-scalar (spin-0), EM (spin-1), and fermionic (spin-1/2)-probe distinct aspects of spacetime geometry and can show signatures of exotic matter fields or quantum gravitational effects \cite{isz36,isz37}. The study of QNMs has gained particular relevance in the era of gravitational wave astronomy, as these modes contribute to the ringdown phase of BH mergers and can potentially distinguish between different gravity theories \cite{isz38,isz39}.

While the individual effects of LQG corrections, quintessence fields (QFs), and CS on BH spacetimes have been investigated, there exist strong theoretical and astrophysical motivations to consider their simultaneous presence. In realistic cosmological environments, BHs do not evolve in isolation but are embedded in backgrounds influenced by both quantum-gravitational effects at short distances and exotic matter components at large scales. LQG-inspired corrections are expected to play a role in the strong-field regime near the event horizon, whereas quintessence provides an effective description of dark-energy-like fields permeating the Universe. Moreover, CS, which may form during early-universe phase transitions, can survive to late times as large-scale topological defects and potentially interact with or surround astrophysical BHs. In such scenarios, these ingredients naturally coexist rather than act independently. Investigating their combined impact is therefore essential for capturing non-linear and synergistic effects that cannot be inferred from models incorporating only a single modification. This unified approach enables a more physically realistic assessment of observable signatures-such as photon sphere, BH shadows, QNM spectra, and thermodynamic properties-and enhances the potential for distinguishing multi-component effects in future high-precision observations. The objectives of the present study are multifold. First, we investigate the geodesic motion in the chosen QOS BH spacetime in the presence of QFs and a CS. In particular, we analyze how the geometric and model parameters influence photon motion-including photon trajectories, the stability of circular null orbits, the photon sphere, and the resulting BH shadow-as well as massive particle dynamics, with special emphasis on the innermost stable circular orbits (ISCOs). Second, we carry out a detailed perturbative analysis of fields with different spins, including scalar, electromagnetic, and fermionic fields, and examine the corresponding effective potentials that govern their dynamics. Finally, we explore the thermodynamic properties of the modified BH spacetime, focusing on how the presence of exotic matter components and topological defects affects the Hawking temperature, entropy, and thermal stability of the system.

Two points about the standing of the model should be stated at the outset. The metric used below is not obtained by quantizing gravitational collapse in the presence of a QF and a CS. It is an effective geometry: the quantum correction is the one derived for the QOS interior in Ref.~\cite{JL}, and the exotic matter is added at the level of the effective field equations. Section~\ref{sec02a} shows that this addition is exact rather than heuristic, since the Einstein tensor of the full metric splits into the three separate sources with no cross terms. That property follows from the mass-function form of the line element and is not an extra assumption. What the construction does not supply is a derivation of the quantum deformation with the extra fields switched on, which would require quantizing the collapsing dust with CS and QF backreaction included. That problem lies outside the present scope. The fields we perturb are, moreover, test fields on a fixed background, so nothing here bears on spin-2 gravitational perturbations.

The paper is organized as follows. Section~\ref{sec02} sets up the QOS BH spacetime with QF and CS, derives the effective source, and classifies the horizons. Geodesic motion, both null and timelike, is treated in Sec.~\ref{sec03}, along with photon trajectories, circular-orbit stability, and the Lyapunov exponent. We turn next to the photon sphere and the shadow (Sec.~\ref{sec04}). Section~\ref{sec05} carries out the perturbation analysis for scalar, EM, and Dirac fields, proves mode stability, and reports the QNM frequencies. Thermal properties, namely the Hawking temperature, the specific heat, and the Gibbs free energy, occupy Sec.~\ref{sec06}. Section~\ref{sec07} closes with a summary and open problems.

\section{QOS BH spacetime with QF and CS} \label{sec02}

Modified BH solutions carrying exotic matter have been built along several routes. Following Letelier \cite{PSL} for CS configurations and Kiselev \cite{VVK} for QF environments, and taking the quantum correction from the QOS construction, we assemble a spacetime that carries all three ingredients at once. The interest is in what the combination does that the separate pieces do not.

 The first study of a BH solution with a CS as the source within the framework of GR was conducted by Letelier \cite{PSL}. In that work, Letelier derived a generalization of the Schwarzschild BH by considering it surrounded by a spherically symmetric CS. The action that describes the CS is given by \cite{PSL}
\begin{equation}
S_{CS} = \int d^4x \sqrt{-g}\, L_{CS}
       = \int (-\gamma)^{1/2} \mathcal{M}\, d\lambda^0 d\lambda^1
       = \int \mathcal{M} \left( -\frac{1}{2}\Sigma^{\mu\nu}\Sigma_{\mu\nu} \right)^{1/2}
       d\lambda^0 d\lambda^1 ,
\label{zz1}
\end{equation}
with $\mathcal{M}$ being a dimensionless constant associated with each string, and
$\gamma$ being the determinant of the induced metric, $\gamma_{AB}$, which is given by
\begin{equation}
\gamma_{AB} = g_{\mu\nu}
\frac{\partial x^\mu}{\partial \lambda^A}
\frac{\partial x^\nu}{\partial \lambda^B} ,
\label{zz2}
\end{equation}
where $x^\mu = x^\mu(\lambda^A)$ describes the string world sheet, with $\lambda^0$
and $\lambda^1$ being time-like and space-like parameters, respectively \cite{JLG}.
The bivector $\Sigma^{\mu\nu}$ can be associated with the string world sheet, and is
written as
\begin{equation}
\Sigma^{\mu\nu} = \epsilon^{AB}
\frac{\partial x^\mu}{\partial \lambda^A}
\frac{\partial x^\nu}{\partial \lambda^B} ,
\label{zz3}
\end{equation}
where $\epsilon^{AB}$ is the Levi-Civita symbol, such that
$\epsilon^{01} = -\epsilon^{10} = 1$. The stress–energy tensor for the CS is given by
\begin{equation}
    T^{\rm CS}_{\mu\nu}=\frac{\rho_c\,\Sigma^{\tau}_{\mu}\,\Sigma_{\tau\nu}}{(-\gamma)^{1/2}}.\label{zz4}
\end{equation}
Due to the spacetime symmetry, the only non-null component of the bivector
$\Sigma^{\mu\nu}$ is $\Sigma^{01} = \frac{\alpha}{\rho_c r^2}$, which depends only on the radial
coordinate. In this identity, $\alpha$ is an integration constant related to the CS, which assumes values in the range $0 < \alpha < 1$ and $\rho_c$ is the energy density of the string cloud.
This configuration is characterized by an energy-momentum tensor of the form  \cite{PSL}:
\begin{equation}
    T^{t}_{t}=T^{r}_{r}=\rho_c=\frac{\alpha}{r^2},\quad T^{\theta}_{\theta}=T^{\phi}_{\phi}=0.\label{pp1}
\end{equation}
The BH metric with CS is described by the following line-element
\begin{equation}
    ds^2=-\left(1-\alpha-2\,M/r\right)\,dt^2+\left(1-\alpha-2\,M/r\right)^{-1}\,dr^2+r^2\,(d\theta^2+\sin^2 \theta\,d\phi^2).\label{pp2}
\end{equation}
Unlike conventional mass-induced curvature, a CS does not generate a standard Newtonian gravitational potential at large distances. Instead, it induces a conical spacetime geometry characterized by a deficit solid angle proportional to the CS parameter $\alpha$, rendering the spacetime asymptotically non-flat. This geometric modification effectively reduces the surface area of spherical hypersurfaces compared to flat spacetime, leading to non-trivial alterations in the angular sector of the metric. Such global geometric effects have important consequences for geodesic motion, influencing photon propagation, particle orbits, and particle dynamics. As a result, observable signatures-including BH shadows, QNM spectra, and gravitational lensing patterns-can exhibit distinctive deviations from their Schwarzschild counterparts. From a physical standpoint, the Letelier spacetime therefore provides a valuable phenomenological framework for investigating how extended one-dimensional matter distributions, such as string clouds arising from symmetry-breaking processes in the early Universe, affect BH physics. Furthermore, when considered in conjunction with other exotic components, such as QFs or quantum-gravity-inspired corrections, the CS can give rise to non-linear and synergistic effects, thereby enriching the structure and observational relevance of modified BH spacetimes.

On the other hand, the study of the QF as a matter content within GR was carried out by Kiselev \cite{VVK}. He obtained a generalization of the Schwarzschild solution describing a BH surrounded by a QF, with the corresponding energy-momentum tensor given by:
\begin{equation}
    T^{t}_{t}=T^{r}_{r}=\rho_q,\quad T^{\theta}_{\theta}=T^{\phi}_{\phi}=-\frac{1}{2}\,\rho_q\,(3\,w_q+1).\label{pp3}
\end{equation}
where $\rho_q$ denotes the energy density, and $p_q$ is the pressure which is related to the density via the equation of state $p_q = w_q \rho_q$, with $w_q$ being the quintessence state parameter. The corresponding line element is given by \cite{VVK}:
\begin{equation}
    ds^2=-\left(1-2\,M/r-N/r^{3\,w_q+1}\right)\,dt^2+\left(1-2\,M/r-N/r^{3\,w_q+1}\right)^{-1}\,dr^2+r^2\,(d\theta^2+\sin^2 \theta\,d\phi^2).\label{pp4}
\end{equation}
The QF energy density follows from the field equations once the metric is written in mass-function form. Throughout we use units with $8\pi G=c=1$, in which Eq.~(\ref{pp1}) holds as printed. Setting $f=1-2m(r)/r$, the Einstein equations give $\rho=2m'(r)/r^2$, $p_r=-\rho$, and $p_t=-m''(r)/r$. For the Kiselev term $2m_q=N\,r^{-3w_q}$ this yields
\begin{equation}
    \rho_q=-\frac{3\,w_q\,N}{r^{3\,(w_q+1)}},\qquad p_t=\frac{(3\,w_q+1)}{2}\,\rho_q ,
\label{kiselev_density}
\end{equation}
which reproduces Eq.~(\ref{pp3}). Here $N>0$ is a constant normalization, and the radial dependence sits in $\rho_q$ rather than in $N$. Taking $w_q=-2/3$ gives $\rho_q=2N/r$, so $N=r\,\rho_q(r)/2$ takes the same value at every radius while the density itself falls off. Since $w_q<0$, the energy density is positive. The state parameter lies in $-1<w_q<-1/3$ for quintessence; the phantom range $w_q<-1$ is not considered here.

In Ref. \cite{JL}, the authors presented a QOS model. The static and spherically symmetric metric is described by the following line-element:
\begin{equation}
    ds^2=-\left(1-2\,M/r+\lambda\,M^2/r^4\right)\,dt^2+\left(1-2\,M/r+\lambda\,M^2/r^4\right)^{-1}\,dr^2+r^2\,(d\theta^2+\sin^2 \theta\,d\phi^2),\label{pp5}
\end{equation}
where $\lambda=16\,\sqrt{3}\,\pi\,\gamma^3\,\ell^2_p$ in which $\ell_{P}$ represents the Planck length and $\gamma$ is the Barbero-Immirzi parameter.

Motivated by the above studies, we consider a static, spherically symmetric model of a QOS BH solution coupled with CS surrounded by QF. We assume that the QF and CS do not interact directly with the effects of LQG, yet their presence leads to non-trivial modifications of the spacetime geometry. Taking these effects into account, the spacetime line element in Schwarzschild coordinates $(t, r, \theta, \phi)$ is given by:
\begin{equation}
    ds^2=-f(r)\,dt^2+\frac{1}{f(r)}\,dr^2+r^2\,(d\theta^2+\sin^2 \theta\,d\phi^2),\label{bb1}
\end{equation}
where the metric function $f(r)$ is given by
\begin{equation}
    f(r)=1-\alpha-\frac{2\,M}{r}+\frac{\lambda\,M^2}{r^4}-\frac{N}{r^{3\,w+1}}\quad\quad (-1 < w < -1/3),\label{bb2}
\end{equation}
where $(N, w)$ represents the quintessential field parameters. Its behavior across the full radial range, both for varying $(\alpha,\lambda,N)$ at $w=-2/3$ and for varying $w$, is collected in Fig.~\ref{fig:horizon-structure} below, which is discussed after the horizon structure has been set out.

\subsection{Effective field equations and additivity of the sources}\label{sec02a}

A metric of the form (\ref{bb1}) is fixed by a single mass function through $f=1-2\,m(r)/r$. For the choice (\ref{bb2}) at $w=-2/3$,
\begin{equation}
2\,m(r)=\alpha\,r+2\,M-\frac{\lambda\,M^2}{r^3}+N\,r^2 .\label{massfunc}
\end{equation}
The Einstein tensor of such a metric is linear in $m$ and its derivatives,
\begin{equation}
G^{t}_{\ t}=G^{r}_{\ r}=-\frac{2\,m'(r)}{r^2},\qquad
G^{\theta}_{\ \theta}=G^{\phi}_{\ \phi}=-\frac{m''(r)}{r},\label{einstein_mass}
\end{equation}
so the effective source separates term by term. Explicitly,
\begin{equation}
\rho_{\rm eff}=\underbrace{\frac{\alpha}{r^2}}_{\rm CS}+\underbrace{\frac{3\,\lambda\,M^2}{r^6}}_{\rm quantum}+\underbrace{\frac{2\,N}{r}}_{\rm QF},\qquad
p_r=-\rho_{\rm eff},\qquad
p_t=\frac{6\,\lambda\,M^2}{r^6}-\frac{N}{r}.\label{effsource}
\end{equation}
The CS piece reproduces Eq.~(\ref{pp1}), the QF piece reproduces Eq.~(\ref{kiselev_density}) at $w=-2/3$, and no cross terms appear at any order. This is worth stating plainly, because the objection that two metrics have simply been superposed would be fatal if the field equations were nonlinear in $m$. They are not, for this symmetry class. What is added are the sources, and the geometry (\ref{bb2}) solves the effective equations with the sum of them.

The $\lambda$ term is not matter in the ordinary sense. It encodes the LQG correction obtained for the QOS interior \cite{JL}, and reading it as an anisotropic fluid with $p_t=2\rho_\lambda$ is a bookkeeping device for checking that the field equations close, not a claim about a physical fluid. Two limitations follow. The deformation parameter was derived for collapsing dust without a CS or a QF, so using it here assumes that the exotic components do not feed back on the quantum correction itself. That assumption is uncontrolled, and we do not defend it beyond noting that $\lambda$ enters at order $r^{-4}$ while the QF enters at order $r$, which keeps the two effects in well separated radial regions.

Using the above metric (\ref{bb1}), we will investigate the geodesic motion, field perturbations, and thermodynamic properties of the associated spacetime. In particular, we aim to analyze how the various parameters characterizing the geometry-such as the CS, quintessential field, and quantum deformation-affect the physical behavior of test particles, photon trajectories, BH shadows, stability of circular orbits, perturbations, and the thermodynamic quantities of BH solutions. The aim is to see how topological defects and gravitational dynamics act together in this extended spacetime.

The horizons can be determined using the condition:
\begin{equation}
    1-\alpha-\frac{2\,M}{r}+\frac{\lambda\,M^2}{r^4}-\frac{N}{r^{3\,w+1}}=0. \label{bb3} 
 \end{equation}
 Throughout this paper, we fix the state parameter $w=-2/3$ for the QF. Therefore, we find
\begin{equation}
    1-\alpha-\frac{2\,M}{r}+\frac{\lambda\,M^2}{r^4}-N\,r=0. \label{bb4} 
 \end{equation}
Multiplying Eq.~(\ref{bb4}) by $r^4$ puts the horizon condition in polynomial form,
\begin{equation}
-N\,r^5+(1-\alpha)\,r^4-2\,M\,r^3+\lambda\,M^2=0 ,\label{bb4a}
\end{equation}
a quintic, not a sixth-order equation as stated in the previous version of this work. Its real positive roots have to be found numerically and then classified, since for $N>0$ there is always more than one.

The classification follows from the behavior of $f$ at the two ends of the radial range. Near the origin the quantum term $\lambda M^2/r^4$ dominates and $f\to+\infty$ for $\lambda>0$; at large $r$ the quintessence term dominates and $f\to-N r\to-\infty$. A generic parameter choice therefore gives three roots: an inner horizon $r_-$ produced by the quantum deformation, the BH horizon $r_+$, and an outer quintessence horizon $r_q$ beyond which $f<0$ again. The static region available to an external observer is $r_+<r<r_q$, and every quantity computed in this paper, including the shadow, the perturbation potentials, and the thermodynamic variables, refers to that region. For $\lambda=0$ the inner root disappears and two horizons remain; for $N=0$ the outer root runs off to infinity and the geometry becomes conical rather than flat, with $f\to1-\alpha$.

\begin{figure}[ht!]
    \centering
    \includegraphics[width=0.99\linewidth]{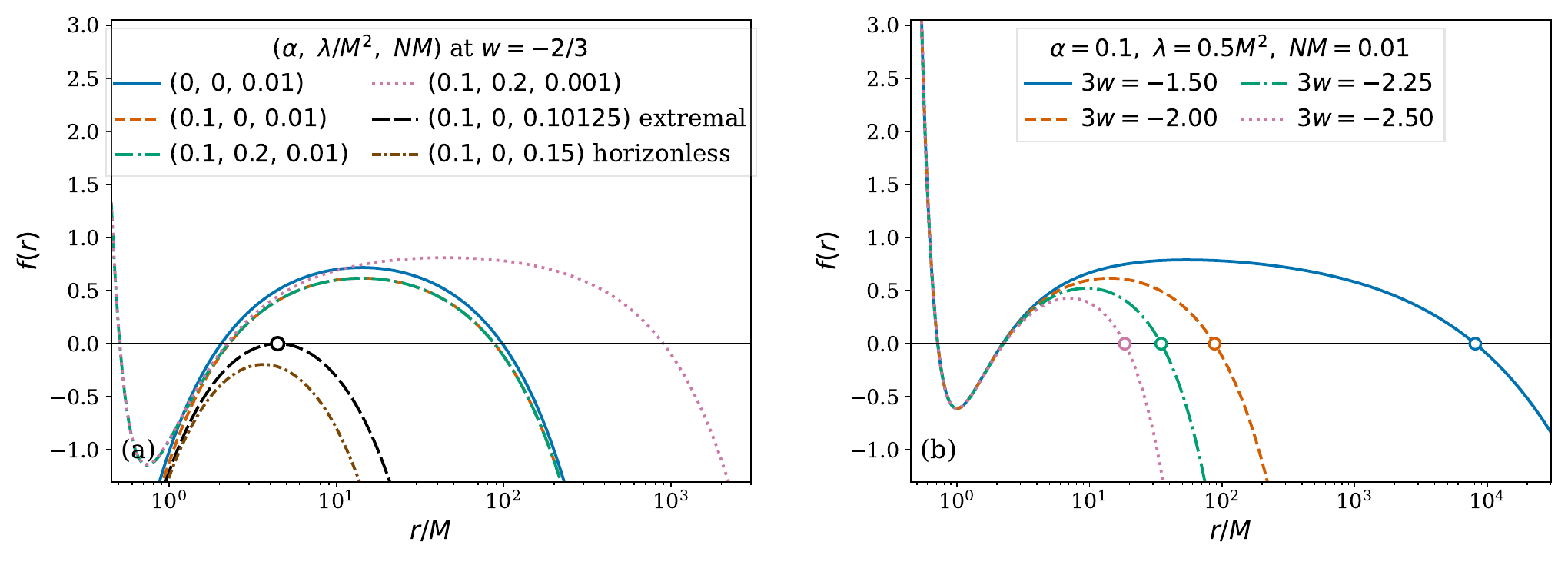}
    \caption{Metric function of Eq.~(\ref{bb2}) against $r/M$ on a logarithmic radial scale, with $M=1$. Panel (a) varies $(\alpha,\ \lambda/M^2,\ N M)$ at $w=-2/3$; panel (b) varies the state parameter at fixed $\alpha=0.1$, $\lambda=0.5M^2$, and $N M=0.01$. The second zero crossing at large radius is the quintessence horizon $r_q$, marked in panel (b) by open circles, and at $\lambda>0$ a third crossing appears near $r/M\simeq0.5$. The last two entries of panel (a) are the limiting cases of Eq.~(\ref{extremal}): the extremal curve touches $f=0$ at $r_{\rm ext}=4.4444M$, marked by the open circle, while the horizonless one stays negative at every radius. The logarithmic scale is what makes the outer crossing visible; on the linear range $r\lesssim10M$ used in the previous version of this work it lies beyond the plotted domain.}
    \label{fig:horizon-structure}
\end{figure}

Table~\ref{istab_new} lists all real positive roots for the parameter sets considered here. The outer roots are large but finite, and they were omitted from the earlier version: at $\alpha=0$, $\lambda=0$, and $N=0.01/M$, for instance, Eq.~(\ref{bb4}) reduces to $1-2M/r-Nr=0$, whose roots are $r_+/M=2.0416848$ and $r_q/M=97.958315$. Figure~\ref{fig:horizon-structure} shows $f(r)$ on a logarithmic radial scale, where the second zero crossing is visible directly.

Two limiting configurations should be separated from the generic three-horizon case, and they are drawn in the same figure. As $N$ grows at fixed $\alpha$ and $\lambda$, the BH horizon and the quintessence horizon approach one another and merge. At $\lambda=0$ the merger follows in closed form from the double root of $1-\alpha-2M/r-Nr=0$,
\begin{equation}
N_{\rm ext}=\frac{(1-\alpha)^2}{8\,M},\qquad r_{\rm ext}=\frac{1-\alpha}{2\,N_{\rm ext}}=\frac{4\,M}{1-\alpha},\label{extremal}
\end{equation}
so that $\alpha=0.1$ and $M=1$ give $N_{\rm ext}=0.10125/M$ and $r_{\rm ext}=4.4444M$. There $f$ touches zero without crossing, the static region has closed up, and the surface gravity vanishes. This degenerate configuration is of Nariai type \cite{Nariai}, and it returns in Sec.~\ref{sec06} as the endpoint of the thermodynamic branch. For $N>N_{\rm ext}$ the quintic has no real positive root: at $\alpha=0.1$ and $N=0.15/M$ the maximum of $f$ sits at $r=\sqrt{2M/N}=3.65M$ with $f=-0.195$, so $f<0$ at every radius and no horizon forms. With $\lambda=0$ the curvature singularity at the origin is then uncovered, and we exclude such parameters rather than describe them as a BH. Everything reported below uses parameters for which the ordering $r_-<r_+<r_q$ holds strictly.

The state parameter controls how far out the quintessence horizon sits, and panel (b) of Fig.~\ref{fig:horizon-structure} shows the effect across the quintessence range. At $\alpha=0.1$, $\lambda=0.5M^2$, and $N M=0.01$, the outer horizon moves from $r_q=8095.55M$ at $3w=-1.5$ to $87.72M$ at $3w=-2$, then to $34.71M$ at $3w=-2.25$ and $18.43M$ at $3w=-2.5$. Two points follow. The outer horizon is present for every $w$ in $-1<w<-1/3$, not only for the value $w=-2/3$ fixed later, because the QF term $-N\,r^{-(3w+1)}$ grows without bound throughout that range. And the static region contracts sharply as $w$ decreases: over the interval plotted, $r_q$ falls by more than two orders of magnitude while $r_+$ moves only from $2.207M$ to $2.259M$. The choice $w=-2/3$ adopted from here on is thus a middle case rather than an extreme one.

\setlength{\tabcolsep}{12pt}
\begin{table}[htbp]
\centering
\begin{tabular}{|c|c|c|c|c|c|}
\hline
\textbf{$\alpha$} & \textbf{$\lambda/M^2$} & \textbf{$N M$} & \textbf{$r_-/M$} & \textbf{$r_+/M$} & \textbf{$r_q/M$} \\
\hline
0.0 & 0.0 & 0.0   & --- & 2.0000000 & --- \\
0.0 & 0.0 & 0.001 & --- & 2.0040161 & 997.99598 \\
0.0 & 0.0 & 0.01  & --- & 2.0416848 & 97.958315 \\
\hline
0.0 & 0.2 & 0.0   & 0.51227404 & 1.9739991 & --- \\
0.0 & 0.2 & 0.001 & 0.51224003 & 1.9780721 & 997.99598 \\
0.0 & 0.2 & 0.01  & 0.51193474 & 2.0162524 & 97.958315 \\
\hline
0.1 & 0.0 & 0.0   & --- & 2.2222222 & --- \\
0.1 & 0.0 & 0.001 & --- & 2.2277365 & 897.77226 \\
0.1 & 0.0 & 0.01  & --- & 2.2799813 & 87.720019 \\
\hline
0.1 & 0.2 & 0.0   & 0.50589768 & 2.2013919 & --- \\
0.1 & 0.2 & 0.001 & 0.50586671 & 2.2069611 & 897.77226 \\
0.1 & 0.2 & 0.01  & 0.50558864 & 2.2596991 & 87.720019 \\
\hline
\end{tabular}
\caption{Real positive roots of the horizon quintic (\ref{bb4a}) for $M=1$ and $w=-2/3$: the inner horizon $r_-$, the BH horizon $r_+$, and the quintessence horizon $r_q$. A dash marks a root that is absent, which happens for $r_-$ when $\lambda=0$ and for $r_q$ when $N=0$. The static region is $r_+<r<r_q$. The spacetime is static and spherically symmetric but not asymptotically flat.}
\label{istab_new}
\end{table}

\begin{figure}[ht!]
    \centering
    \setlength{\tabcolsep}{0pt}
    \begin{minipage}{0.25\textwidth}
        \centering
        \includegraphics[width=\textwidth]{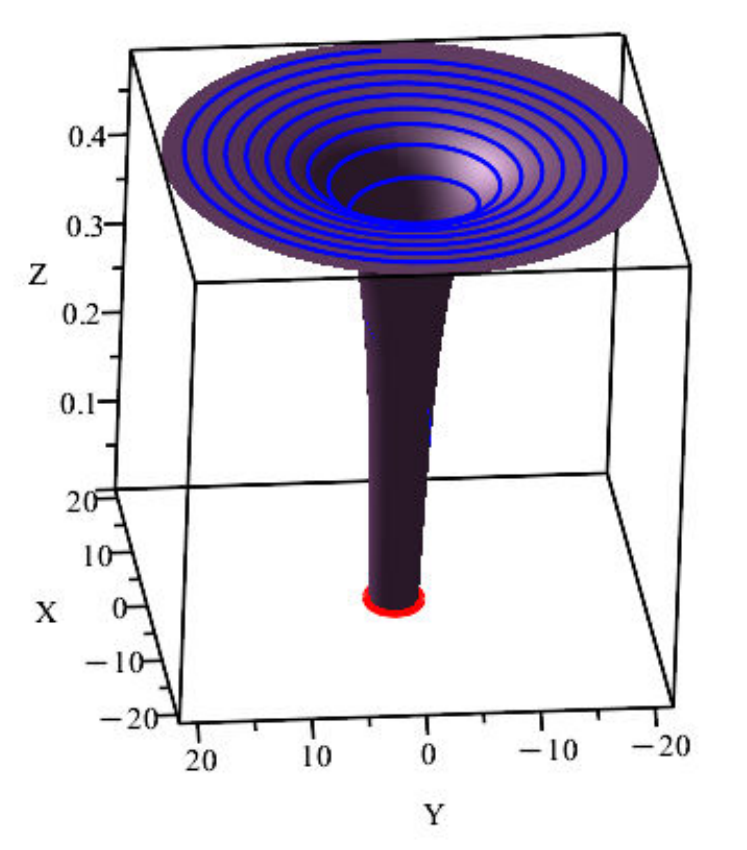}
        \subcaption{\footnotesize [$\alpha=0$,  $\lambda=0$, $N=0$] \newline Schwarzschild BH case with $r_{+}=2.0$.}
        \label{fig:a}
    \end{minipage}
    \begin{minipage}{0.25\textwidth}
        \centering
        \includegraphics[width=\textwidth]{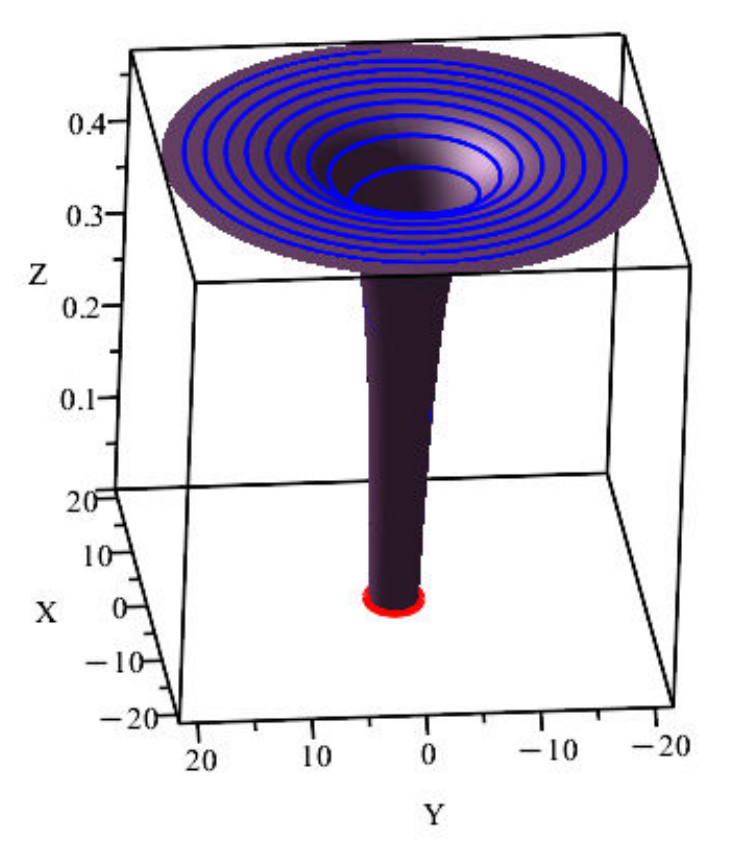}
        \subcaption{\footnotesize [$\alpha=0$,  $\lambda=0$, $N=0.001$] \newline QOS BH with $r_{+}=2.0040161$.}
        \label{fig:b}
    \end{minipage}
    \begin{minipage}{0.25\textwidth}
        \centering
        \includegraphics[width=\textwidth]{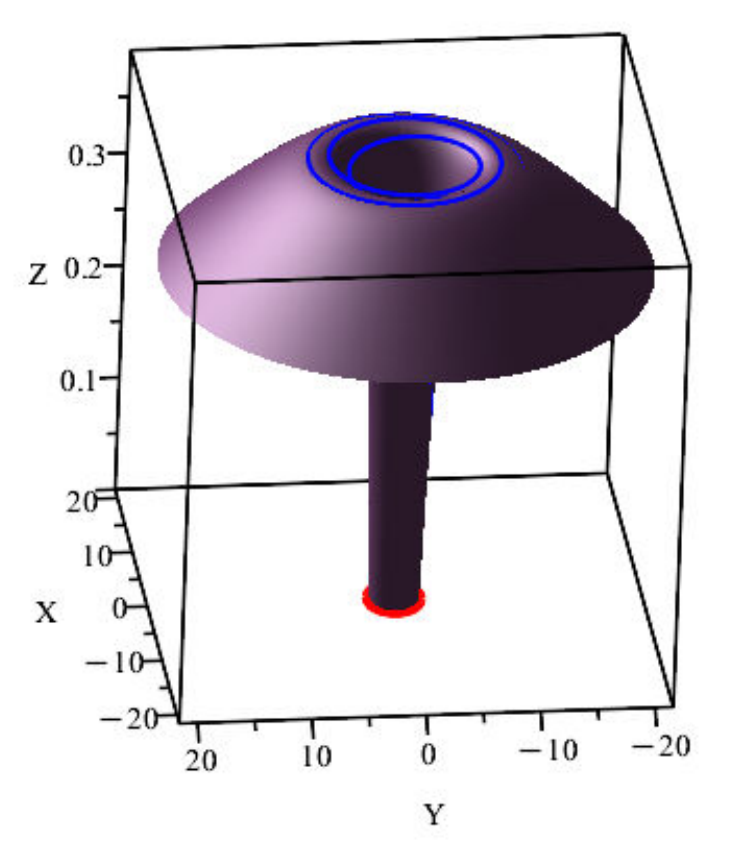}
        \subcaption{\footnotesize [$\alpha=0$,  $\lambda=0.2$, $N=0.01$] \newline QOS BH with $r_{+}=2.0416848$.}
        \label{fig:c}
    \end{minipage}
    \vspace{0.5em} 
    \begin{minipage}{0.25\textwidth}
        \centering
        \includegraphics[width=\textwidth]{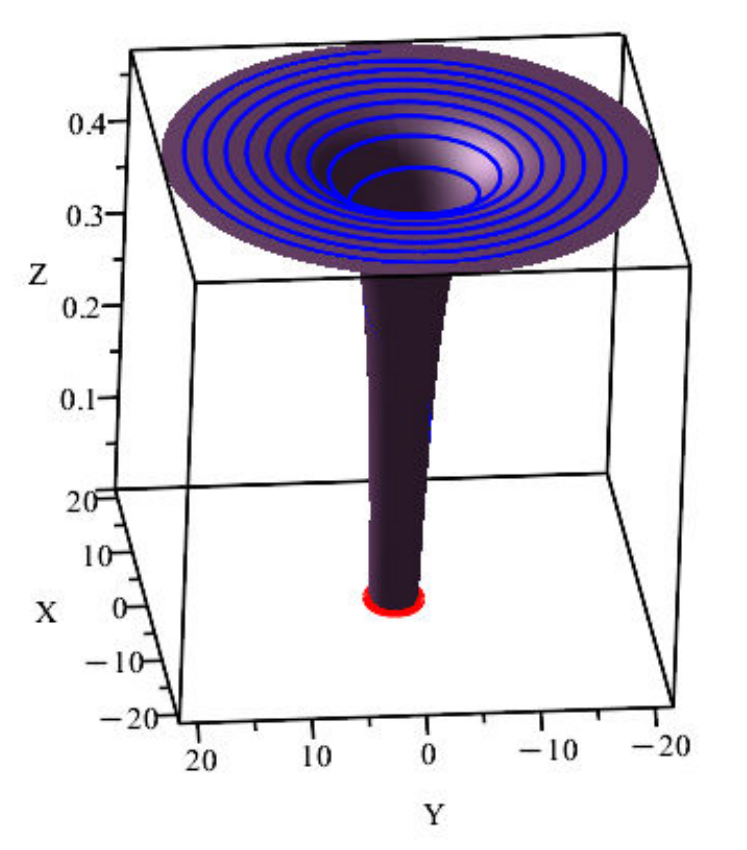}
        \subcaption{\footnotesize [$\alpha=0$,  $\lambda=0.2$, $N=0.001$] \newline QOS BH with $r_{+}=1.9780721$.}
        \label{fig:d}
    \end{minipage}
    \begin{minipage}{0.25\textwidth}
        \centering
        \includegraphics[width=\textwidth]{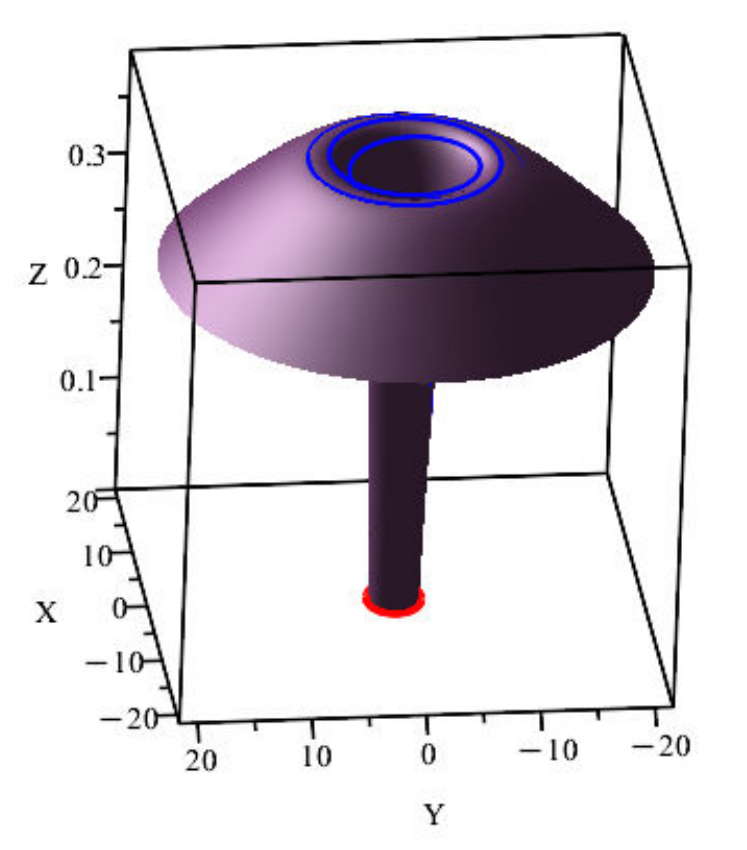}
        \subcaption{\footnotesize [$\alpha=0$,  $\lambda=0.2$, $N=0.01$] \newline QOS BH with $r_{+}=2.0162524$.}
        \label{fig:e}
    \end{minipage}
    \begin{minipage}{0.25\textwidth}
        \centering
        \includegraphics[width=\textwidth]{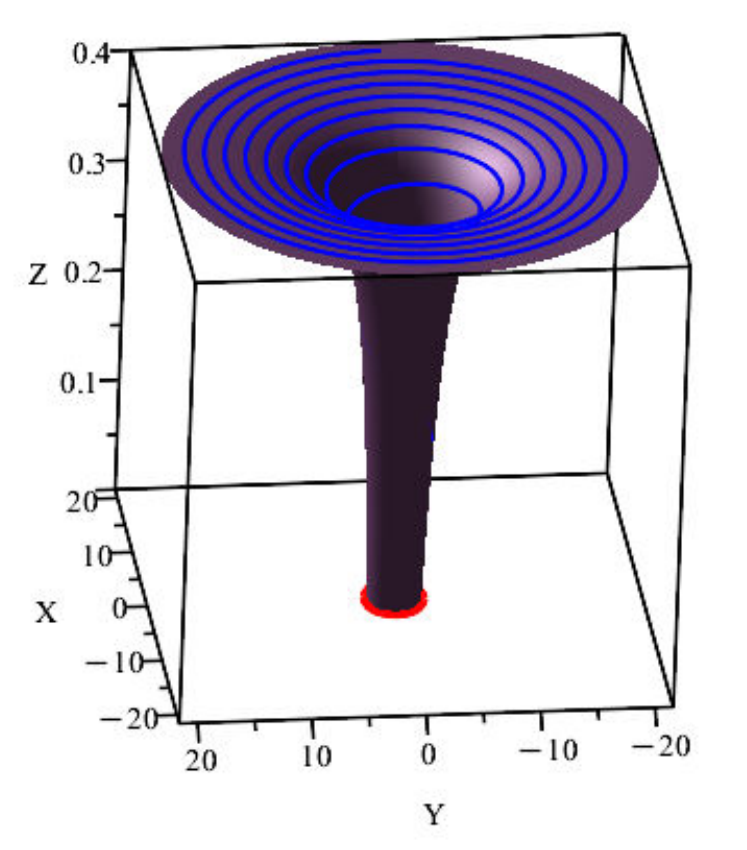}
         \subcaption{\footnotesize [$\alpha=0.1$,  $\lambda=0$, $N=0$] \newline QOS BH with $r_{+}=2.2222222]$.}
        \label{fig:f}
    \end{minipage}
    \vspace{0.5em}
    \begin{minipage}{0.25\textwidth}
        \centering
        \includegraphics[width=\textwidth]{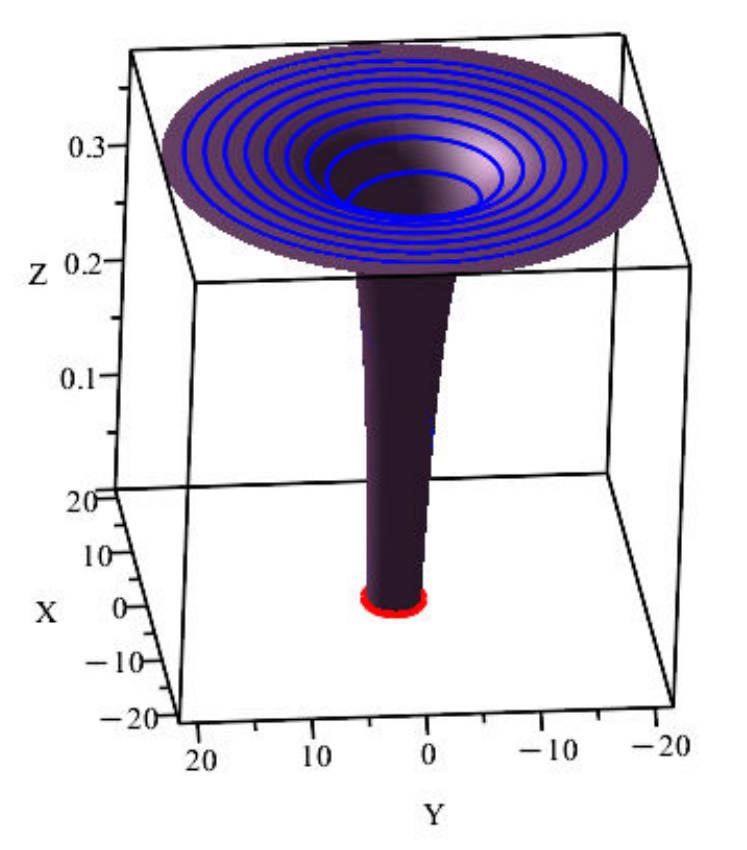}
        \subcaption{\footnotesize [$\alpha=0.1$,  $\lambda=0.2$, $N=0$] \newline QOS BH with $r_{+}=2.2013919]$.}
        \label{fig:g}
    \end{minipage}
    \begin{minipage}{0.25\textwidth}
        \centering
        \includegraphics[width=\textwidth]{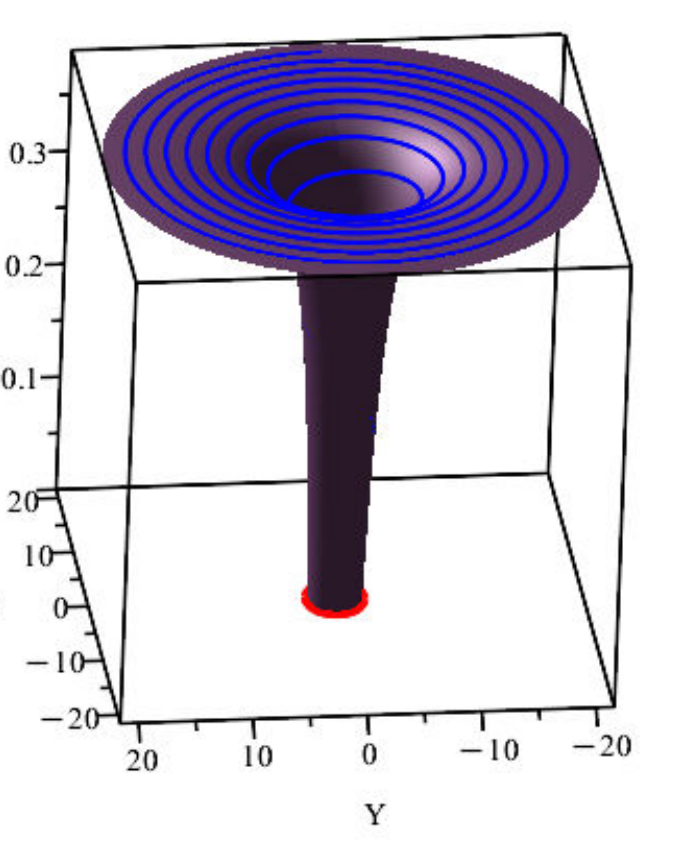}
        \subcaption{\footnotesize [$\alpha=0.1$,  $\lambda=0.2$, $N=0.001$] \newline QOS BH with $r_{+}=2.2069611]$.}
        \label{fig:h}
    \end{minipage}
    \begin{minipage}{0.25\textwidth}
        \centering
        \includegraphics[width=\textwidth]{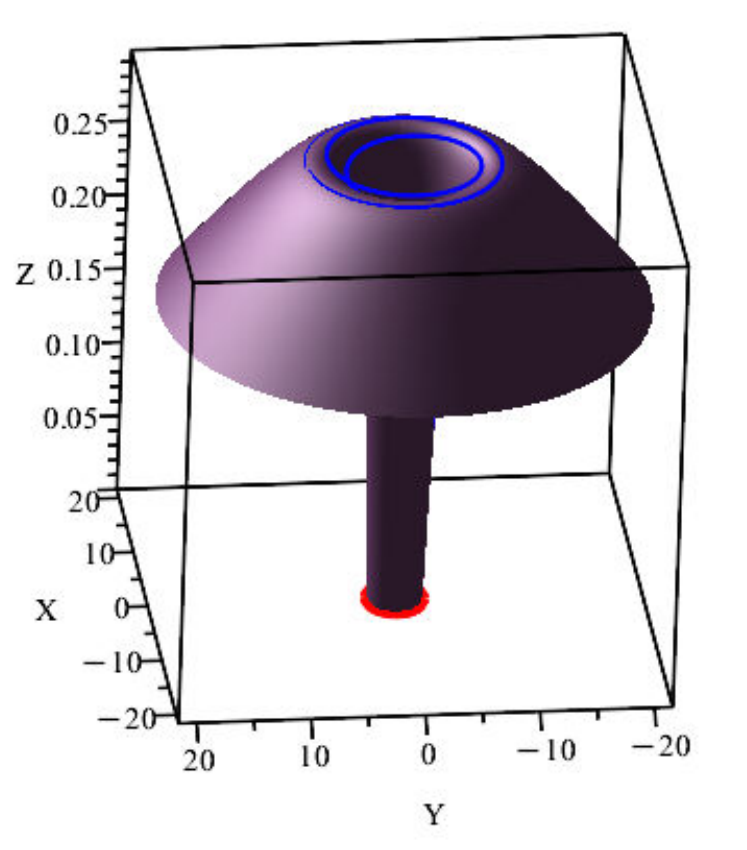}
        \subcaption{\footnotesize [$\alpha=0.1$,  $\lambda=0.2$, $N=0.01$] \newline QOS BH with $r_{+}=2.2596991]$.}
        \label{fig:i}
    \end{minipage}
    \caption{Embedding diagrams of the QOS BH with QF and CS for various parameter values of $\alpha$, $\lambda$, and $N$ values. The parameters are set to $M=1$ and $w=-2/3$. The event horizons (the red rings) are governed by Table \ref{istab_new}.
    }
    \label{fig:isfull_embedding}
\end{figure}

The embedding diagrams presented in Figure~\ref{fig:isfull_embedding} visualize the geometric structure of the QOS BH with QF and CS across different parameter configurations. Panel (a) shows the reference Schwarzschild BH case with $r_+ = 2.0$, displaying the characteristic funnel-like geometry with a well-defined throat structure. As we introduce the various exotic field contributions, the modifications to the spacetime geometry become visible. The QF parameter $N = 0.001$ in panel (b) produces a slight geometric modification, with $r_+ = 2.0040161$, while the quantum deformation parameter $\lambda = 0.2$ combined with the QF parameter $N = 0.01$ in panel (c) results in $r_+ = 2.0416848$, indicating the influence of quantum corrections on the near-horizon geometry. The CS parameter $\alpha = 0.1$ demonstrates significant effects on the BH structure, as seen in panel (f), where the horizon radius increases substantially to $r_+ = 2.2222222$. This expansion reflects the additional energy density contribution from the CS, which effectively increases the gravitational mass of the system. The combined effects of all parameters are most clearly illustrated in panels (g), (h), and (i), where the interplay between quantum corrections ($\lambda = 0.2$), string cloud ($\alpha = 0.1$), and QF ($N$ ranging from $0$ to $0.01$) produces a more involved geometry. Notably, the throat region exhibits varying degrees of curvature, and the horizon positions range from $r_+ = 2.2013919$ to $r_+ = 2.2596991$, demonstrating how these exotic matter fields collectively modify the classical Schwarzschild geometry. The embedding surfaces are drawn over a radial window that reaches a few tens of $M$, well inside the quintessence horizon listed in Table~\ref{istab_new}, and the flattening visible away from the throat is a statement about that window only. It is not asymptotic flatness. Pushing the same construction toward $r_q$ would show the surface turning over, since $f$ decreases through zero there, and at $N=0$ the large-$r$ limit is conical rather than flat because $f\to1-\alpha$ leaves a solid-angle deficit $4\pi\alpha$.

\section{Geodesic Motions: Null and Time-like Geodesic } \label{sec03}

In this section, we investigate the geodesic motion of test particles in the spacetime geometry of a quantum-corrected Oppenheimer-Snyder BH, incorporating the effects of QF and CS. Our primary objective is to analyze how the Barbero-Immirzi parameter, QF characteristics, and the string cloud influence key aspects of particle dynamics-particularly particle trajectories, the stability of circular orbits, and light deflection (photon orbits).

The study of geodesic motion plays a fundamental role in understanding the physical and geometrical properties of BH spacetimes. By examining the behavior of test particles and light rays in strong gravitational fields, we learn about the underlying metric structure. Such analyses also reveal potential observational signatures, including gravitational lensing, BH shadow formation, and accretion disk dynamics. Furthermore, the stability of circular orbits matters for astrophysical processes such as jet formation and radiation emission from compact objects, making it essential for interpreting high-energy phenomena near BHs. Recent studies on geodesic motion in modified BH spacetimes have demonstrated rich and complex dynamics, particularly in scenarios involving exotic matter fields and alternative gravity theories.

We employ the Lagrangian method to discuss geodesic motion and analyze the outcomes. The Lagrangian density function using the metric (\ref{bb1}) is given by
\begin{equation}
   \mathcal{L}=\frac{1}{2}\,\left[-f(r)\,\left(\frac{dt}{d\tau}\right)^2+\frac{1}{f(r)}\,\left(\frac{dr}{d\tau}\right)^2+r^2\,\left(\frac{d\theta}{d\tau}\right)^2+r^2\,\sin^2 \theta\,\left(\frac{d\phi}{d\tau}\right)^2\right],\label{mm1}
\end{equation}
where $\tau$ represents an affine parameter.

Considering the geodesic motion in the equatorial plane defined by $\theta=\pi/2$ and $\frac{d\theta}{d\tau}=0$, the Lagrangian density function (\ref{mm1}) reduces as
\begin{equation}
   \mathcal{L}=\frac{1}{2}\,\left[-f(r)\,\left(\frac{dt}{d\tau}\right)^2+\frac{1}{f(r)}\,\left(\frac{dr}{d\tau}\right)^2+r^2\,\left(\frac{d\phi}{d\tau}\right)^2\right].\label{mm1a}
\end{equation}

Since the chosen spacetime is static and spherically symmetric, it admits two Killing vector fields: $\xi_{(t)} \equiv \partial_{t}$, corresponding to time translation symmetry, and $\xi_{(\phi)} \equiv \partial_{\phi}$, corresponding to rotational symmetry about the azimuthal axis. The conserved quantities associated with these symmetries are the energy $\mathrm{E}$ and the angular momentum $\mathrm{L}$ of test particles, respectively. Explicitly, these are given by
\begin{equation}
   \mathrm{E}=f(r)\,\left(\frac{dt}{d\tau}\right)\quad,\quad \mathrm{L}=r^2\,\frac{d\phi}{d\tau}.\label{mm2}
\end{equation}

Substituting Eq. (\ref{mm2}) into the Eq. (\ref{mm1}) and after rearranging, we find the equation of motion associate with the radial coordinate $r$ given by
\begin{equation}
   \left(\frac{dr}{d\tau}\right)^2+V_\text{eff}(r)=\mathrm{E}^2\label{mm3}
\end{equation}
which is equivalent to the one-dimensional equation of motion of a unit mass particle having energy $\mathrm{E}^2$ and the potential  $V_\text{eff}(r)$. This potential known as the effective potential governs the dynamics of test particles around the BH. The effective potential in our case is given by ($2\,\mathcal{L}=\varepsilon$)
\begin{equation}
   V_\text{eff}(r)=\frac{\mathrm{L}^2}{r^2}\,f(r)=\left(-\varepsilon+\frac{\mathrm{L}^2}{r^2}\right)\,\left(1-\alpha-\frac{2\,M}{r}+\frac{\lambda\,M^2}{r^4}-N\,r\right).\label{mm4}
\end{equation}
Here $\varepsilon=0$ for null geodesics and $-1$ for time-like geodesics.

From expression given in Eq. (\ref{mm4}), it is clear that the effective potential for null or time-like geodesics is influenced by several geometric and physical parameters. These include the CS parameter $\alpha$, the quantum deformation parameter $\lambda$, and the normalization constant $N$ of the quintessential field. Additionally, the conserved angular momentum $\mathrm{L}$ and the BH mass $M$ modify this effective potential.

\subsection{Photon Dynamics: Null Geodesic}

The study of photon geodesics plays a fundamental role in understanding BH physics, particularly in determining key observational features such as the photon sphere and BH shadow. The effective potential for null geodesics around a BH is a tool to describe the behavior of light in curved spacetime near a BH. It helps us understand the motion of light around BHs, particularly in regions where the spacetime is highly curved, such as near the event horizon. The effective potential decides whether photons will be able to escape from the BH's gravitational pull (if they are outside the photon sphere) or whether they will be captured by the BH. With the help of the effective potential, one can determine the photon sphere radius including the BH shadow and others.

For null geodesic, $\varepsilon=0$, the effective potential from Eq. (\ref{mm4}) reduces as,
\begin{equation}
   V_\text{eff}(r)=\frac{\mathrm{L}^2}{r^2}\,\left(1-\alpha-\frac{2\,M}{r}+\frac{\lambda\,M^2}{r^4}-N\,r\right).\label{cc1}
\end{equation}

\begin{figure}[ht!]
   \centering
   \subfloat[$\lambda=0.5\,M^2,\,N=0.02/M$]{\centering{}\includegraphics[width=0.3\linewidth]{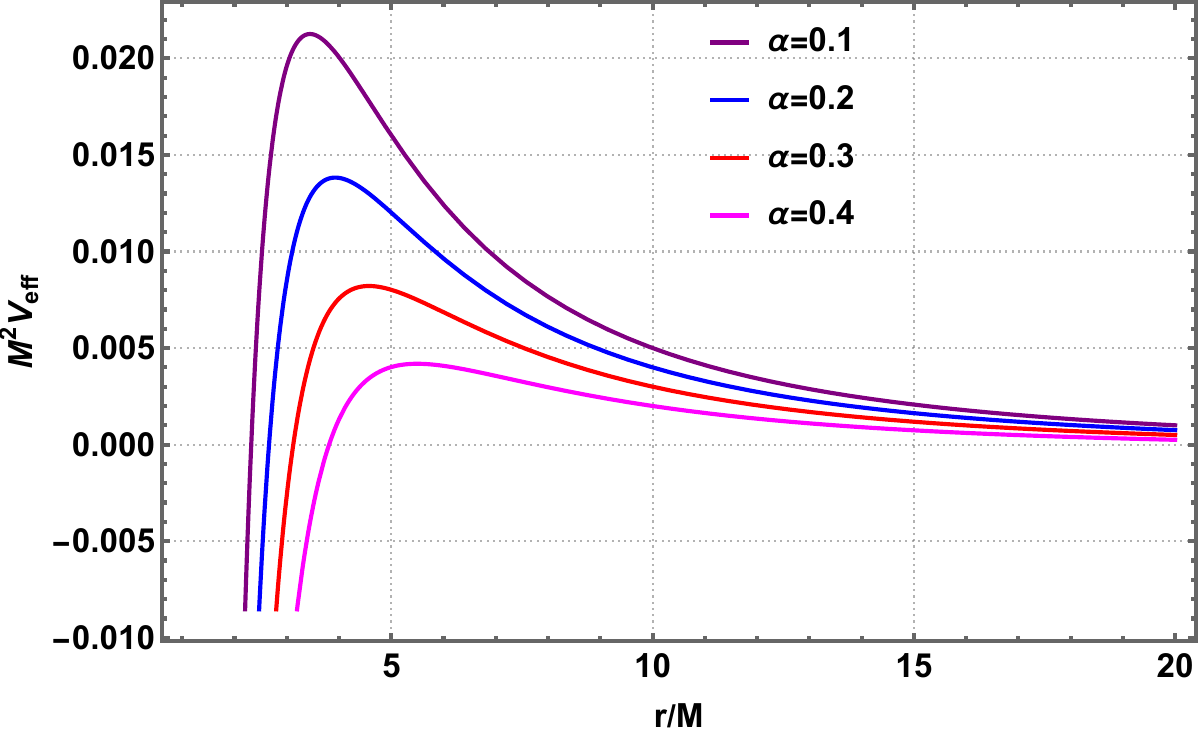}}\quad\quad
   \subfloat[$\alpha=0.5,\,N=0.03/M$]{\centering{}\includegraphics[width=0.3\linewidth]{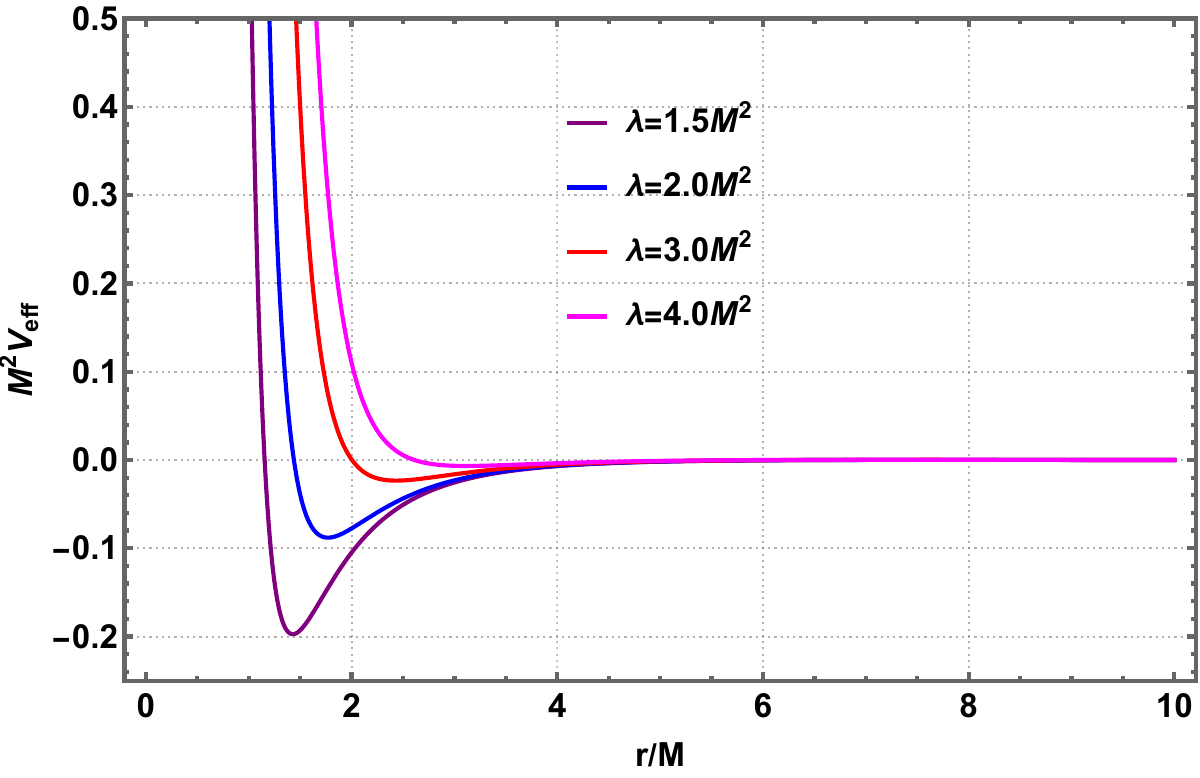}}\quad\quad
   \subfloat[$\alpha=0.1,\lambda=0.5\,M^2$]{\centering{}\includegraphics[width=0.3\linewidth]{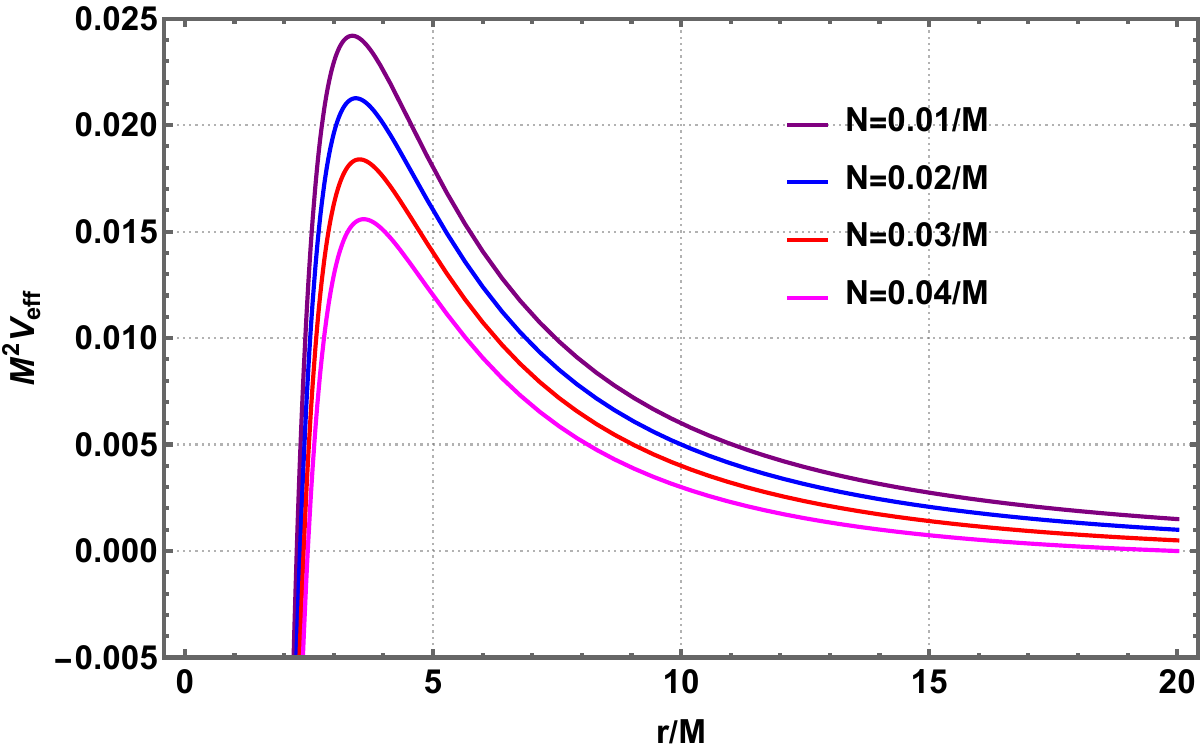}}\\
   \subfloat[$N=0.02/M$]{\centering{}\includegraphics[width=0.3\linewidth]{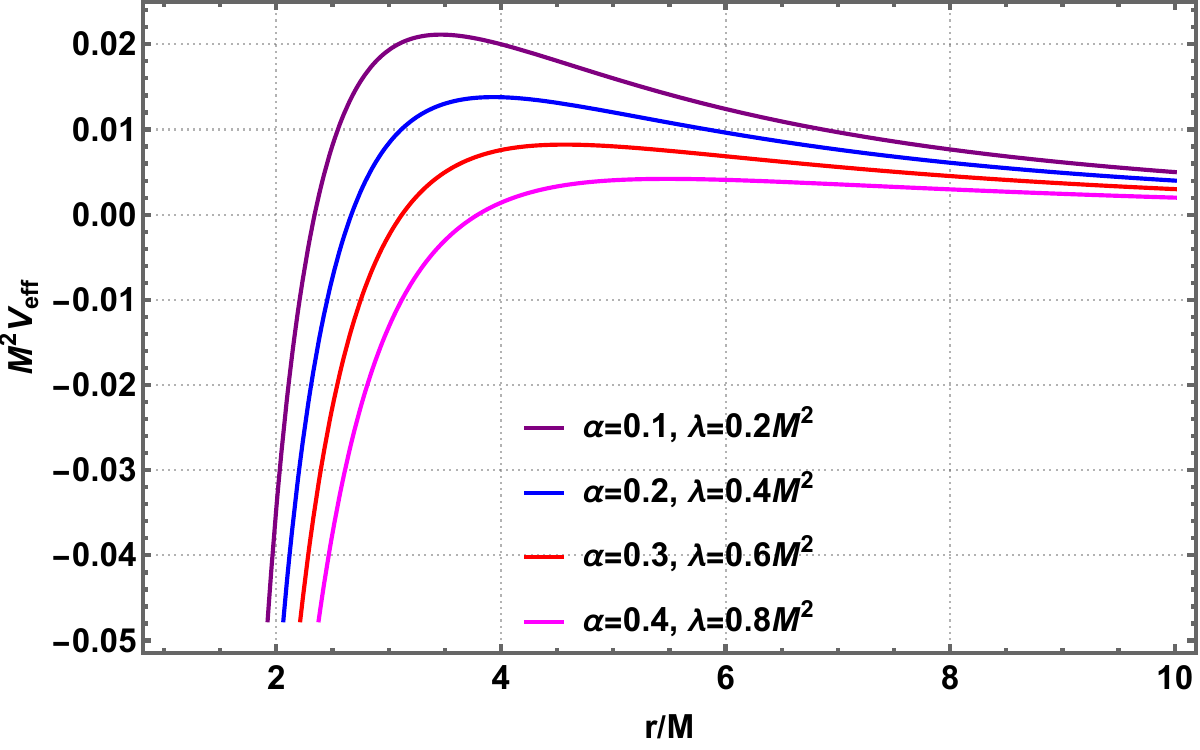}}\quad\quad
   \subfloat[$\lambda=0.5\,M^2$]{\centering{}\includegraphics[width=0.3\linewidth]{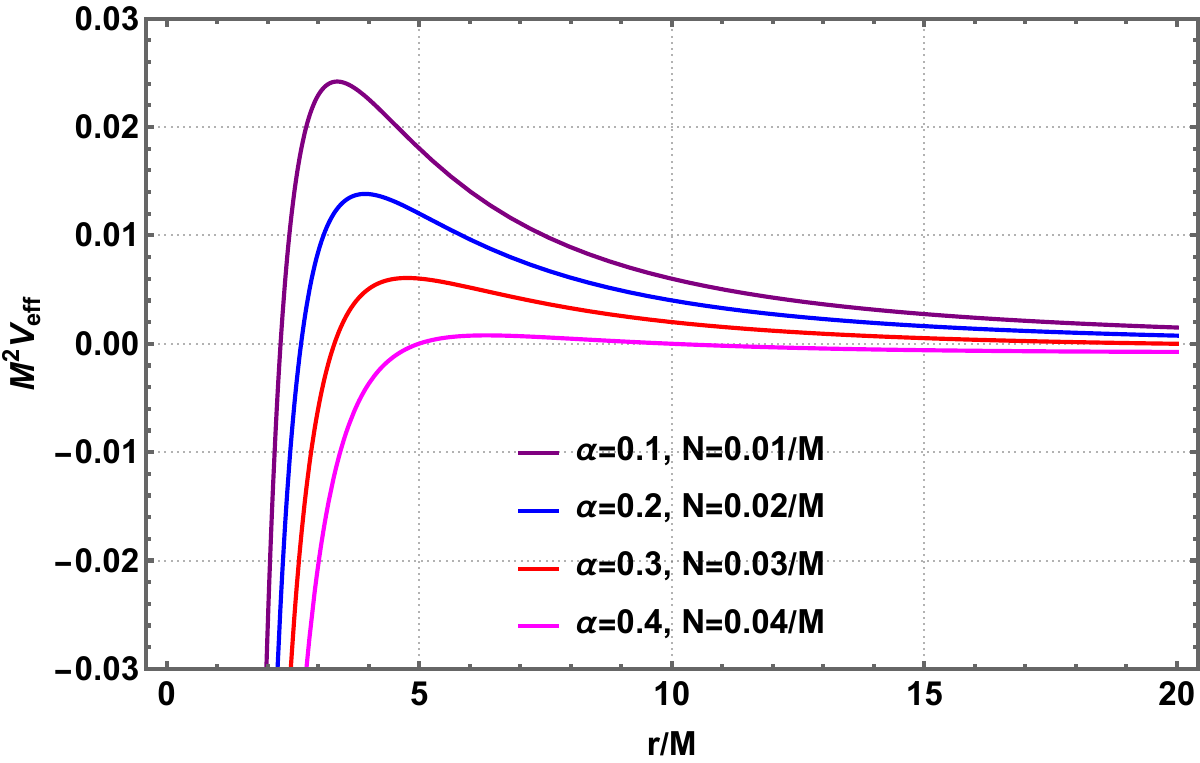}}\quad\quad
   \subfloat[\mbox{all varies}]{\centering{}\includegraphics[width=0.3\linewidth]{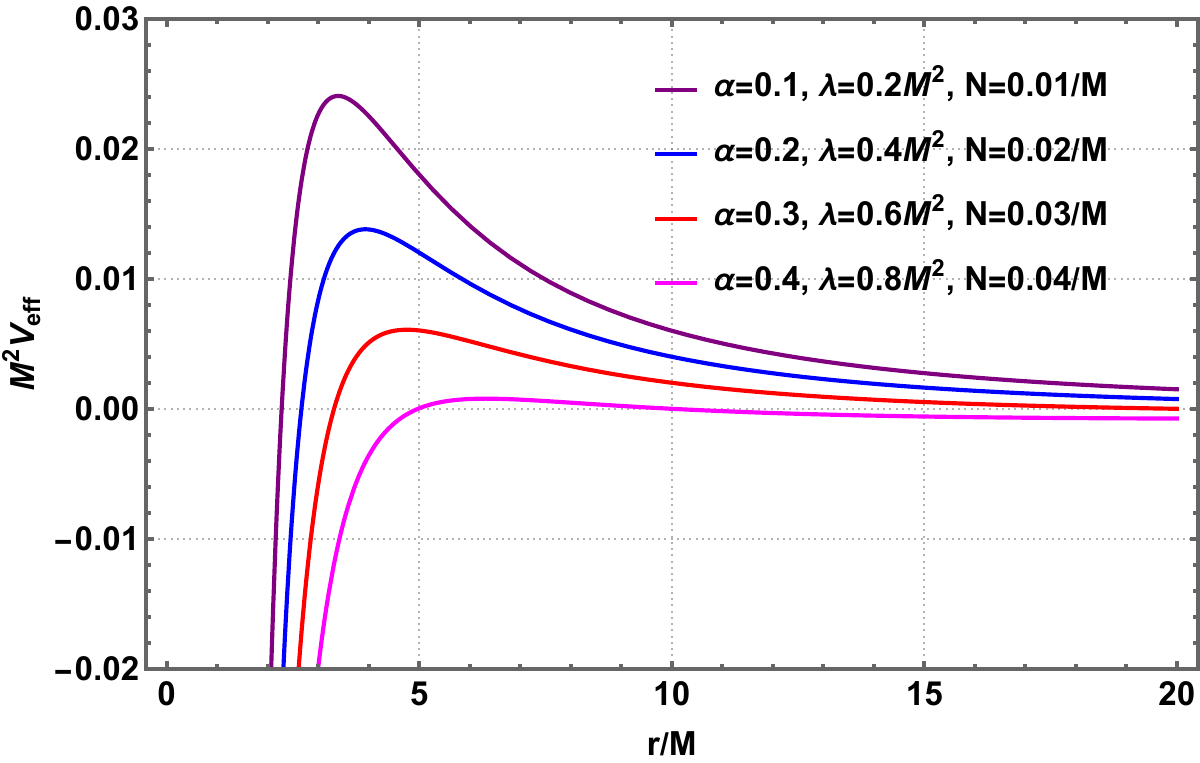}}\
   \caption{Behavior of the effective potential for null geodesics as a function of $r$ for different values of the string parameter $\alpha$, deformation parameter $\lambda$, the normalization constant $N$ of the field, and their combinations. Here, we set the angular momentum $\mathrm{L}=1$.}
   \label{fig:potential-null}
\end{figure}

\begin{figure}[ht!]
   \centering
   \subfloat[$\lambda=0.5\,M^2,N=0.01/M$]{\centering{}\includegraphics[width=0.3\linewidth]{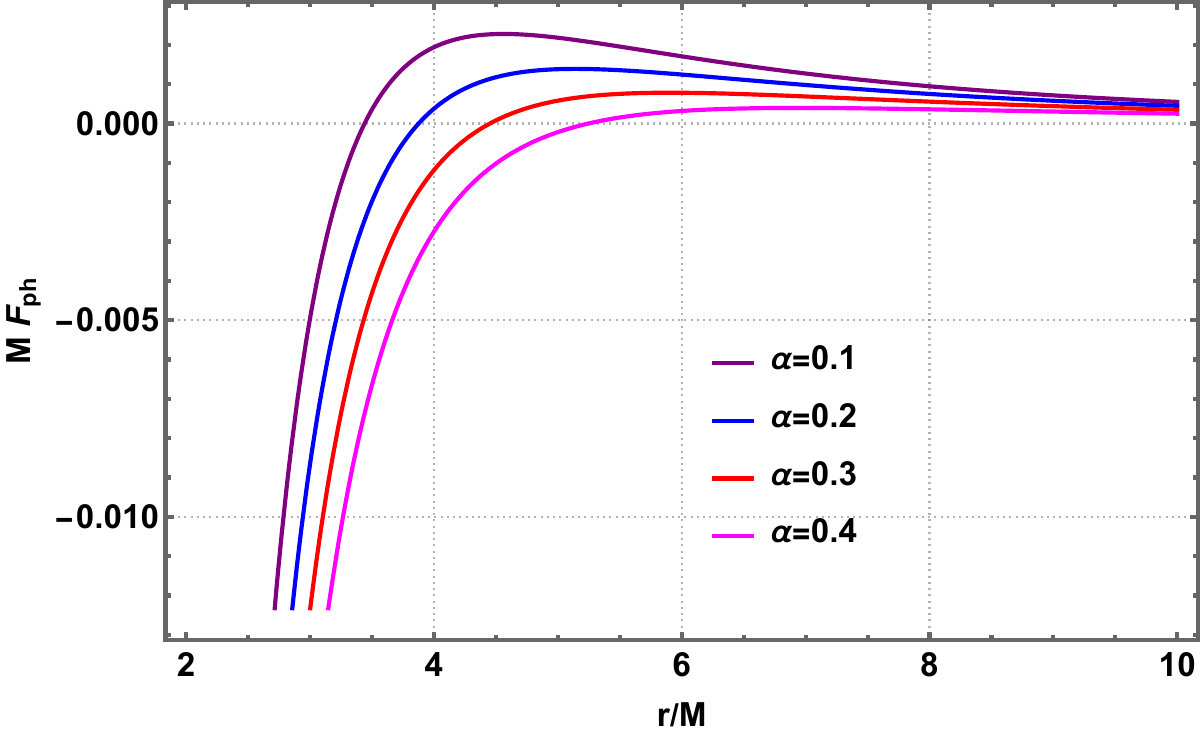}}\quad\quad
   \subfloat[$\alpha=0.5,\,N=0.03/M$]{\centering{}\includegraphics[width=0.3\linewidth]{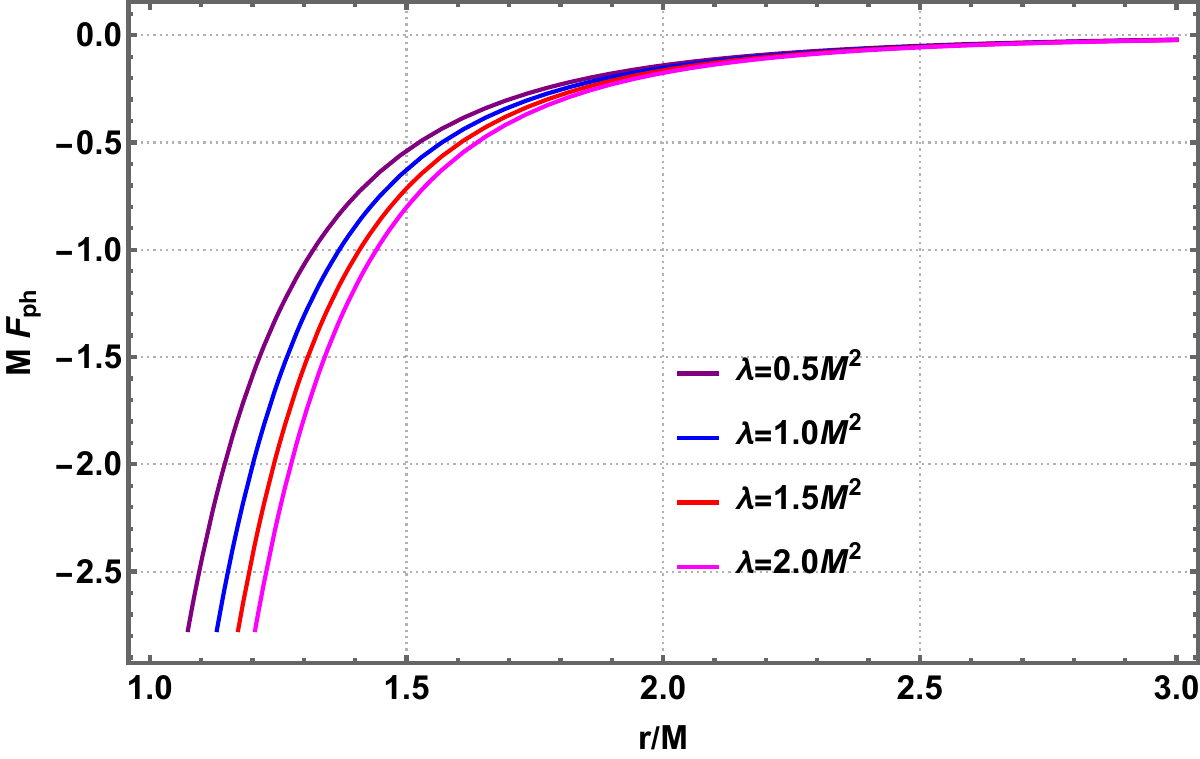}}\quad\quad
   \subfloat[$\alpha=0.1,\lambda=0.5\,M^2$]{\centering{}\includegraphics[width=0.3\linewidth]{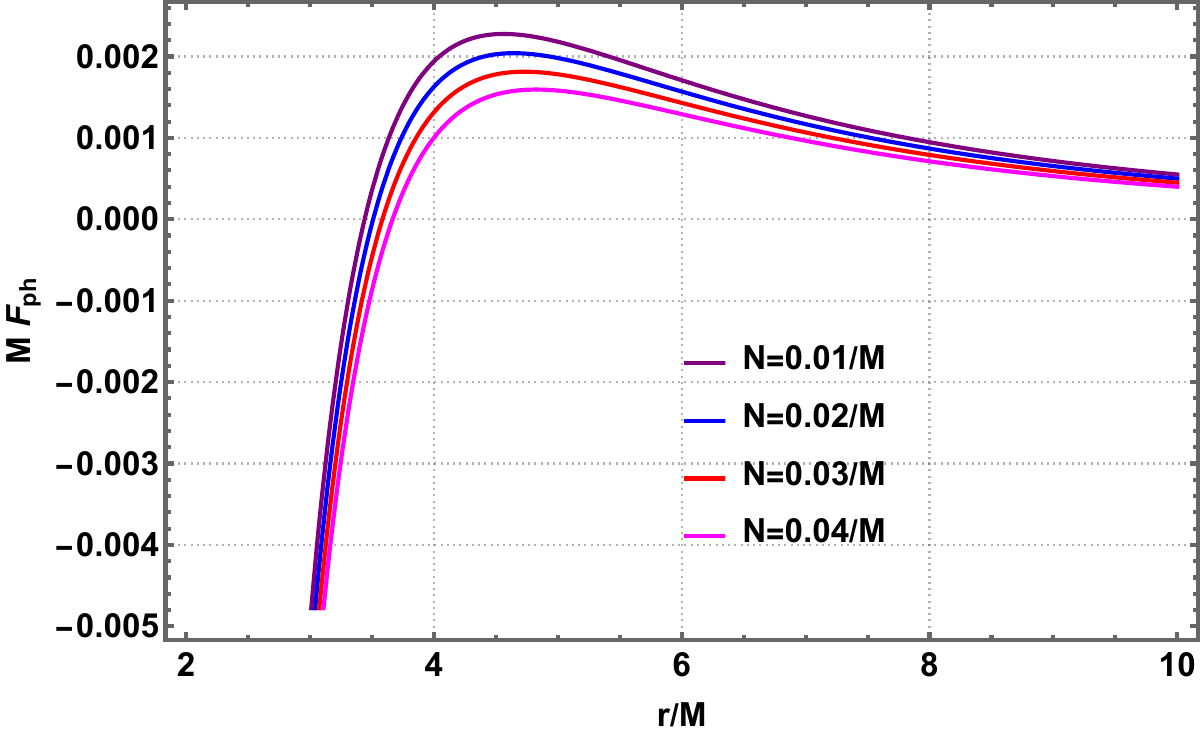}}\\
   \subfloat[$N=0.01/M$]{\centering{}\includegraphics[width=0.3\linewidth]{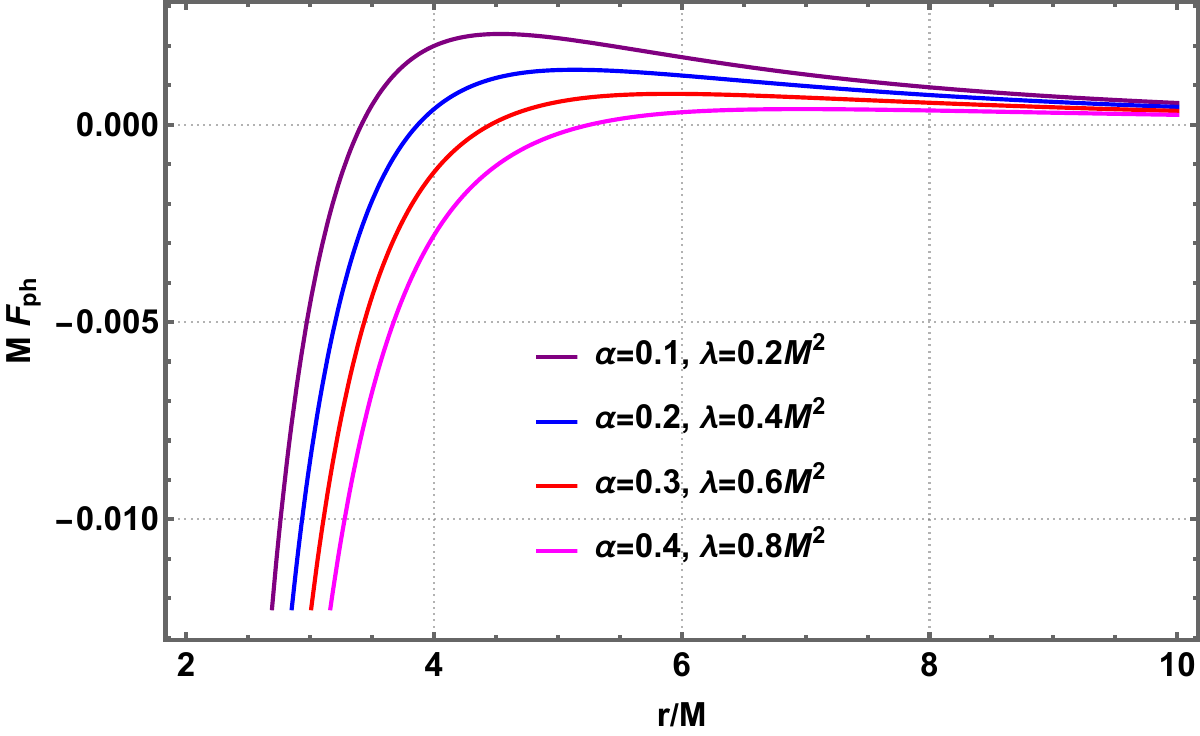}}\quad\quad
   \subfloat[$\lambda=0.05\,M^2$]{\centering{}\includegraphics[width=0.3\linewidth]{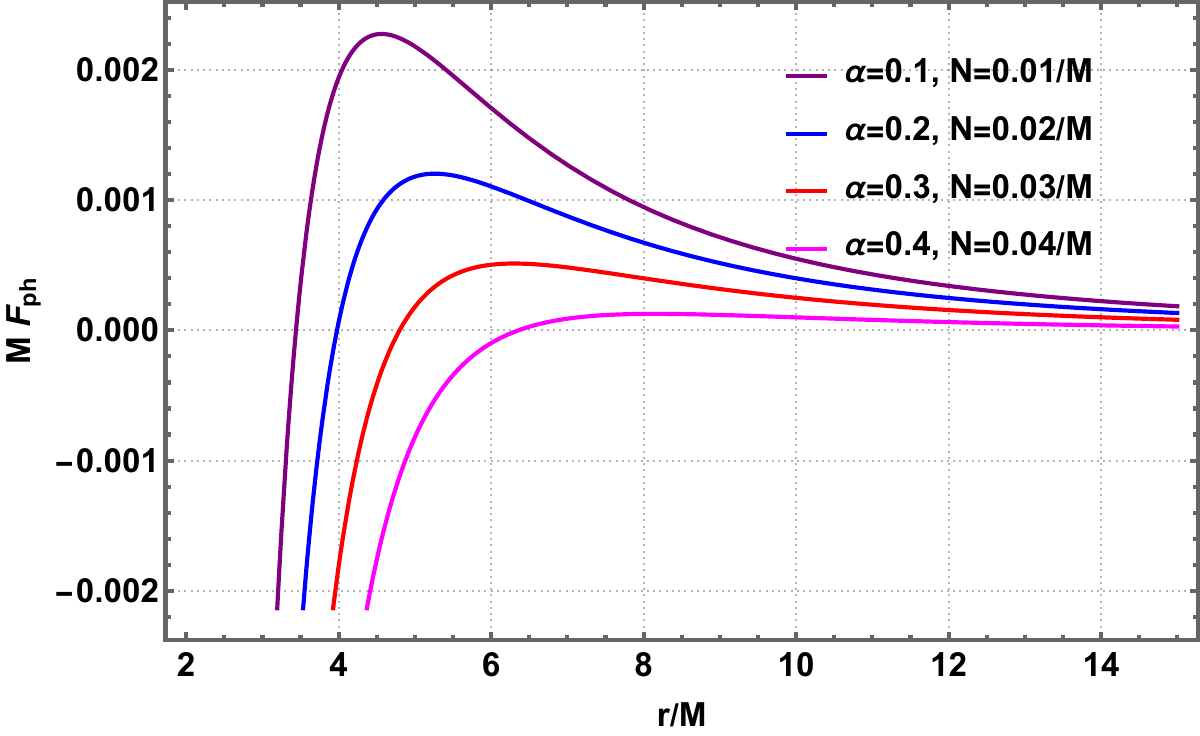}}\quad\quad
   \subfloat[\mbox{all varies}]{\centering{}\includegraphics[width=0.3\linewidth]{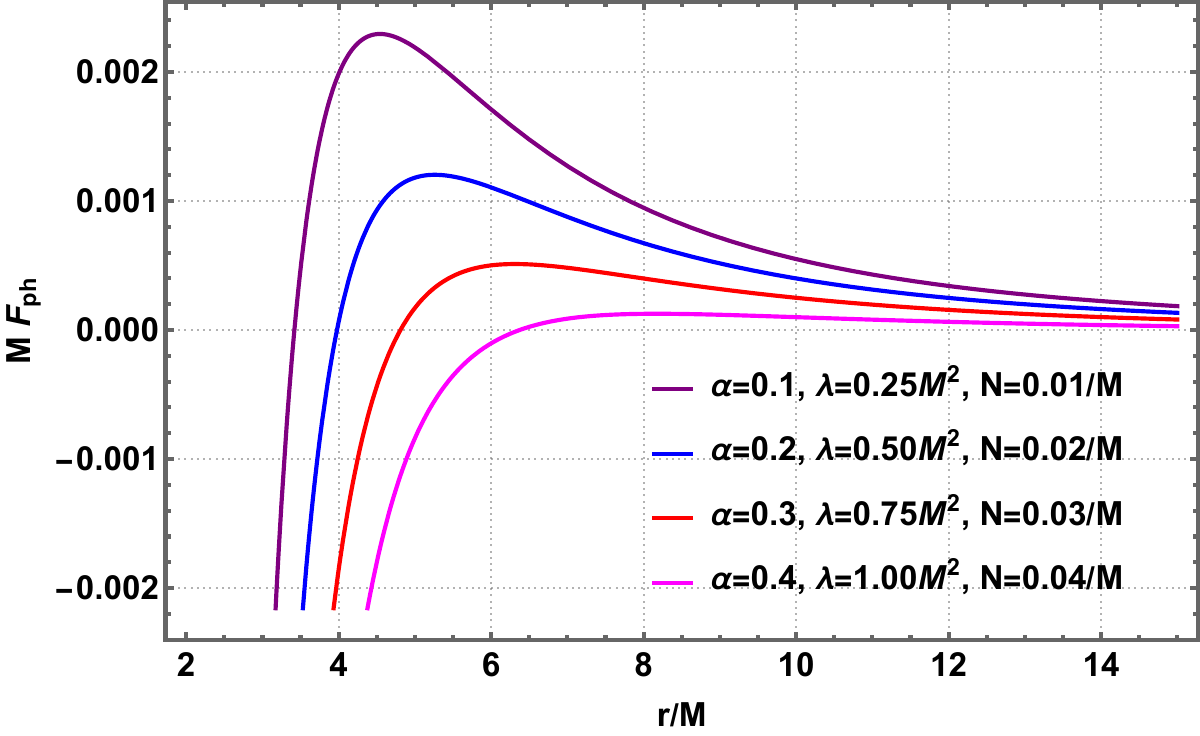}}\
   \caption{Behavior of the force on photon particles as a function of $r$ for different values of the string parameter $\alpha$, deformation parameter $\lambda$, the normalization constant $N$ of the field, and their combinations. Here, we set $\mathrm{L}/M^2=1$.}
   \label{fig:force}
\end{figure}

With the help of effective potential, we now study the dynamics of photon particles in the given gravitational field and show how various parameters affect the effective radial force experienced by the photon particles in the vicinity of the BH. Using the effective potential  given in Eq. (\ref{cc1}), we can determine the effective radial force on as,
\begin{equation}
   \mathrm{F}_\text{ph}=-\frac{1}{2}\,\frac{dV_\text{eff}}{dr}=\frac{\mathrm{L}^2}{r^3}\,\left(1-\alpha-\frac{3\,M}{r}+\frac{3\,\lambda\,M^2}{r^4}-\frac{N\,r}{2}\right).\label{cc2}
\end{equation}

From the above expression Eq. (\ref{cc2}), it becomes clear that the effective radial force experienced by the photon particles is influenced by several factors. These include the string cloud parameter $\alpha$, the quantum deformation parameter $\lambda$, the quintessential normalization constant $N$. Additionally, the conserved angular momentum $\mathrm{L}$ and the BH mass $M$ changes this radial force. In the limit where $\lambda=0$ and $N=0$-corresponding to the absence of quantum deformation and the quintessential field, the above result (\ref{cc2}) reduces to that of the Letelier BH solution, which further simplifies to the standard Schwarzschild BH result when $\alpha=0$.

 Figure~\ref{fig:potential-null} shows the radial behavior of the effective potential
$M^{2}V_{\text{eff}}$ governing null geodesics for different choices of the spacetime
parameters. Panel~\ref{fig:potential-null}(a) illustrates the influence of the CS
parameter $\alpha$, panel~\ref{fig:potential-null}(b) displays the effect of the quantum
deformation parameter $\lambda$, and panel~\ref{fig:potential-null}(c) highlights the
role of the QF parameter $N$. The combined impact of these parameters is shown
in panels~\ref{fig:potential-null}(d)--\ref{fig:potential-null}(f).
In all cases, increasing one or more of the parameters steadily lowers
the effective potential profile for null geodesics. This reduction modifies the height
and location of the potential barrier associated with circular photon orbits, thereby
affecting the allowed radial motion of light rays. Consequently, the geometric
parameters, singly and together, shape the
propagation of null geodesics and determining the structure of photon trajectories in
the underlying gravitational field.
Figure~\ref{fig:force} presents the behavior of mass times the effective radial force 
$M \mathrm{F}_{\text{ph}}$ acting on photon particles as a function of the dimensionless
radial coordinate $r/M$, for different values of the model parameters. Panel~\ref{fig:force}(a) shows the effect of the CS parameter $\alpha$, panel~\ref{fig:force}(b) illustrates the influence of the quantum deformation parameter
$\lambda$, and panel~\ref{fig:force}(c) displays the role of the QF parameter
$N$. The combined effects of these parameters are shown in panels~\ref{fig:force}(d)--\ref{fig:force}(f).
Consistent with the behavior observed in Figure~\ref{fig:potential-null}, all panels
show the effective radial force falling in magnitude as one or
more of the parameters increase. This reduction indicates a weakening of the radial
influence experienced by photon trajectories while preserving the qualitative structure
of the force profile. These results demonstrate that the geometric parameters of the
spacetime modify the dynamics of photon particles in the given
gravitational field.

\subsubsection{\bf Photon Trajectories}

In this part, we study photon trajectories in the given gravitational field and see how the geometric and physical parameters affect the paths of photon particles.

The equation of orbit using (\ref{mm2}),  (\ref{mm3}) and (\ref{cc1}) is given by
\begin{equation}
   \left(\frac{1}{r^2}\,\frac{dr}{d\phi}\right)^2=\frac{1}{\beta^2}-\frac{1}{r^2}\,\left(1-\alpha-\frac{2\,M}{r}+\frac{\lambda\,M^2}{r^4}-N\,r\right),\label{mm5}
\end{equation}
where $\beta=\mathrm{L}/\mathrm{E}$ is the impact parameter for photon particles.

Transforming to a new variable via $u=\frac{1}{r}$ into Eq. (\ref{mm5}) and after simplification, we find
\begin{equation}
   \left(\frac{du}{d\phi}\right)^2+(1-\alpha)\,u^2=\frac{1}{\beta^2}+2\,M\,u^3-\lambda\,M^2\,u^6+N\,u.\label{mm6}
\end{equation}
In order to derive a second-order differential equation for photon trajectory, differentiating both sides of Eq. (\ref{mm6}) w. r. to $\phi$ and after simplification, we find the following differential equation:
\begin{equation}
   \frac{d^2u}{d\phi^2}+(1-\alpha)\,u=3\,M\,u^2-3\,\lambda\,M^2\,u^5+N/2.\label{mm7}
\end{equation}
Equation (\ref{mm7}) is a non-linear, second-order partial differential equation representing photon trajectory in the gravitational field produced by QOS BH model with a QF and surrounded by a CS.

\begin{figure}[ht!]
   \centering
   \subfloat[$\alpha=0.2$]{\centering{}\includegraphics[width=0.24\linewidth]{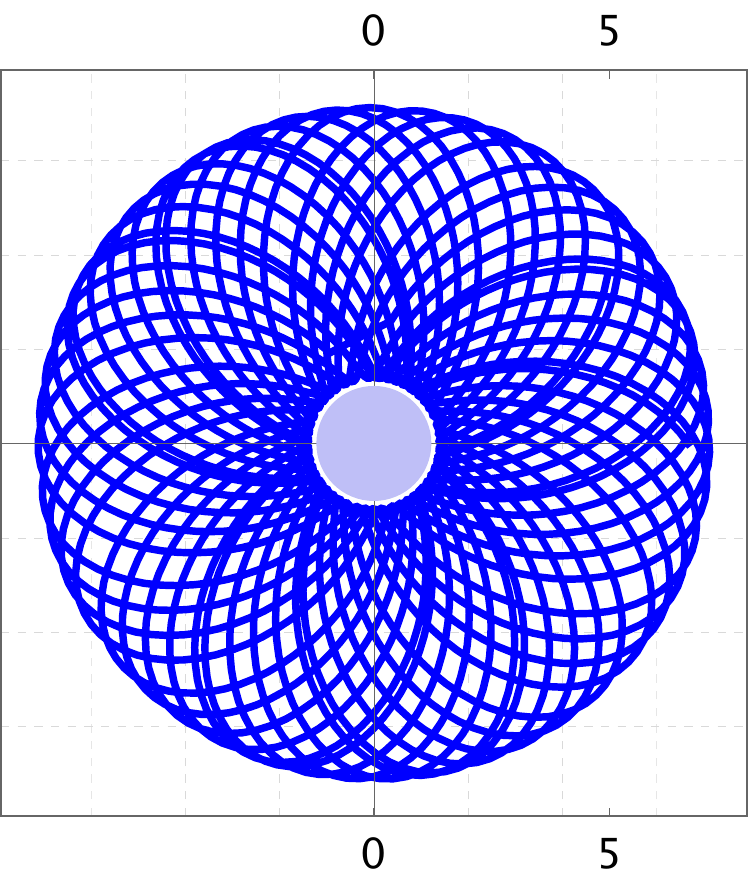}}\quad\quad
   \subfloat[$\alpha=0.4$]{\centering{}\includegraphics[width=0.24\linewidth]{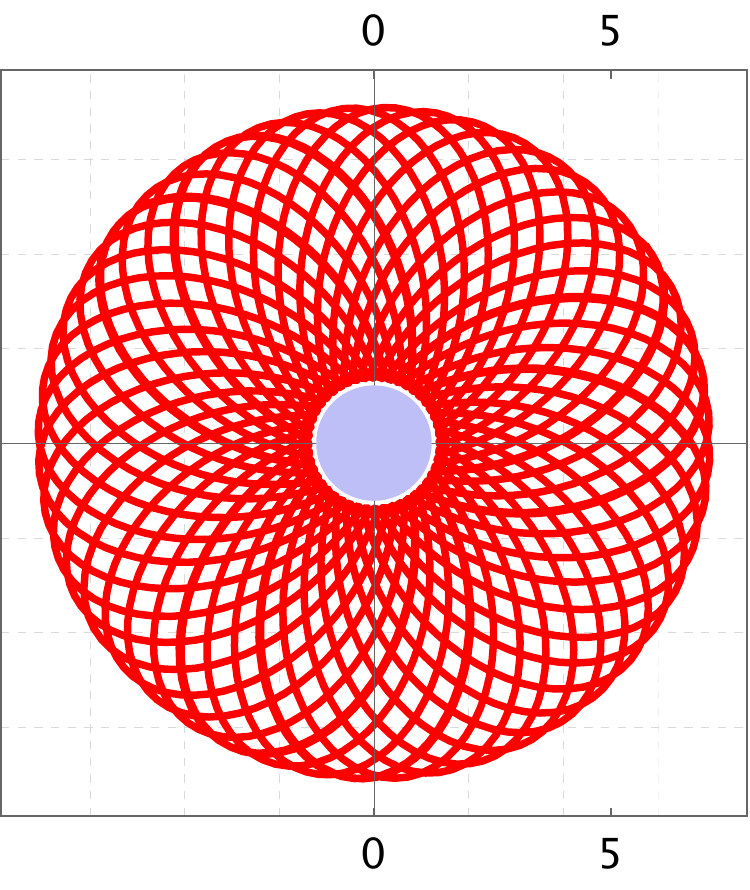}}\quad\quad
   \subfloat[$\alpha=0.6$]{\centering{}\includegraphics[width=0.24\linewidth]{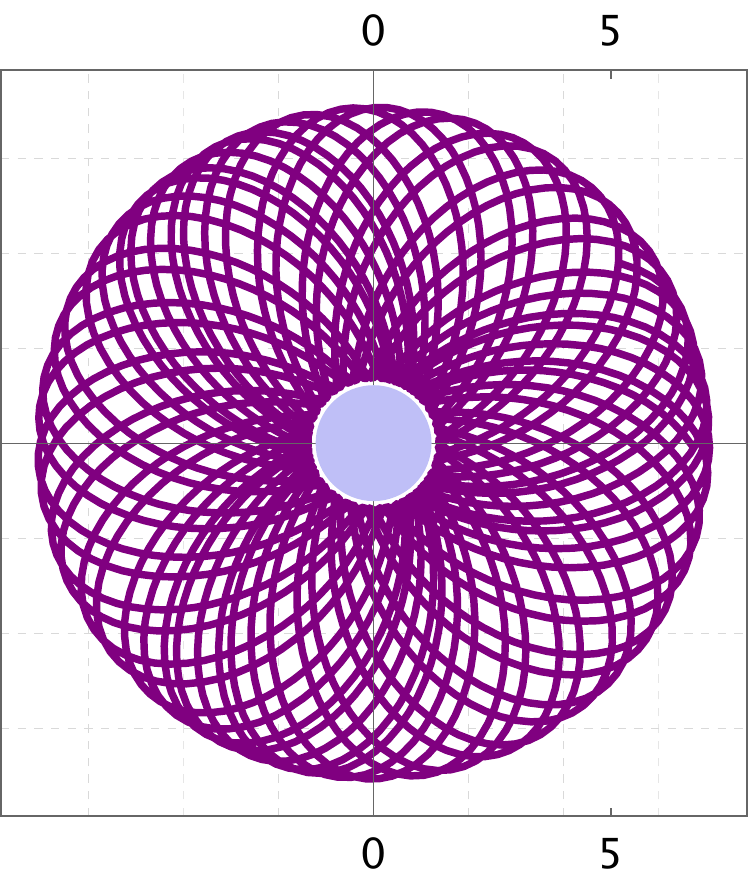}}
   \caption{Orbital trajectories projected onto the \( x\text{--}y \) plane for varying values of the CS parameter $\alpha$, while keeping the quantum deformation parameter $\lambda=0.5$, the BH mass $M=10$ and the normalization constant $N=0.02$ fixed. Here, we set the initial conditions $u(0)=0.15$ and $u'(0)=0.1$.}
   \label{fig:plot-1}
   \centering
   \subfloat[$\lambda=0.2$]{\centering{}\includegraphics[width=0.24\linewidth]{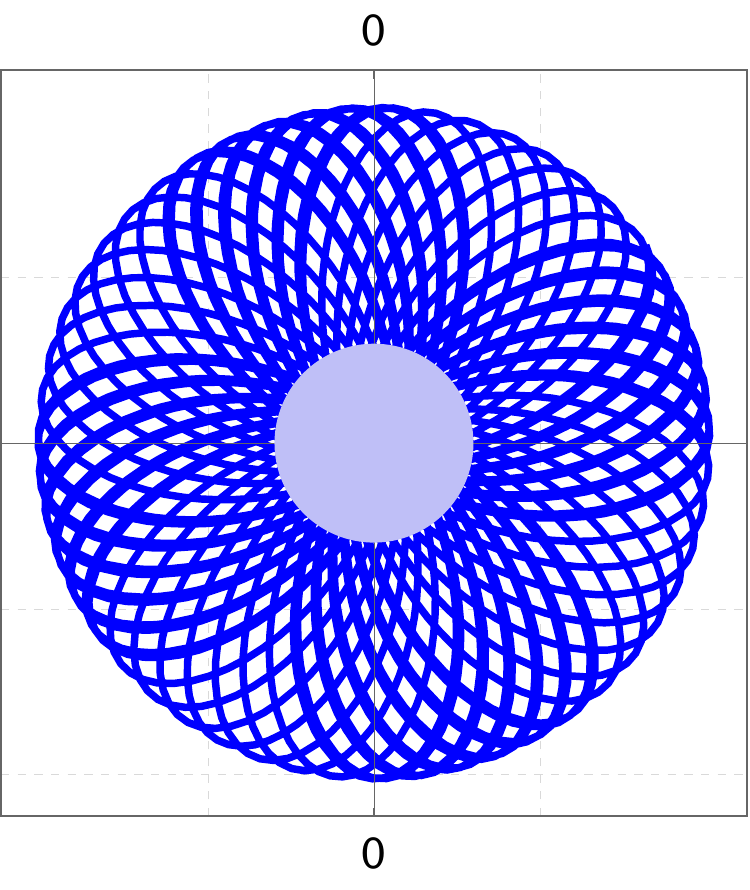}}\quad\quad
   \subfloat[$\lambda=0.4$]{\centering{}\includegraphics[width=0.24\linewidth]{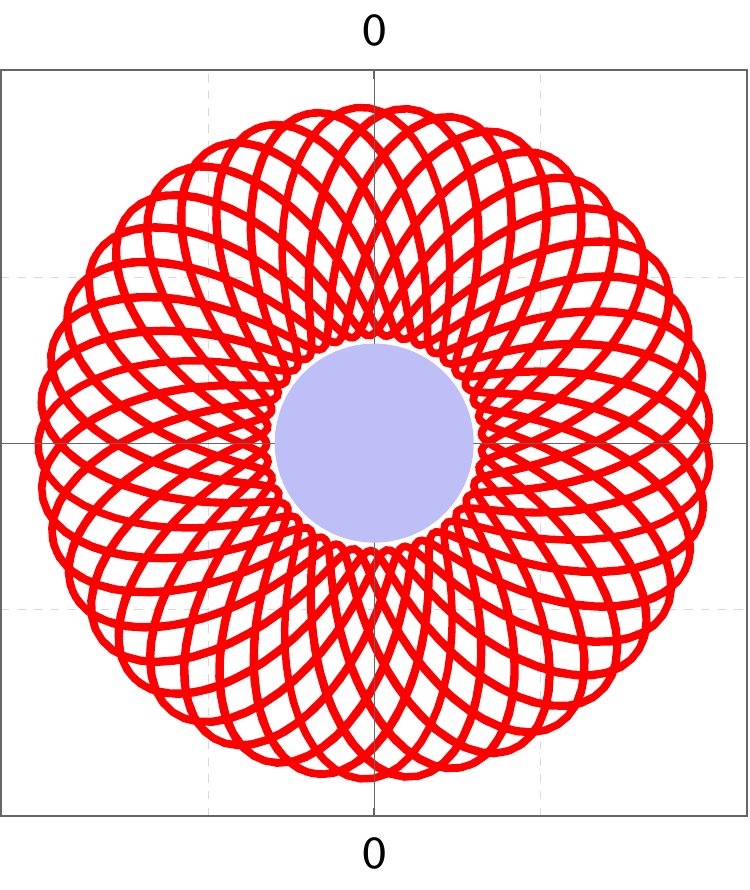}}\quad\quad
   \subfloat[$\lambda=0.6$]{\centering{}\includegraphics[width=0.24\linewidth]{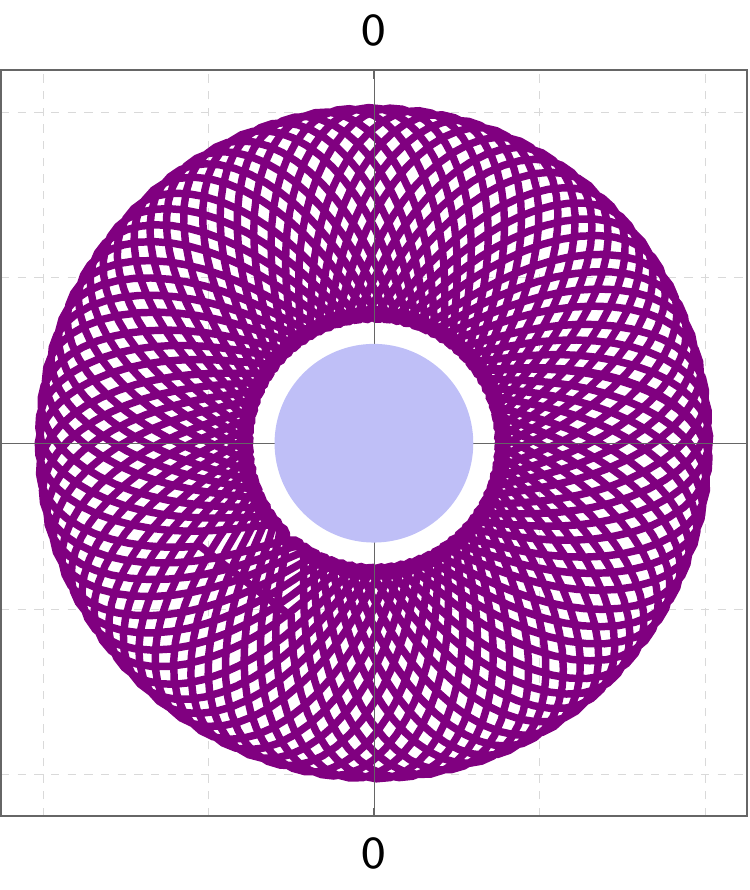}}
   \caption{Orbital trajectories projected onto the \( x\text{--}y \) plane for varying values of the quantum deformation parameter $\lambda$, while keeping the CS parameter $\alpha=0.1$, the BH mass $M=10$ and the normalization constant $N=0.02$ fixed. Here, we set the initial conditions $u(0)=0.15$ and $u'(0)=0.1$.}
   \label{fig:plot-2}
   \centering
   \subfloat[$N=0.03$]{\centering{}\includegraphics[width=0.24\linewidth]{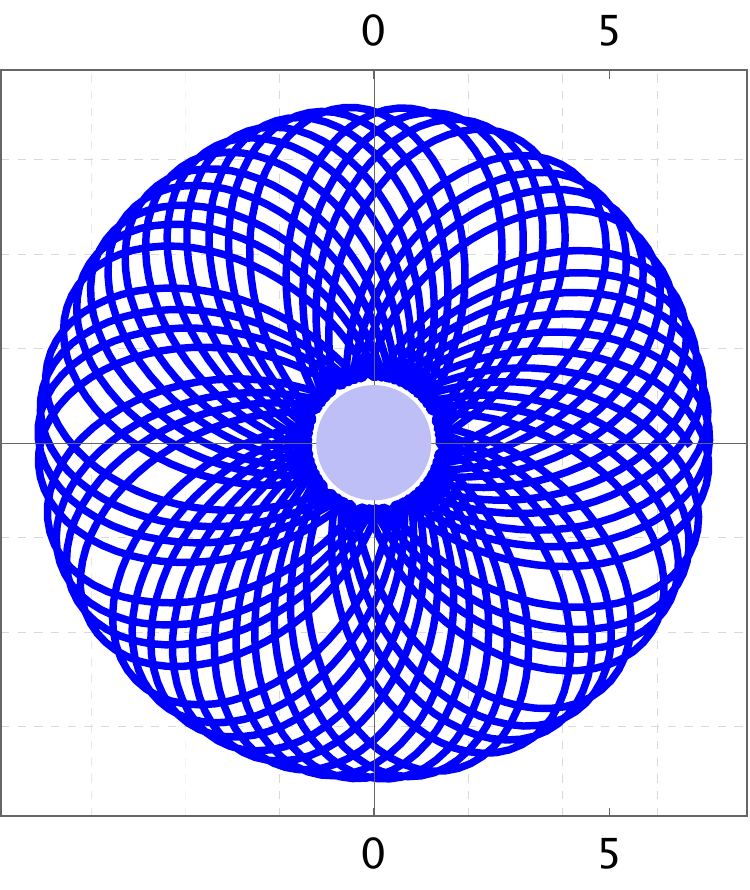}}\quad\quad
   \subfloat[$N=0.05$]{\centering{}\includegraphics[width=0.24\linewidth]{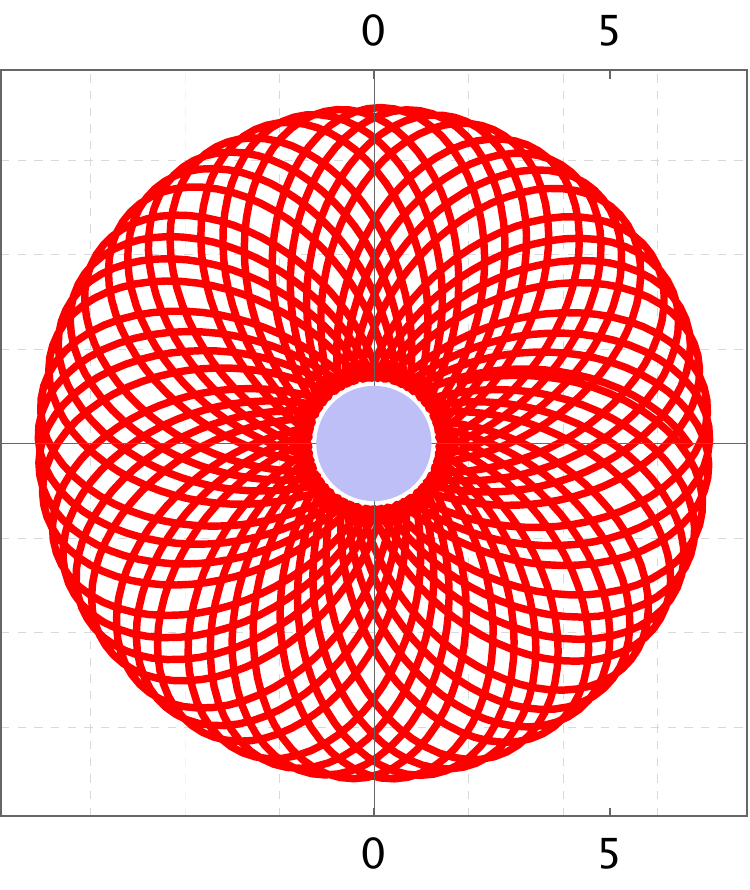}}\quad\quad
   \subfloat[$N=0.07$]{\centering{}\includegraphics[width=0.24\linewidth]{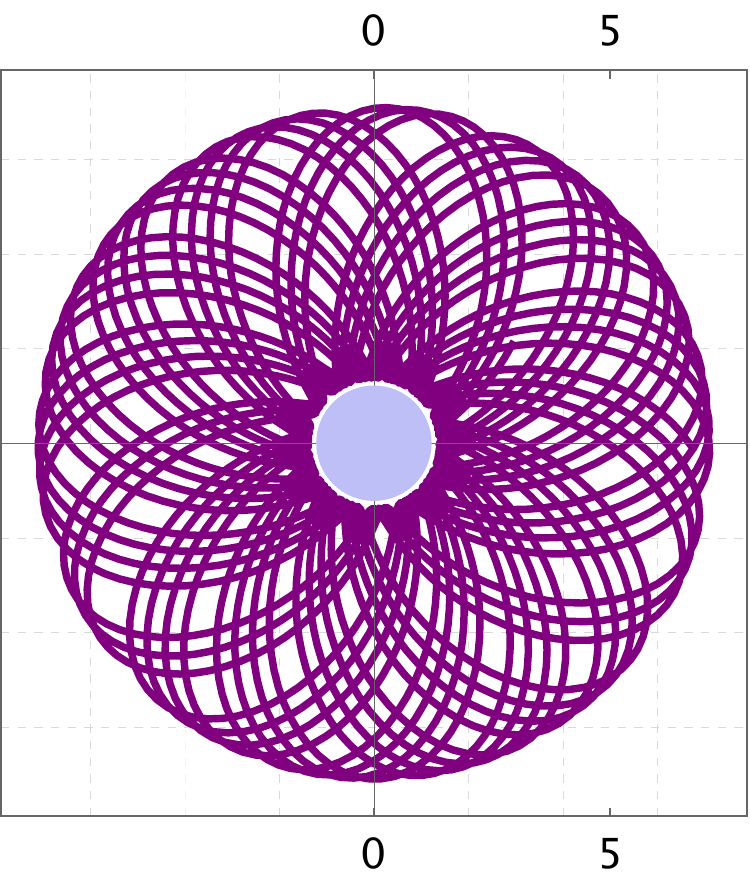}}
   \caption{Orbital trajectories projected onto the \( x\text{--}y \) plane for varying values of the normalization constant $N$ of the QF, while keeping the quantum deformation parameter $\lambda=0.5$, the CS parameter $\alpha=0.2$, the BH mass $M=10$ fixed. Here, we set the initial conditions $u(0)=0.15$ and $u'(0)=0.1$.}
   \label{fig:plot-3}
\end{figure}

We observe that the photon trajectory is influenced by various factors: the CS parameter $\alpha$, the quantum deformation parameter $\lambda$, the quintessential normalization constant $N$. In the limit where $\lambda=0$ and $N=0$-corresponding to the absence of quantum deformation and the quintessential field-the above result (\ref{mm7}) reduces to that of the Letelier BH solution, which further simplifies to the standard Schwarzschild BH result when $\alpha=0$.

Moreover, the radial motion along the time coordinate using Eqs. (\ref{mm2}) and (\ref{cc1}), is given by:
\begin{equation}
   \left(\frac{dr}{dt}\right)^2=f^2\,(r)\,\left(1-\frac{\beta^2}{r^2}\,f(r)\right),\quad\quad f(r)=1-\alpha-\frac{2\,M}{r}+\frac{\lambda\,M^2}{r^4}-N\,r.\label{mm8}
\end{equation}

The radial geodesic motion corresponds to the vanishing of the angular momentum ($\mathrm{L}=0$). In that case, there exist two possible trajectories: photons that escape away from the singularity, and those inevitably falling into it. From Eqs. (\ref{mm2}) and (\ref{mm3}), the geodesic equations governing the $t$ and $r$ coordinates can be derived as follows:
\begin{equation}
   \frac{dt}{d\tau}=\frac{\mathrm{E}}{f(r)}\quad\quad,\quad\quad \frac{dr}{d\tau}=\pm\,\mathrm{E}.\label{mm9}
\end{equation}
And the radial motion along the time coordinate is given by
\begin{equation}
   \frac{dr}{dt}=\pm\,f\,(r)\Rightarrow t(r(\tau))=\pm\,\int \frac{dr(\tau)}{\left(1-\alpha-\frac{2\,M}{r(\tau)}+\frac{\lambda\,M^2}{r^4(\tau)}-N\,r(\tau)\right)}.\label{mm10}
\end{equation}

We observe that the radial geodesic motion of photons is also influenced by the CS parameter $\alpha$, the quantum deformation parameter $\lambda$, the quintessential normalization constant $N$. In the limit where $\lambda=0$ and $N=0$-corresponding to the absence of quantum deformation and the quintessential field-the above results (\ref{mm9})--(\ref{mm10}) reduce to that of the Letelier BH solution, which further simplifies to the standard Schwarzschild BH result when $\alpha=0$.

Figures~\ref{fig:plot-1}, \ref{fig:plot-2}, and \ref{fig:plot-3} display the photon trajectories projected onto the $x$--$y$ plane, obtained by numerically integrating Eq.~(\ref{mm7}) with initial conditions $u(0)=0.15$ and $u'(0)=0.1$. In Figure~\ref{fig:plot-1}, we observe that increasing the CS parameter $\alpha$ (from 0.2 to 0.6) causes the photon orbits to wind more tightly around the central BH. This behavior arises because larger $\alpha$ values enhance the effective gravitational pull through the deficit angle contribution, making photon capture more likely. Figure~\ref{fig:plot-2} illustrates the effect of the quantum deformation parameter $\lambda$. As $\lambda$ increases from 0.2 to 0.6, the trajectories become progressively more open, indicating that quantum corrections weaken the gravitational confinement at small radii. This is consistent with the LQG-inspired regularization of the central singularity. Figure~\ref{fig:plot-3} shows the influence of the QF parameter $N$. Larger values of $N$ produce more loosely wound orbits, reflecting the repulsive dark energy contribution that becomes significant at larger radial distances. Collectively, these trajectory plots provide a visual confirmation of the competing effects encoded in the effective potential: the CS and mass terms attract photons toward the BH, while quantum deformation and QF tend to weaken this attraction.

\subsubsection{\bf Circular Orbits and stability or instability conditions}

In this section, we investigate the motion of photon particles in circular orbits around the BH, examining how geometric and physical parameters influence the photon sphere radius and the criteria for the stability or instability of null geodesics.

For circular null orbits of radius $r=r_c$, the conditions $\frac{dr}{d\tau}=0$ and $\frac{d^2r}{d\tau^2}=0$ must be satisfied. Using the equation of motion given in Eq. (\ref{mm3}), we find the following relations
\begin{equation}
   \mathrm{E}^2=V_\text{eff}(r).\label{cc3}
\end{equation}
And
\begin{equation}
    V'_\text{eff}(r)=0.\label{cc3a}
\end{equation}
Here prime denotes partial derivative w. r. t. $r$.

Substituting the effective potential $V_\text{eff}$ given in Eq. (\ref{cc1}) into the relation (\ref{cc3}), we find the critical impact parameter for photon particles at radius $r=r_c$. This parameter is given by
\begin{equation}
   \beta_c=\frac{\mathrm{L}_\text{ph}}{\mathrm{E}_\text{ph}}=\frac{r}{\sqrt{1-\alpha-\frac{2\,M}{r}+\frac{\lambda\,M^2}{r^4}-N\,r}}\Big{|}_{r=r_{c}}.\label{cc4}
\end{equation}

From the expression given in Eq.~(\ref{cc4}), we observe that the critical impact parameter for photon particles orbiting in the circular null geodesics at radius $r=r_c$ is influenced by the CS parameter $\alpha$, the quantum deformation parameter $\lambda$, the quintessential normalization constant $N$ as well as the BH mass $M$.

Depending on the relation between the impact parameter $\beta=\mathrm{L}/\mathrm{E}$ and the critical impact parameter $\beta_c$, photon particles are either captured or scattered. If $\beta<\beta_c$ the photon is captured by the BH. If $\beta>\beta_c$ it is deflected and escapes toward the outer region of the static domain, reaching an observer at finite $r_O<r_q$ rather than an observer at infinity, which this geometry does not admit. If $\beta=\beta_c$ the photon spirals onto an unstable circular orbit, the photon sphere.

We now turn our attention to the behavior of circular orbits, with particular emphasis on their stability or instability. This behavior of such orbits can be analyzed using a key physical quantity known as the Lyapunov exponent, which provides a measure of how rapidly a photon deviates from a circular orbit, thereby distinguishing between stable and unstable configurations. For null geodesics, the Lyapunov exponent can be expressed in terms of the second derivative of the effective potential $V_\text{eff}$, as follows \cite{VC}:
\begin{equation}
   \lambda^\text{null}_L=\sqrt{-\frac{V''_\text{eff}(r)}{2\,\dot{t}^2}},\label{cc6}
\end{equation}
where $\dot{t}$ is given in Eq. (\ref{mm2}).

Now, $V''_\text{eff}(r)$ in terms of the metric function $f(r)$ is given by
\begin{equation}
   V''_\text{eff}(r)=\frac{\mathrm{L}^2}{r^4}\,\left(r\,f''(r)-4\,r\,f'(r)+6\,f(r)\right).\label{cc7}
\end{equation}
For photon sphere, we have the relation $r\,f'(r)=2\,f(r)$ which we will discuss in detail later on. Therefore, the final expression from Eq. (\ref{cc7}) simplifies as
\begin{equation}
   V''_\text{eff}(r)=\frac{\mathrm{L}^2}{r^4}\,\left(r\,f''(r)-2\,f(r)\right).\label{cc8}
\end{equation}

Substituting Eqs. (\ref{mm2}) and (\ref{cc8}) in the equation (\ref{cc6}) and finally using the given metric function $f(r)$ results:
\begin{eqnarray}
   \lambda^\text{null}_L=\frac{1}{r}\,\sqrt{f(r)\,\left(f(r)-\frac{r^2\,f''(r)}{2}\right)}=\frac{1}{r}\,\sqrt{\left(1-\alpha-\frac{9\,\lambda\,M^2}{r^4}-N\,r\right)\,\left(1-\alpha-\frac{2\,M}{r}+\frac{\lambda\,M^2}{r^4}-N\,r\right)}.\label{cc9}
\end{eqnarray}

\begin{figure*}[tbhp]
   \centering
   \subfloat[$\lambda=0.5\,M^2,N=0.02/M$]{\centering{}\includegraphics[width=0.3\linewidth]{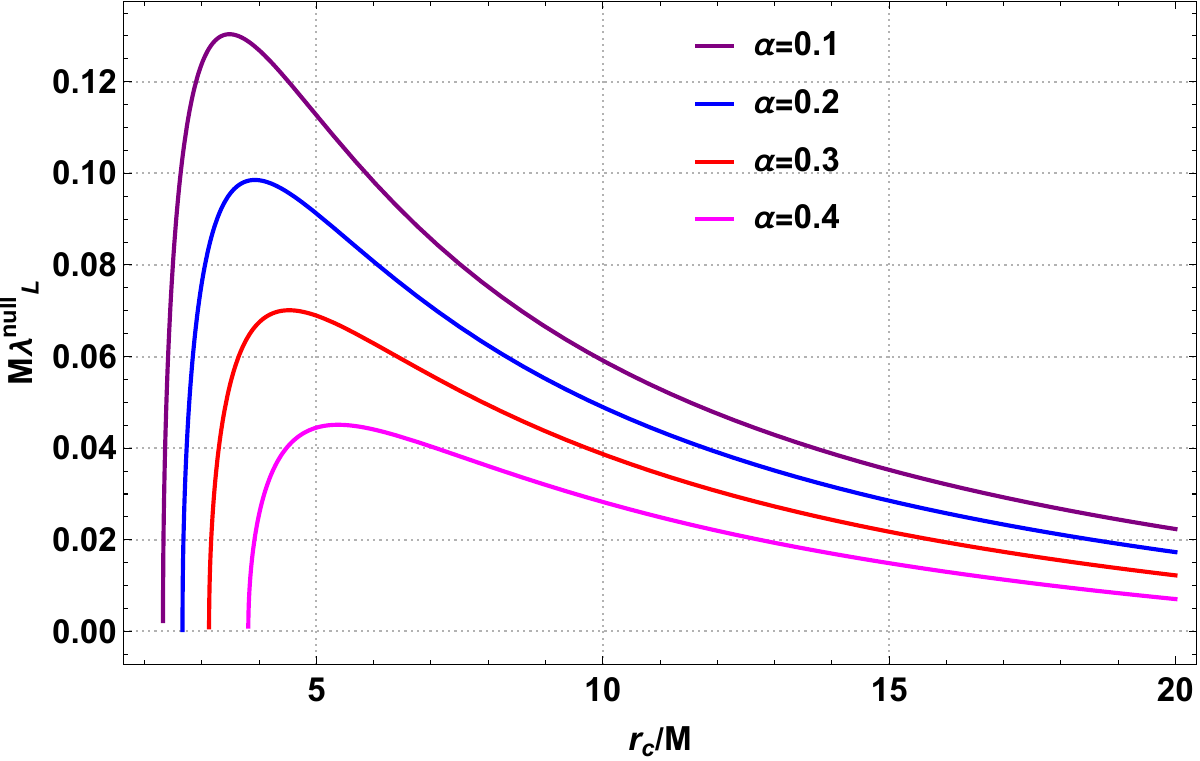}}\quad\quad
   \subfloat[$\alpha=0.1,\lambda=0.5\,M^2$]{\centering{}\includegraphics[width=0.3\linewidth]{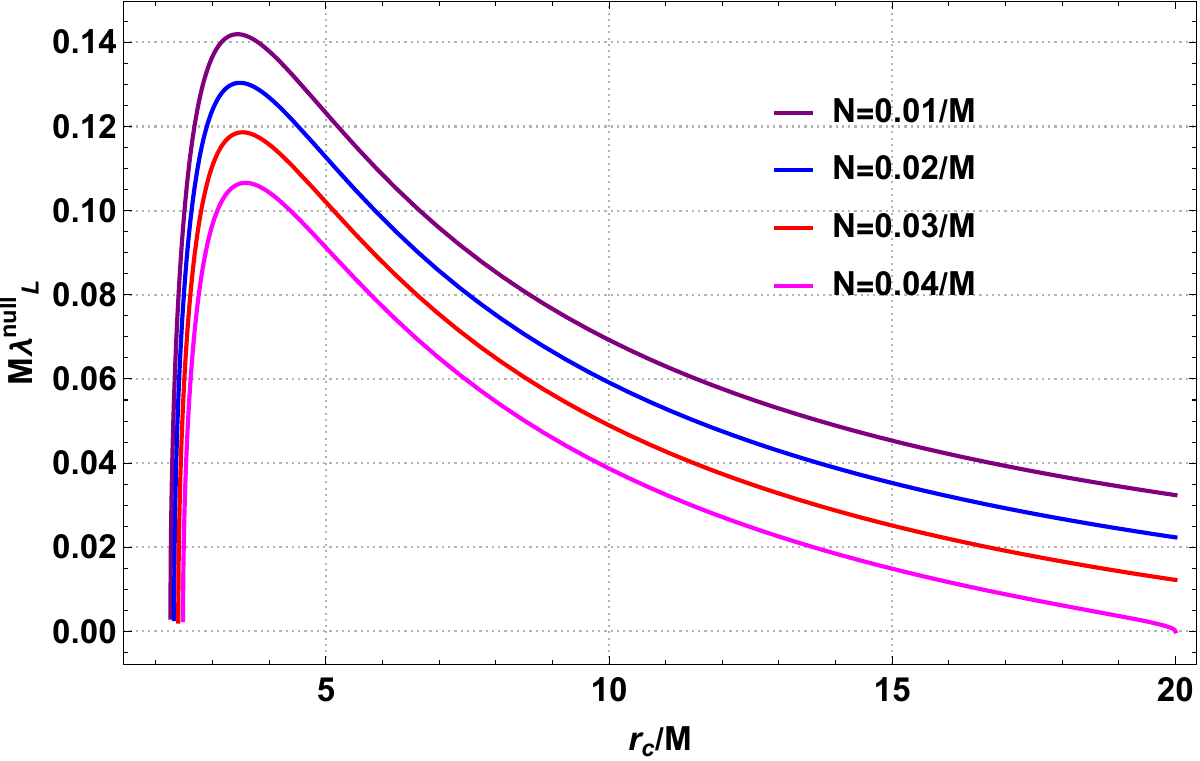}}\quad\quad
   \subfloat[$N=0.02/M$]{\centering{}\includegraphics[width=0.3\linewidth]{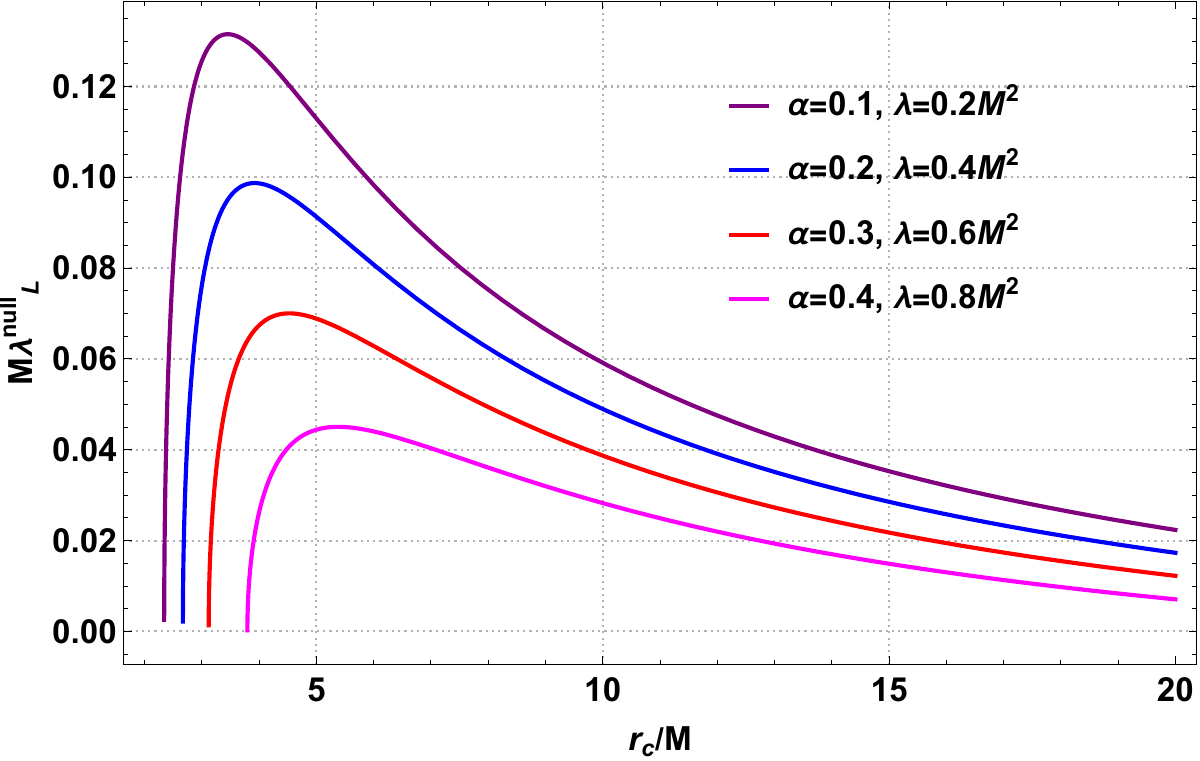}}\\
   \subfloat[$\lambda=0.2\,M^2$]{\centering{}\includegraphics[width=0.32\linewidth]{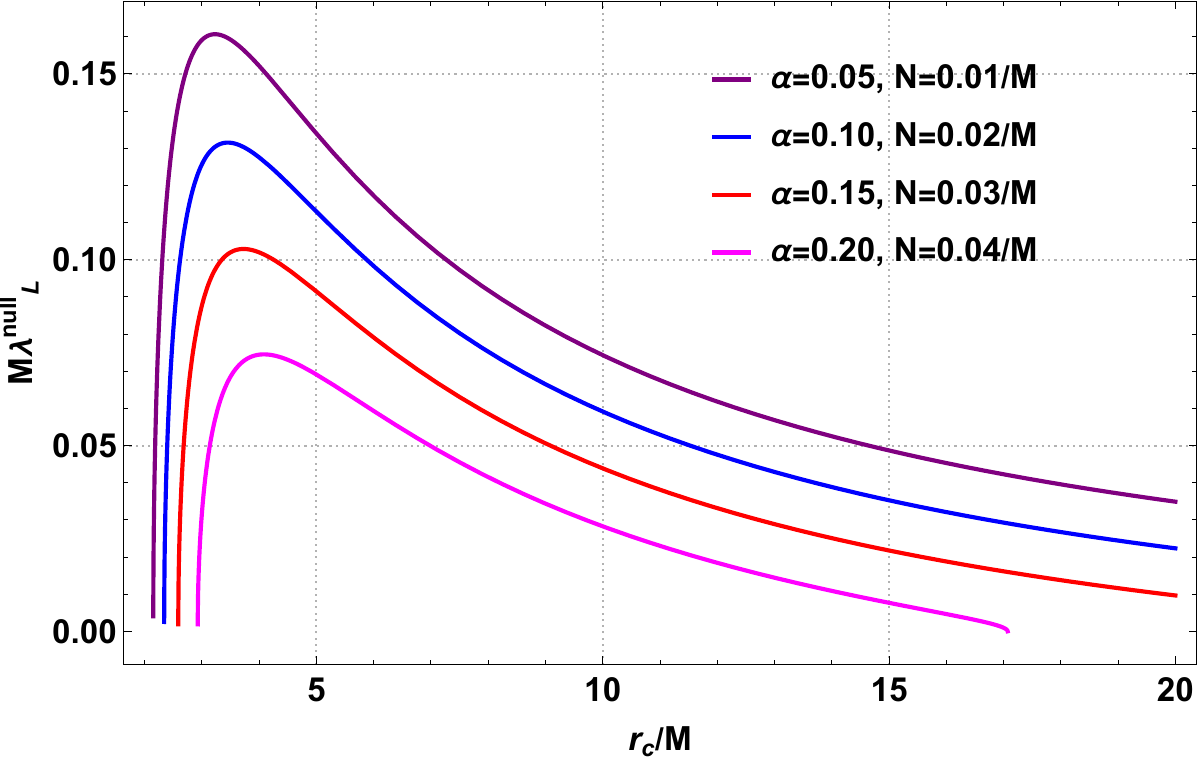}}\quad\quad
   \subfloat[\mbox{all varies}]{\centering{}\includegraphics[width=0.32\linewidth]{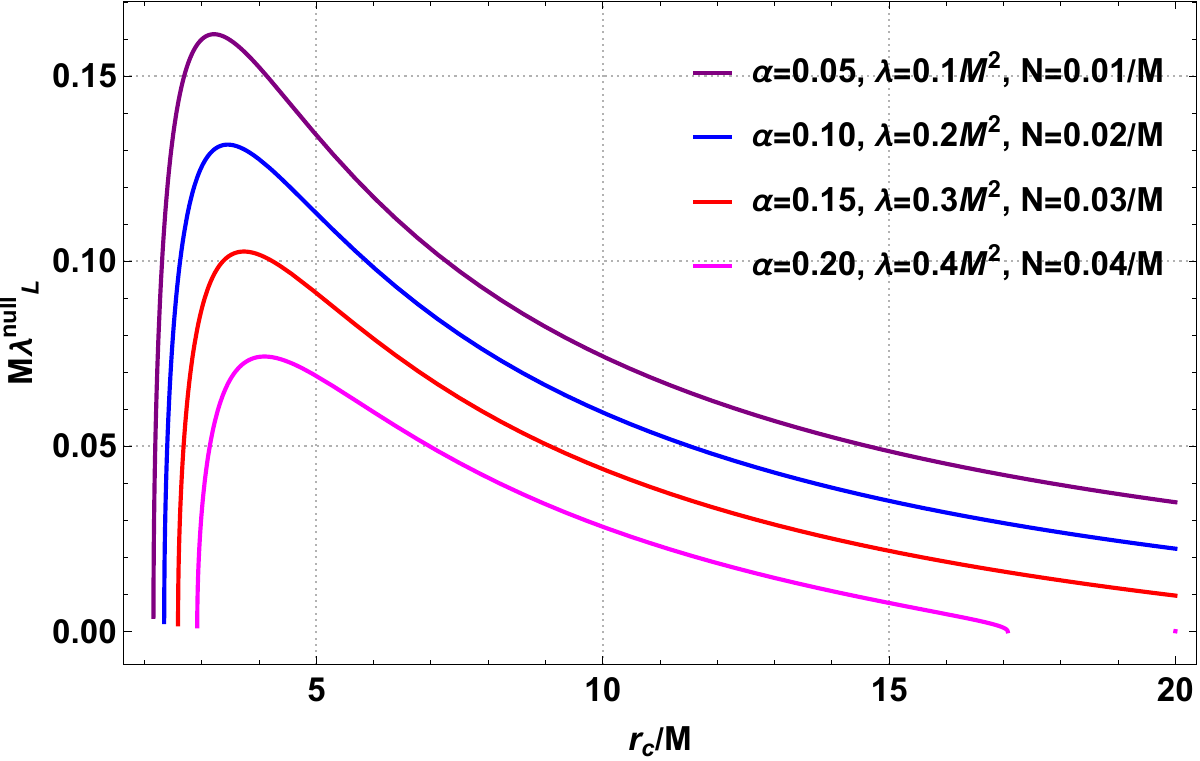}}
   \caption{Behavior of the Lyapunov exponent as a function of the circular null orbits radius $r_c$ for different values of the cosmic string parameter $\alpha$, the deformation parameter $\lambda$, the normalization constant $N$ of field, and their combination.}
   \label{fig:lyapunov}
\end{figure*}

The expression (\ref{cc9}) shows that the Lyapunov exponent for circular null geodesics is influenced by several parameters. These include the CS parameter $\alpha$, the quantum deformation parameter $\lambda$, the quintessential normalization constant $N$. Additionally, the BH mass $M$ alter the Lyapunov exponent. It is worth mentioning that in the limit where $\lambda=0$ and $N=0$-corresponding to the absence of quantum deformation and the quintessential field-the above result (\ref{cc9}) reduces to that of the Letelier BH solution, which further simplifies to the standard Schwarzschild BH result when $\alpha=0$.

 Figure~\ref{fig:lyapunov} depicts the radial dependence of the Lyapunov exponent term $M\lambda^{\text{null}}_{L}$ for different choices of the model parameters. Panel~\ref{fig:lyapunov}(a) shows the effect of varying the CS parameter $\alpha$, while panel~\ref{fig:lyapunov}(b) illustrates the influence of the quantum deformation parameter $\lambda$. The impact of the QF parameter $N$ is presented in panel~\ref{fig:lyapunov}(c). Panels~\ref{fig:lyapunov}(d) and \ref{fig:lyapunov}(e) display the combined effects of $\alpha$ and $N$, and of $\alpha$, $\lambda$, and $N$,
respectively. In all cases, the Lyapunov exponent remains positive over the relevant range of the radial coordinate $r_c/M$, confirming the instability of circular photon orbits. Increasing any of the parameters-either individually or in combination-steadily reduces the magnitude of $M\lambda^{\text{null}}_{L}$, indicating a weakening of the instability while preserving its qualitative behavior. These results
demonstrate that the geometric parameters of the spacetime significantly influence the degree of orbital instability, but do not alter the fundamentally unstable nature of circular photon orbits.

Now, we define the geodesic angular velocity (coordinate) as follows \cite{VC}:
\begin{equation}
   \Omega^\text{null}=\frac{\dot{\phi}}{\dot{t}}=\frac{f(r)}{r^2}\,\frac{\mathrm{L}}{\mathrm{E}}=\frac{\sqrt{f(r)}}{r}\Bigg{|}_{r=r_c},\label{cc10}
\end{equation}
where we have used the relation (\ref{cc3}).

Substituting the metric function $f(r)$ from Eq. (\ref{bb2}), we find the geodesic angular velocity given by
\begin{equation}
   \Omega^\text{null}=\frac{1}{r}\,\sqrt{1-\alpha-\frac{2\,M}{r}+\frac{\lambda\,M^2}{r^4}-N\,r}\Big{|}_{r=r_c}.\label{cc11}
\end{equation}
The expression (\ref{cc11}) shows that the geodesic angular velocity for circular orbits is influenced by the CS parameter $\alpha$, the quantum deformation parameter $\lambda$, the quintessence normalization constant $N$, and the BH mass $M$. In the limit where $\lambda=0$ and $N=0$-corresponding to the absence of quantum deformation and the quintessential field-the geodesic angular velocity (\ref{cc11}) reduces to that of the Letelier BH solution, which further simplifies to the standard Schwarzschild BH result when $\alpha=0$.  Thus, it becomes evident that changing these parameters modifies
the curvature structure of the spacetime, leading to quantitative shifts in the geodesic angular velocity. As a result, the photon's orbital motion responds sensitively to the combined influence of these geometric contributions, reflecting how departures from the standard Schwarzschild geometry alter the dynamics of null circular orbits.
At the radius $r_c=r_{\rm ph}$, equal to the photon sphere radius, the geodesic angular velocity is given by
\begin{equation}
   \Omega^\text{null}=\frac{1}{r_{\rm ph}}\,\sqrt{1-\alpha-\frac{2\,M}{r_{\rm ph}}+\frac{\lambda\,M^2}{r^4_{\rm ph}}-N\,r_{\rm ph}},\label{cc11a}
\end{equation}
where $r_{\rm ph}$ can be determined using relation (\ref{cc3a}), which will be discussed in a subsequent section.

\subsection{Test Particles Dynamics: Time-like Geodesic }

Timelike geodesics represent the natural paths followed by particles such as planets, stars, and observers in curved spacetime. Studying these geodesics tells us about the structure of spacetime, gravitational potentials, and the dynamical behavior of matter in the vicinity of compact objects such as BHs. In static and spherically symmetric spacetimes, the motion along timelike geodesics is typically analyzed using the effective potential method, which allows classification of possible orbits (e.g., bound, unbound, circular, or plunging) depending on the particle's energy and angular momentum.

For timelike particles, $\epsilon=-1$, the effective potential from Eq. (\ref{mm4}) will be
\begin{equation}
   V_\text{eff}(r)=\left(1+\frac{\mathrm{L}^2}{r^2}\right)\,f(r)=\left(1+\frac{\mathrm{L}^2}{r^2}\right)\,\left(1-\alpha-\frac{2\,M}{r}+\frac{\lambda\,M^2}{r^4}-N\,r\right).\label{ss1}
\end{equation}

For circular timelike geodesics, the conditions $\frac{dr}{d\tau}=0$ and $\frac{d^2r}{d\tau^2}=0$ must be satisfied. Upon simplification, these conditions yield the following two physical quantities, namely, the specific angular momentum given by
\begin{eqnarray}
   \mathrm{L}_\text{sp}(r)=\sqrt{\frac{r^3\,f'(r)}{2\,f(r)-r\,f'(r)}}=r\,\sqrt{\frac{\frac{M}{r}-\frac{2\,\lambda\,M^2}{r^4}-\frac{N\,r}{2}}{1-\alpha-\frac{3\,M}{r}+\frac{3\,\lambda\,M^2}{r^4}-\frac{N\,r}{2}}}.\label{ss2}
\end{eqnarray}
And the particle's specific energy given by
\begin{eqnarray}
   \mathrm{E}_\text{sp}(r)=\pm\,\sqrt{\frac{2}{2\,f(r)-r\,f'(r)}}\,f(r)=\pm\,\frac{\left(1-\alpha-\frac{2\,M}{r}+\frac{\lambda\,M^2}{r^4}-N\,r\right)}{\sqrt{1-\alpha-\frac{3\,M}{r}+\frac{3\,\lambda\,M^2}{r^4}-\frac{N\,r}{2}}}.\label{ss3}
\end{eqnarray}

From the expressions given in Eqs. (\ref{ss2}) and (\ref{ss3}), it is evident that the specific angular momentum and energy of test particles traversing in circular timelike geodesics are influenced by several key parameters: the string parameter $\alpha$, the quantum deformation parameter $\lambda$, the quintessence normalization constant $N$, and the BH mass $M$.

In the limit where $N =0$, corresponds to the absence of the QF, the specific angular momentum and energy becomes:
\begin{eqnarray}
   \mathrm{L}_\text{sp}(r)=r\,\sqrt{\frac{\frac{M}{r}-\frac{2\,\lambda\,M^2}{r^4}}{1-\alpha-\frac{3\,M}{r}-\frac{3\,\lambda\,M^2}{r^4}}}\quad,\quad \mathrm{E}_\text{sp}(r)=\pm\,\frac{\left(1-\alpha-\frac{2\,M}{r}+\frac{\lambda\,M^2}{r^4}\right)}{\sqrt{1-\alpha-\frac{3\,M}{r}-\frac{3\,\lambda\,M^2}{r^4}}}.\label{ss3a}
\end{eqnarray}
Moreover, in the limit where $\lambda=0$, corresponds to the absence of the quantum deformation in the BH solution, the specific angular momentum and energy becomes:
\begin{eqnarray}
   \mathrm{L}_\text{sp}(r)=r\,\sqrt{\frac{\frac{M}{r}-\frac{N\,r}{2}}{1-\alpha-\frac{3\,M}{r}-\frac{N\,r}{2}}}\quad,\quad \mathrm{E}_\text{sp}(r)=\pm\,\frac{\left(1-\alpha-\frac{2\,M}{r}-N\,r\right)}{\sqrt{1-\alpha-\frac{3\,M}{r}-\frac{N\,r}{2}}}.\label{ss3b}
\end{eqnarray}
Finally, in the limit $\lambda=0$ and $N=0$-which corresponds to the absence of quantum deformation and the quintessential field-the expressions for specific angular momentum and energy reduce to those of the Letelier BH solution given by
\begin{eqnarray}
   \mathrm{L}_\text{sp}(r)=r\,\sqrt{\frac{\frac{M}{r}}{1-\alpha-\frac{3\,M}{r}}}\quad,\quad \mathrm{E}_\text{sp}(r)=\pm\,\frac{\left(1-\alpha-\frac{2\,M}{r}\right)}{\sqrt{1-\alpha-\frac{3\,M}{r}}}.\label{ss3c}
\end{eqnarray}
Furthermore, setting $\alpha=0$ into the above result recovers the standard Schwarzschild BH result.

\subsubsection{\bf Speed of the test particles at large distance, $r>>r_{+}$}

We now aim to determine the orbital speed of a timelike particle moving in a circular orbit around the BH at a very large radial distance, {\it i.e.,} in the asymptotic region far from the event horizon. This scenario is analogous to a distant star orbiting the central BH of a galaxy along a circular trajectory. In the zeroth-order (Newtonian) approximation, where relativistic corrections are negligible and the influence of the BH's spacetime curvature becomes weak, the orbital speed can be estimated using the centripetal acceleration.

In the zeroth-order approximation, one can write the metric function as
\begin{equation}
   f(r)=1+2\,\Phi(r),\label{ss5}
\end{equation}
where $\Phi(r)$ is the Newtonian gravitational potential for the test particle of unit mass. Explicitly, we get
\begin{equation}
   \Phi(r)=\frac{1}{2}\,\left(-\alpha-\frac{2\,M}{r}+\frac{\lambda\,M^2}{r^4}-N\,r\right).\label{ss6}
\end{equation}

The effective gravitational force acting on a test particle in orbit is directed toward the center of the BH and is given by the negative gradient of the Newtonian gravitational potential. Mathematically, it is expressed as:
\begin{equation}
   \mathrm{F}_c=-\frac{\partial \Phi(r)}{\partial r}.\label{ss0}
\end{equation}
Evaluating this derivative using (\ref{ss6}) allows us to determine the central force responsible for maintaining circular motion around the BH. Accordingly, we obtain:
\begin{eqnarray}
   \mathrm{F}_c=-\frac{M}{r^2}+\frac{2\,\lambda\,M^2}{r^5}+\frac{N}{2}.\label{ss7}
\end{eqnarray}

This central force can be equated with centripetal acceleration, i.e., $|\mathrm{F}_c|=\frac{v^2}{r}$ in which $v$ is the speed of unit mass test particle in the orbit. This equation results an expression for the circular speed $v$ given by
\begin{eqnarray}
   v=\sqrt{\left|-\frac{M}{r}+\frac{2\,\lambda\,M^2}{r^4}+\frac{N\,r}{2}\right|}.\label{ss8}
\end{eqnarray}

The expression (\ref{ss8}) shows that the speed of timelike particles orbiting in circular path at large distances is influenced by various parameters. These include the quantum deformation parameter $\lambda$, the quintessence normalization constant $N$, and the BH mass $M$, but is independent of CS.

\begin{figure}[ht!]
    \centering
    \includegraphics[width=0.3\linewidth]{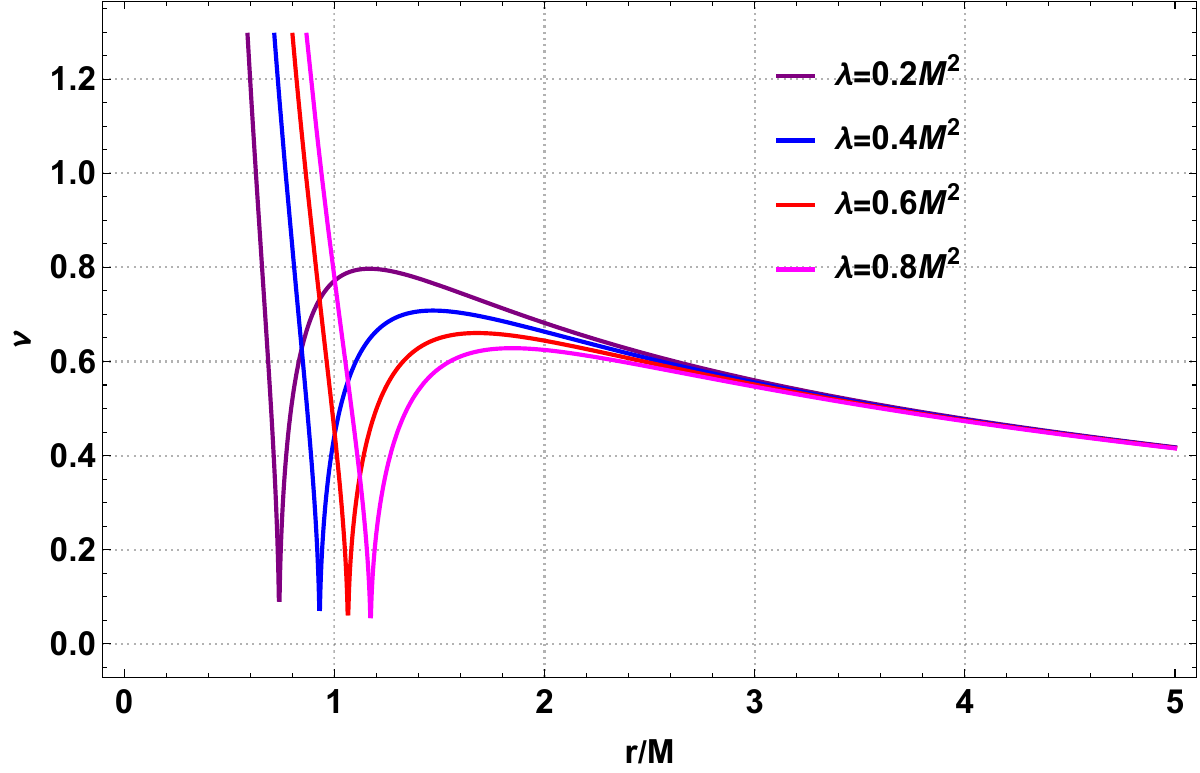}\quad
    \includegraphics[width=0.3\linewidth]{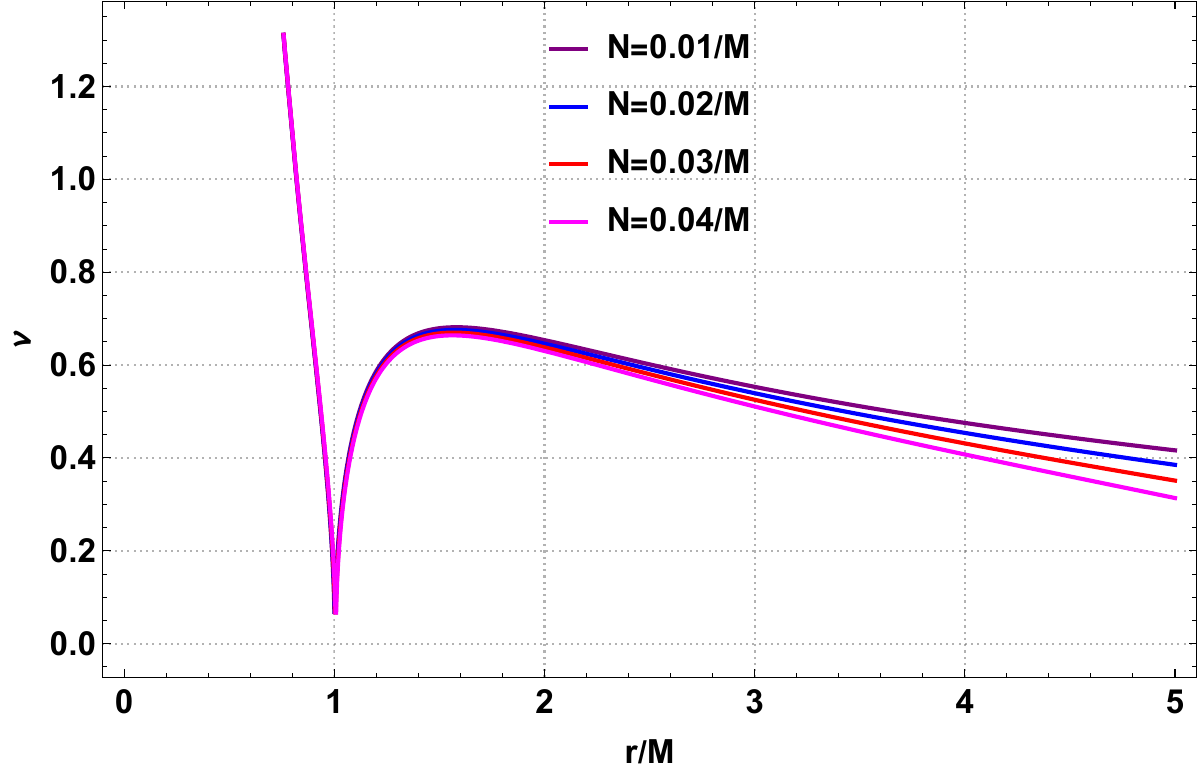}\quad
    \includegraphics[width=0.3\linewidth]{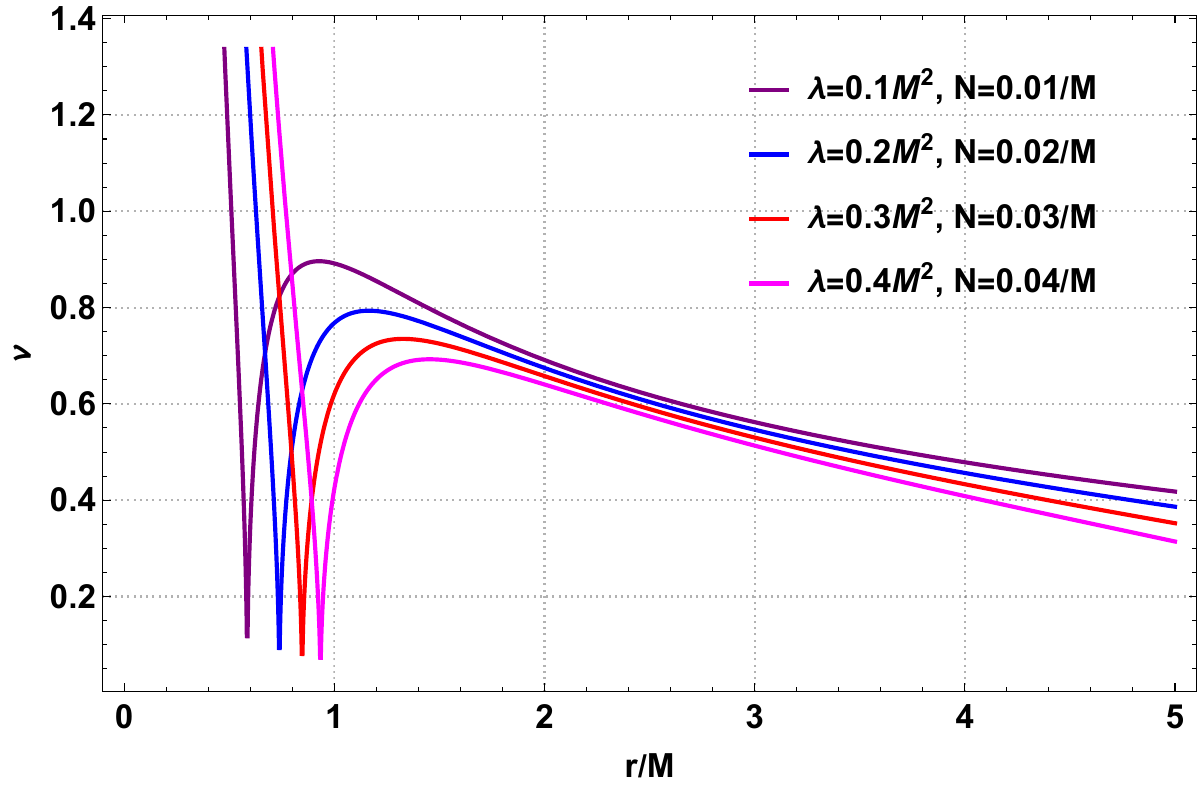}\\
    (a) $N=0.01/M$ \hspace{4cm} (b) $\lambda=0.5 M^2$ \hspace{4cm} (c) all varies
    \caption{Variation of the orbital speed $v$ as a function of $r/M$ for different values of the normalization parameter $N$, the quantum deformation $\lambda$, and their combined effects.}
    \label{fig:speed}
\end{figure}

 Figure~\ref{fig:speed} illustrates the variation of the orbital speed $v$ of a timelike particle at distances far from the BH horizon, plotted as a function of $r/M$ for different values of the QF parameter $N$, the quantum deformation parameter $\lambda$, and their combined effects. To better capture the behavior, we consider the range $0 \leq r/M \leq 5$. In all panels, we observe that as the values of these parameters increase, the orbital speed of the test particle decreases, indicating that the motion of test particles slows down in the presence of larger $N$ and $\lambda$ values. This behavior reflects the influence of the exotic matter field and quantum corrections on the dynamics of test particles in the modified BH spacetime.

In the limit where $N=0$ corresponds to the absence of the QF, the orbital speed becomes:
\begin{eqnarray}
   v=\sqrt{\left|-\frac{M}{r}+\frac{2\,\lambda\,M^2}{r^4}\right|}.\label{ss8a}
\end{eqnarray}
Moreover, in the limit where $\lambda=0$ corresponds to the absence of the quantum deformation in the BH solution, the orbital speed becomes
\begin{eqnarray}
   v=\sqrt{\left|-\frac{M}{r}+\frac{N\,r}{2}\right|}.\label{ss8b}
\end{eqnarray}
Finally, for $\lambda=0$ and $N=0$, which removes both the quantum deformation and the quintessential field, the orbital speed reduces to the Schwarzschild result
\begin{eqnarray}
   v=\sqrt{\left|-\frac{M}{r}\right|}.\label{ss8c}
\end{eqnarray}

The comparison with Schwarzschild needs care, and the earlier version of this work stated it the wrong way round. Because the bracket in Eq.~(\ref{ss8}) carries an absolute value, the sign of $-M/r+2\lambda M^2/r^4+N r/2$ decides the direction of the shift. Inside the radius $r_c$ at which that combination vanishes the bracket is negative, so $v^2=M/r-2\lambda M^2/r^4-N r/2<M/r$ and the orbital speed falls below the Schwarzschild value, which is what Figure~\ref{fig:speed} actually shows over $0\le r/M\le5$. Only beyond $r_c$ does the quintessence term take over and push $v$ above $v_{\rm Sch}$. For $M=1$, $\lambda=0.5M^2$, and $N=0.01/M$ the crossover sits at $r_c\simeq13.9M$, which is inside the static region since $r_q\simeq98M$ for that $N$. The statement $v>v_{\rm Sch}$ therefore holds only in the outer part of the static domain, and the sign of the effect reverses closer in.

\subsubsection{\bf Measurement of geodesic precession frequency (GPF)}

In this part, we turn our attention to the GPF for timelike particles and show how various parameters, such as CS, the quantum deformation, and the QF influence it. Before that, we determine the orbital angular velocity of time-like particle in circular orbits in the vicinity of the BH. The orbital angular velocity is defined as follows:
\begin{equation}
   \Omega^\text{timelike}=\frac{\dot{\phi}}{\dot{t}}=\frac{f(r)}{r^2}\,\frac{\mathrm{L}}{\mathrm{E}}=\sqrt{\frac{f'(r)}{2\,r}}=\sqrt{\frac{M}{r^3}-\frac{2\,\lambda\,M^2}{r^6}-\frac{N}{2\,r}}.\label{ss4}
\end{equation}
where we have used the relations (\ref{ss2}) and (\ref{ss3}).

From the above expression, it becomes evident that the orbital angular velocity of time-like particles in circular orbits around the BH is influenced by several parameters. Specifically, the quantum deformation parameter $\lambda$, the quintessence normalization constant $N$, and the BH mass $M$ all contribute to the dynamics of the test particles. Notably, the expression is independent of CS, indicating that string clouds effects do not directly alter the angular velocity in this configuration. In the limiting case where $\lambda = 0$ and $N = 0$ corresponding to the absence of quantum deformation and the QF, the expression of angular velocity reduces to the result known for the Schwarzschild BH.

Now, the proper angular velocity of time-like particle is defined by $\mathrm{L}=r^2\,\dot{\phi}=r^2\,\omega$. Thus, we have that
\begin{equation}
   \omega=\Omega^\text{timelike}\,\sqrt{\frac{2}{2\,f-r\,f'}},\label{ss9}
\end{equation}
where we have used Eq. (\ref{ss2}) and (\ref{ss4}).

To determine the GPF $\Theta_{GPF}$ for a gyroscope within the GP-B, we use the equation (\ref{ss4}) and follow the methodology described in \cite{RS}. This yields
\begin{equation}
   \Theta_{GPF}=\Omega-\Omega_{GPF}\quad,\quad \Omega_{GPF}=\Omega^\text{timelike}\,\sqrt{f(r)-\left(\Omega^\text{timelike}\right)^2\,r^2}.\label{ss10}
\end{equation}

Using (\ref{ss4}) and substituting the metric function $f(r)$, we find
\begin{eqnarray}
   \Theta_{GPF}=\sqrt{\frac{f'(r)}{2\,r}}\left[1-\sqrt{\frac{2\,f(r)-r\,f'(r)}{2}}\right]=\sqrt{\frac{M}{r^3}-\frac{2\,\lambda\,M^2}{r^6}-\frac{N}{2\,r}}\left[1-\sqrt{1-\alpha-\frac{3\,M}{r}+\frac{3\,\lambda\,M^2}{r^4}-\frac{N\,r}{2}}\right].\label{ss11}
\end{eqnarray}

Equation (\ref{ss11}) is the required expression of the GPF for timelike particles in the circular orbits. From the above expression, we observe that the GPF is influenced by multiple parameters. These include the CS parameter $\alpha$, the quantum deformation parameter $\lambda$, the quintessence normalization constant $N$, and the BH mass $M$.

In the limit $N=0$ corresponds to the absence of QF, the GPF reduces as,
\begin{eqnarray}
   \Theta_{GPF}=\sqrt{\frac{M}{r^3}-\frac{2\,\lambda\,M^2}{r^6}}\left[1-\sqrt{1-\alpha-\frac{3\,M}{r}-\frac{3\,\lambda\,M^2}{r^4}}\right].\label{ss11a}
\end{eqnarray}
Moreover, in the limit where $\lambda=0$ corresponds to the absence of the quantum deformation, the GPF reduces as,
\begin{eqnarray}
   \Theta_{GPF}=\sqrt{\frac{M}{r^3}-\frac{N}{2\,r}}\left[1-\sqrt{1-\alpha-\frac{3\,M}{r}-\frac{N\,r}{2}}\right].\label{ss11b}
\end{eqnarray}
Finally, in the limit $\lambda=0$ and $c=0$ which corresponds to the absence of quantum deformation and the quintessential field, the expressions for GPF reduce to that of the Letelier BH solution given by
\begin{eqnarray}
   \Theta_{GPF}=\sqrt{\frac{M}{r^3}}\left[1-\sqrt{1-\alpha-\frac{3\,M}{r}}\right].\label{ss11c}
\end{eqnarray}
Furthermore, setting $\alpha=0$ into the above result recovers the standard Schwarzschild BH case.

\begin{figure}[ht!]
   \centering
   \includegraphics[width=0.45\linewidth]{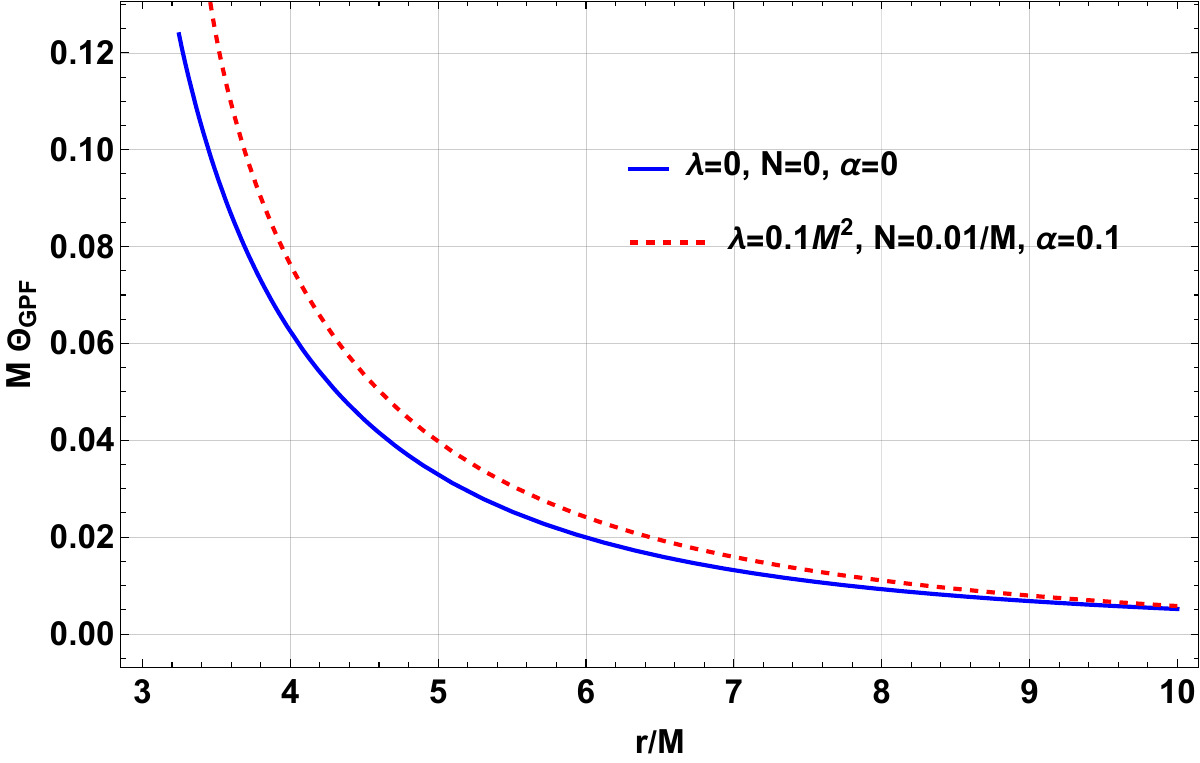}
   \caption{Comparison of the GPF $\Theta_{\text{GPF}}$ as a function of the radial coordinate $r$, shown both without and with the effects of quantum deformation ($\lambda$), CS ($\alpha$), and the normalization constant $N$ of the QF.}
   \label{fig:GPF}
\end{figure}

The precession frequency is shifted relative to Schwarzschild by three groups of terms: $-2\lambda M^2/r^6$ and $3\lambda M^2/r^4$ from the quantum deformation, $-Nr/2$ and $-N/(2r)$ from the QF, and $-\alpha$ from the CS. Their net effect depends on the radius. For the parameter set of Fig.~\ref{fig:GPF}, namely $\lambda=0.1M^2$, $N=0.01/M$, and $\alpha=0.1$, we find $\Theta_{\rm GPF}>\Theta_{\rm GPF}^{\rm Sch}$ throughout $3\lesssim r/M\lesssim10$, with the gap widening as $r$ decreases. We do not extrapolate the comparison outside that window, and we make no claim that the shift is within reach of present gyroscope experiments. 

 To clearly demonstrate the deviation from the standard Schwarzschild result, we present Fig.~\ref{fig:GPF}, which depicts the behavior of the mass-scaled geodetic precession frequency, $M\,\Theta_{\text{GPF}}$, as a function of the dimensionless radial coordinate $r/M$. The figure compares the cases with and without the effects of a CS, quantum deformation, and a QF. The red dotted curve corresponds to the baseline case with $\lambda = 0$, $N = 0$, and $\alpha = 0$, representing the absence of quantum deformation, QF, and topological defects. As expected, this curve reproduces the standard result for a Schwarzschild BH. In contrast, the blue solid curve corresponds to the parameter choice $\lambda = 0.1M^{2}$, $N = 0.01/M$, and $\alpha = 0.1$, thereby incorporating the combined effects of quantum deformation, the QF, and the CS.
From Fig.~\ref{fig:GPF}, it is evident that the inclusion of these additional physical effects leads to noticeable deviations in the geodetic precession frequency of timelike test particles when compared to the Schwarzschild case. In particular, the modified spacetime geometry induced by the QF and the topological defects alters the orbital precession behavior, highlighting the sensitivity of the GPF to departures from classical GR. These deviations become more pronounced at smaller radial distances, which is where the quantum deformation and the matter content weigh most.

\section{Photon sphere and BH Shadow} \label{sec04}

This section investigates how model parameters for the QOS BH with QF and CS affect the shadow radius. The BH shadow represents one of the most direct observational signatures of strong gravitational fields and provides a unique opportunity to test theoretical predictions against high-resolution astronomical observations. With the advent of the EHT and future space-based interferometry missions, precise measurements of BH shadows have become feasible, so predictions of shadow properties bear on distinguishing between different gravity theories and exotic matter configurations.

With a photon sphere in spherically symmetric and static metrics, an observer can see an endless number of pictures of a light source. The photon orbit radius can be calculated using the following relation:
\begin{equation}
r_\text{ph}\,f'(r_\text{ph})=2\,f(r_\text{ph}).\label{shadeq0}
\end{equation}
Inserting the metric function $f(r)$ in the Eq. (\ref{bb2}), the equation for $r_{\rm ph}$ becomes:
\begin{equation}
3\,M\,r_\text{ph}^3+r_\text{ph}^4\,(N\,r_\text{ph}/2+\alpha-1)-3\,M^2\,\lambda=0. \label{eps1}
\end{equation}

 The above expression represents a fifth-order nonlinear algebraic equation in $r$, whose real root corresponds to the photon sphere radius $r = r_{\mathrm{ph}}$. Since Eq.~(\ref{eps1}) does not admit an analytical solution, we employ numerical methods to determine the photon sphere radius $r_{\mathrm{ph}}$ for selected values of the QF parameter $N$, the CS parameter $\alpha$, and the quantum deformation parameter $\lambda$. The resulting numerical values of $r_{\mathrm{ph}}$ for representative choices of $N$, $\alpha$, and $\lambda$ are summarized in Tables~\ref{tab:photon-radius-1}--\ref{tab:photon-radius-3}. From Eq.~(\ref{eps1}), it is evident that the photon sphere radius is influenced by the presence of the CS, the quantum deformation parameter $\lambda$, and the QF parameter $N$.

\begin{table}[ht!]
\centering
\caption{Numerical values of $r_{\rm ph}/M$ for varying $N$ and $\lambda$ with $\alpha=0.1$ fixed.}
\begin{tabular}{|c|c|c|c|c|c|}
\hline
$\lambda/M^2$ & $N=0.01/M$ & $N=0.02/M$ & $N=0.03/M$ & $N=0.04/M$ & $N=0.05/M$ \\
\hline
0.1 & 3.38856 & 3.45815 & 3.53392 & 3.61702 & 3.70891 \\
0.2 & 3.37951 & 3.44928 & 3.52524 & 3.60851 & 3.70057 \\
0.3 & 3.37032 & 3.44028 & 3.51642 & 3.59988 & 3.69213 \\
0.4 & 3.36097 & 3.43113 & 3.50747 & 3.59112 & 3.68356 \\
0.5 & 3.35146 & 3.42183 & 3.49839 & 3.58224 & 3.67488 \\
\hline
\end{tabular}
\label{tab:photon-radius-1}
\end{table}

\begin{table}[ht!]
\centering
\caption{Numerical values of the photon sphere radius $r_{\rm ph}/M$ for varying $\lambda$ and $\alpha$ with $N=0.01/M$ fixed.}
\begin{tabular}{|c|c|c|c|c|c|}
\hline
$\lambda/M^2$ & $\alpha=0.00$ & $\alpha=0.05$ & $\alpha=0.10$ & $\alpha=0.15$ & $\alpha=0.20$ \\
\hline
0.1 & 3.03534 & 3.20225 & 3.38856 & 3.59798 & 3.83529 \\
0.2 & 3.02403 & 3.19211 & 3.37951 & 3.58997 & 3.82823 \\
0.3 & 3.01245 & 3.18177 & 3.37032 & 3.58184 & 3.82109 \\
0.4 & 3.00060 & 3.17122 & 3.36097 & 3.57360 & 3.81387 \\
0.5 & 2.98845 & 3.16045 & 3.35146 & 3.56524 & 3.80657 \\
\hline
\end{tabular}
\label{tab:photon-radius-2}
\end{table}

\begin{table}[ht!]
\centering
\caption{Photon sphere radius $r_{\rm ph}/M$ for different values of $N$, with varying
$\lambda$ and $\alpha$.}
\begin{tabular}{|c|c|c|c|c|c|c|}
\hline
$\lambda/M^2$ & $\alpha$ & $N=0.01/M$ & $N=0.02/M$ & $N=0.03/M$ & $N=0.04/M$ & $N=0.05/M$ \\
\hline
0.1 & 0.05 & 3.20225 & 3.26070 & 3.32372 & 3.39203 & 3.46655 \\
0.1 & 0.10 & 3.38856 & 3.45815 & 3.53392 & 3.61702 & 3.70891 \\
0.1 & 0.15 & 3.59798 & 3.68182 & 3.77423 & 3.87704 & 3.99275 \\
0.1 & 0.20 & 3.83529 & 3.93767 & 4.05225 & 4.18213 & 4.33177 \\

0.2 & 0.05 & 3.19211 & 3.25074 & 3.31394 & 3.38243 & 3.45713 \\
0.2 & 0.10 & 3.37951 & 3.44928 & 3.52524 & 3.60851 & 3.70057 \\
0.2 & 0.15 & 3.58997 & 3.67398 & 3.76656 & 3.86954 & 3.98543 \\
0.2 & 0.20 & 3.82823 & 3.93079 & 4.04554 & 4.17559 & 4.32539 \\

0.3 & 0.05 & 3.18177 & 3.24060 & 3.30399 & 3.37267 & 3.44755 \\
0.3 & 0.10 & 3.37032 & 3.44028 & 3.51642 & 3.59988 & 3.69213 \\
0.3 & 0.15 & 3.58184 & 3.66604 & 3.75880 & 3.86197 & 3.97802 \\
0.3 & 0.20 & 3.82109 & 3.92383 & 4.03877 & 4.16898 & 4.31895 \\

0.4 & 0.05 & 3.17122 & 3.23026 & 3.29386 & 3.36274 & 3.43782 \\
0.4 & 0.10 & 3.36097 & 3.43113 & 3.50747 & 3.59112 & 3.68356 \\
0.4 & 0.15 & 3.57360 & 3.65799 & 3.75095 & 3.85430 & 3.97054 \\
0.4 & 0.20 & 3.81387 & 3.91680 & 4.03192 & 4.16232 & 4.31246 \\

0.5 & 0.05 & 3.16045 & 3.21971 & 3.28353 & 3.35263 & 3.42792 \\
0.5 & 0.10 & 3.35146 & 3.42183 & 3.49839 & 3.58224 & 3.67488 \\
0.5 & 0.15 & 3.56524 & 3.64984 & 3.74299 & 3.84654 & 3.96297 \\
0.5 & 0.20 & 3.80657 & 3.90970 & 4.02501 & 4.15560 & 4.30592 \\
\hline
\end{tabular}
\label{tab:photon-radius-3}
\end{table}

Once we have these data, we can compute the shadow radii using \cite{Perlick2022}: 
\begin{equation}
    R_s=r_\text{ph}\sqrt{\frac{f(r_O)}{f(r_\text{ph})}}, \label{shadeq1}
\end{equation}
where $r_O$ is the position of a static observer. For the present case, we find the following expression for the shadow radius
\begin{equation}
    R_s=r_\text{ph}\sqrt{\frac{1-\alpha-\frac{2\,M}{r_O}+\frac{\lambda\,M^2}{r^4_O}-N\,r_O}{1-\alpha-\frac{2\,M}{r_{\rm ph}}+\frac{\lambda\,M^2}{r^4_{\rm ph}}-N\,r_{\rm ph}}}.\label{shadeq2}
\end{equation}

When the quintessence field is switched off, $N=0$, the static region extends to arbitrarily large radius and $f\to1-\alpha$, so a distant observer is available even though the geometry stays conical. In that case
\begin{equation}
    R_s=r_\text{ph}\sqrt{\frac{1-\alpha}{1-\alpha-\frac{2\,M}{r_{\rm ph}}+\frac{\lambda\,M^2}{r^4_{\rm ph}}}},\label{shadeq3}
\end{equation}
where the $N r_{\rm ph}$ term that appeared in the denominator of this expression in the previous version has been dropped, since it is inconsistent with the limit being taken,
where $r_{\rm ph}$ can now be determined using the following relation:
\begin{equation}
r_\text{ph}^4\left\{1-\alpha-\frac{3\,M}{r_\text{ph}}\right\}+3\,\lambda\,M^2=0. \label{eps2}
\end{equation}

Equation~(\ref{shadeq1}) gives the apparent angular size of the BH for an observer at rest at $r=r_O$ in the equatorial plane, and it is the critical impact parameter that separates capture from scattering. The observer cannot be placed at infinity here. With $N>0$ the static region terminates at $r_q$, and $f(r_O)$ is the redshift factor of a physical static observer inside it, so $r_O$ has to be specified as part of the observable. We take $r_O=10M$ throughout, which lies well inside $r_q$ for every parameter set in Tables~\ref{tab:photon-radius-1}--\ref{table:1}: the smallest $r_q$ occurring there is $45.6M$, at $\alpha=0.5$ and $N=0.01/M$. Shadow radii quoted for different $r_O$ are not directly comparable, since the factor $\sqrt{f(r_O)}$ rescales all of them.

Table~\ref{table:1} collects the shadow radii, recomputed from Eq.~(\ref{shadeq2}) with $r_O=10M$. The values replace those of the previous version, which could not be reproduced from the metric function (\ref{bb2}); the cosmological constant $\Lambda=-0.03$ that survived in the caption of the corresponding figure is a trace of the model those numbers came from, and no such term appears anywhere in the present solution.

The dependence on $\alpha$ is monotonic and strong: at $N=0.001/M$ and $\lambda=0.2M^2$ the shadow grows from $R_s/M=5.06972$ at $\alpha=0.1$ to $8.05706$ at $\alpha=0.5$, against $4.64758$ for Schwarzschild seen from the same radius. This traces to the solid-angle deficit, which enlarges the photon sphere. The dependence on $\lambda$ is monotonic and weak, a decrease in the fourth digit, because $\lambda M^2/r^4$ is already small at $r\simeq r_{\rm ph}$.

The dependence on $N$ is not monotonic, and this corrects a claim made in the earlier version. Two competing effects are at work: increasing $N$ pushes the photon sphere outward, which enlarges $R_s$, but it also lowers $f(r_O)$ at the fixed observer position, which shrinks the redshift factor in Eq.~(\ref{shadeq2}). At $\alpha=0.1$ the second effect wins and $R_s/M$ falls from $5.06972$ to $4.98347$ as $N$ runs from $0.001/M$ to $0.01/M$; at $\alpha=0.5$ the first wins and $R_s/M$ rises from $8.05706$ to $8.14091$ over the same range. The turnover sits near $\alpha\simeq0.4$ for $\lambda=0.2M^2$.

\begin{figure}[ht!]
    \centering
    \includegraphics[width=0.62\linewidth]{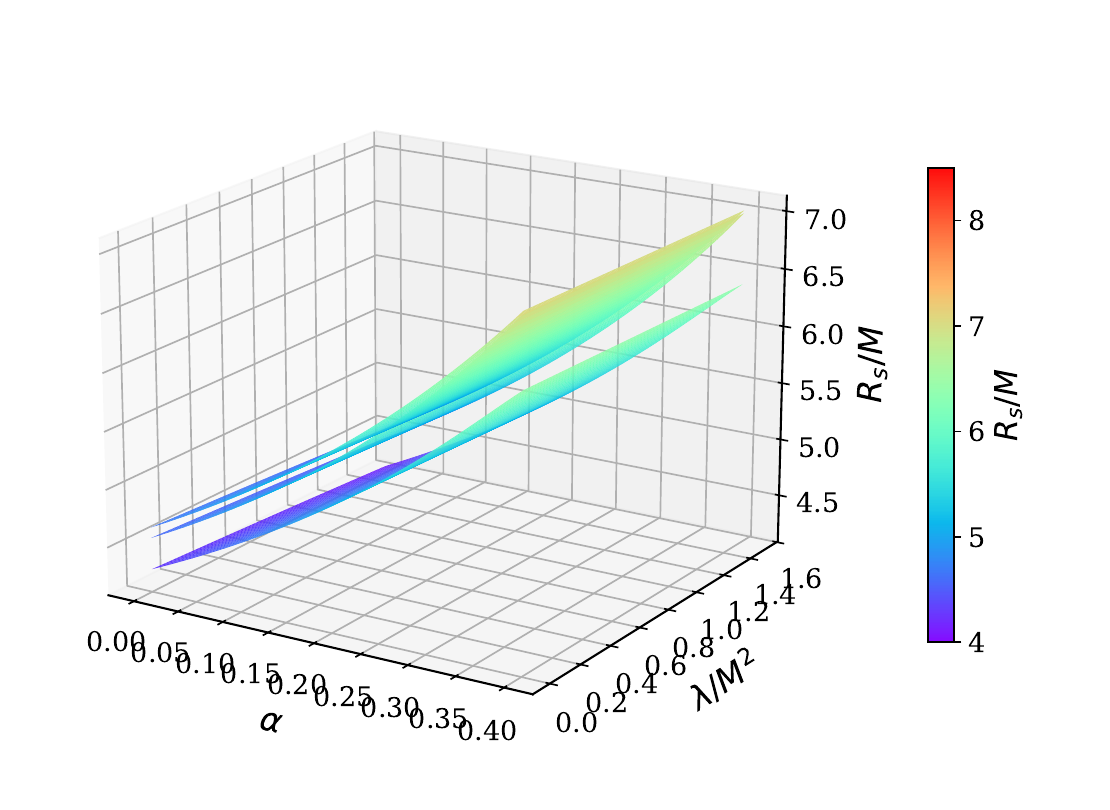}\quad\quad
    \caption{Shadow radius against the CS parameter and the quantum deformation parameter, for $N M=0.0001$, $0.01$, and $0.03$, with $M=1$ and a static observer at $r_O=10M$. The three surfaces are ordered by $N$ at fixed $\alpha$ only for large $\alpha$; they cross near $\alpha\simeq0.4$.}
    \label{figph1}
\end{figure}

\begin{figure}[ht!]
   \includegraphics[width=0.95\linewidth]{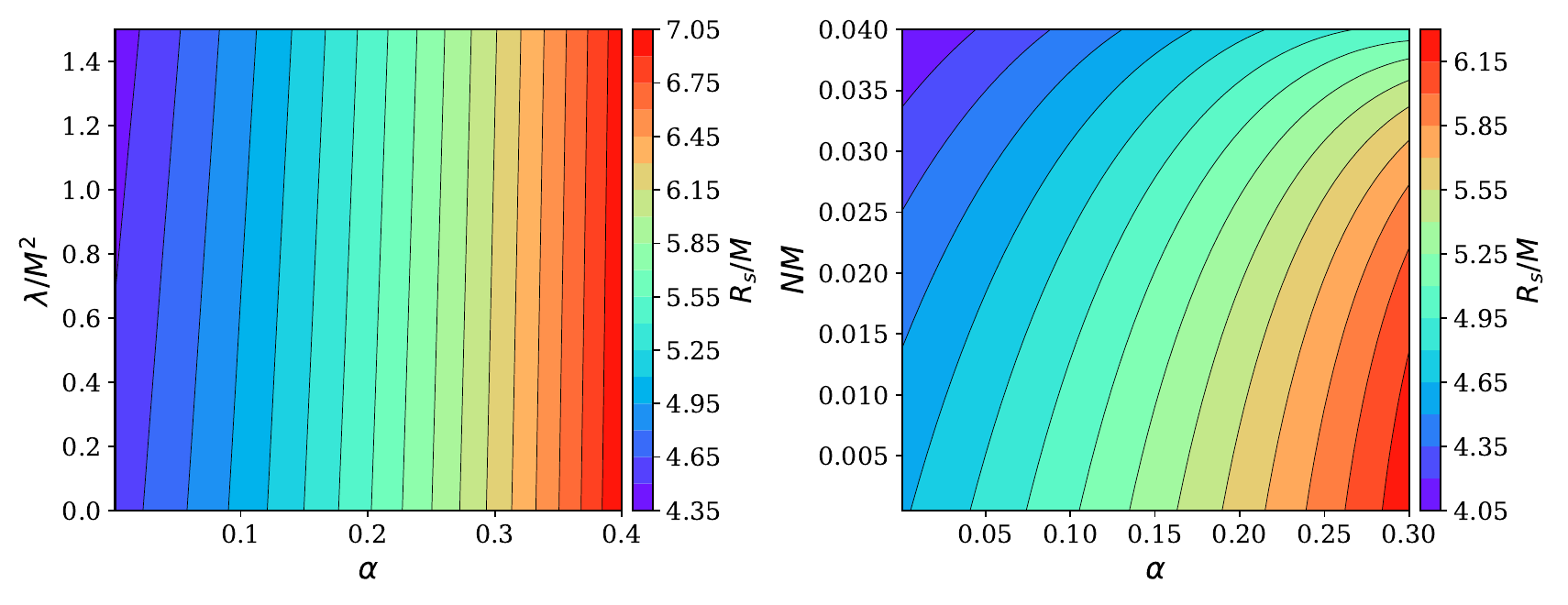}
    \caption{Shadow radius in the $(\alpha,\lambda)$ plane at $N M=0.01$ (left) and in the $(\alpha,N)$ plane at $\lambda=0.2M^2$ (right), for $r_O=10M$. The near-vertical contours on the left show that $\lambda$ barely moves $R_s$. The curvature of the contours on the right is the non-monotonic dependence on $N$ discussed in the text.}
    \label{shad12}
\end{figure}

\begin{table}[ht!]
\centering
\scriptsize
\begin{tabular}{|c|c|c|c|c|c|c|c|c|c|}
 \hline
 & \multicolumn{3}{|c|}{$\bf N M=0.001$} & \multicolumn{3}{|c|}{$\bf N M=0.005$} & \multicolumn{3}{|c|}{$\bf N M=0.01$} \\
 \hline
$\boldsymbol{\alpha}$ & $\boldsymbol{\lambda/M^2=0.2}$ & $\bf 0.6$ & $\bf 1$ & $\boldsymbol{\lambda/M^2=0.2}$ & $\bf 0.6$ & $\bf 1$ & $\boldsymbol{\lambda/M^2=0.2}$ & $\bf 0.6$ & $\bf 1$ \\
 \hline
$0.1$ & $5.06972$ & $5.04099$ & $5.01060$ & $5.03437$ & $5.00585$ & $4.97574$ & $4.98347$ & $4.95525$ & $4.92549$ \\
$0.3$ & $6.26045$ & $6.24441$ & $6.22796$ & $6.23509$ & $6.21912$ & $6.20277$ & $6.19188$ & $6.17599$ & $6.15973$ \\
$0.5$ & $8.05706$ & $8.05008$ & $8.04305$ & $8.09796$ & $8.09100$ & $8.08398$ & $8.14091$ & $8.13389$ & $8.12681$ \\
 \hline
\end{tabular}
\caption{Shadow radius $R_s/M$ from Eq.~(\ref{shadeq2}) for a static observer at $r_O=10M$, with $M=1$ and $w=-2/3$. The Schwarzschild value at the same observer position is $R_s/M=4.64758$.}
\label{table:1}
\end{table}

The surfaces of Fig.~\ref{figph1} and the contours of Fig.~\ref{shad12} display the same behavior over a continuous parameter range. In Fig.~\ref{figph1} the rise along $\alpha$ is steady across the whole plotted interval, while motion along $\lambda$ leaves the surface almost level, which follows from the $r^{-4}$ falloff of the quantum term evaluated at $r_{\rm ph}\gtrsim3M$. The three surfaces belonging to different $N$ are not nested. They intersect, and the ordering of the shadow radii by $N$ flips as $\alpha$ increases past roughly $0.4$.

The left panel of Fig.~\ref{shad12} makes the weak $\lambda$ dependence explicit: contours run nearly parallel to the $\lambda$ axis, so $R_s$ is set almost entirely by $\alpha$ there. The right panel bends, and the bending is the competition described above between the outward shift of the photon sphere and the reduction of $f(r_O)$. Because both effects scale with $N$ but with opposite sign and different $\alpha$ weighting, no single monotonic statement covers the plane. For an observational comparison this matters: at fixed $\alpha$, a measured shadow does not determine the sign of $N$ without an independent handle on $\alpha$.

\section{Perturbations in QOS BH Spacetime} \label{sec05}

Perturbation analysis is a powerful method for exploring the stability and spectral properties of BH spacetimes. By examining the response of BHs to small external disturbances, this approach reveals their dynamical behavior and possible observational signatures. Among the various types of perturbations, the scalar, vector, and tensor-scalar field perturbations are particularly useful because of their relative mathematical simplicity. Although simpler than tensor (gravitational) perturbations, scalar perturbations still capture key features of a BH's dynamical response, making them a valuable approximation for studying wave propagation and stability in curved spacetimes.

Understanding scalar perturbations is essential not only for their simplicity but also for their broad applicability. They have been extensively used to analyze the stability of a wide range of BH solutions in GR and alternative theories of gravity. Notable examples include detailed studies of Schwarzschild, Kerr, and RN BHs, where scalar field perturbations have provided information about matter and field propagation, as well as the geometric response of spacetime \cite{Chandra,isz27}. These analyses have also been extended to BHs in modified gravity frameworks, offering comparative perspectives on stability across diverse gravitational models. The study of perturbations becomes particularly relevant in the context of gravitational wave astronomy, where the QNMs of BHs contribute to the ringdown phase of merger events and can potentially distinguish between different theories of gravity.

\subsection{Scalar Perturbations}

In this section, we investigate scalar-field perturbations in the background of the deformed BH incorporating a quintessential field and a CS. Our analysis begins with the derivation of the massless Klein-Gordon equation, which governs the evolution of a scalar field in this spacetime geometry. The study of perturbations typically starts with identifying the appropriate field equation. For a massless scalar field, the dynamics is described by the Klein-Gordon equation in curved spacetime, expressed in its covariant form as \cite{FA1,FA2,FA3,FA4,FA5,FA6,FA7,FA8,FA9,FA10,FA11,FA12}:
\begin{equation}
\frac{1}{\sqrt{-g}}\,\partial_{\mu}\left[\left(\sqrt{-g}\,g^{\mu\nu}\,\partial_{\nu}\right)\,\Psi\right]=0,\label{ff1}    
\end{equation}
where $\Psi$ is the wave function of the scalar field, $g_{\mu\nu}$ is the covariant metric tensor, $g=\det(g_{\mu\nu})$ is the determinant of the metric tensor, $g^{\mu\nu}$ is the contravariant form of the metric tensor, and $\partial_{\mu}$ is the partial derivative with respect to coordinate systems. The given spacetime (\ref{bb1}) can be expressed as $ds^2=g_{\mu\nu}\,dx^{\mu}\,dx^{\nu}$ with $\mu,\nu=0,1,2,3$. Thus, the covariant and contrvariant forms of the metric tensor $g_{\mu\nu}$ is given by
\begin{equation}
    g_{\mu\nu}=\mbox{diag}\left(-f(r),\,\frac{1}{f(r)},\,r^2\,\,r^2\,\sin^2 \theta\right),\quad
    g_{\mu\nu}=\mbox{diag}\left(-\frac{1}{f(r)},\,f(r),\,\frac{1}{r^2}\,\,\frac{1}{r^2\,\sin^2 \theta}\right).\label{ff2}
\end{equation}
The determinant of the metric tensor is $g=\mbox{det}(g_{\mu\nu})=-r^4\,\sin^2 \theta$.

To explicitly write the above equation (\ref{ff1}), let us consider the following scalar field ansatz form
\begin{equation}
    \Psi(t, r,\theta, \phi)=\exp(-i\,\omega\,t)\,Y^{m}_{\ell} (\theta,\phi)\,\frac{\psi(r)}{r},\label{ff3}
\end{equation}
where $\omega$ is (possibly complex) the temporal frequency, $\psi (r)$ is a propagating scalar field, and $Y^{m}_{\ell} (\theta,\phi)$ is the spherical harmonics characterized by the quantum numbers $\ell$ and $m$.

By employing the metric tensor given in Eq.~(\ref{ff2}) and the scalar field ansatz from Eq.~(\ref{ff3}), the scalar field Eq.~(\ref{ff1}) reduces to a one-dimensional Schrödinger-like equation of the form:
\begin{equation}
    \frac{\partial^2 \psi(r_*)}{\partial r^2_{*}}+\left(\omega^2-\mathcal{V}_\text{scalar}\right)\,\psi(r_*)=0,\label{ff4}
\end{equation}
where $\mathcal{V}_\text{scalar}$ represents the effective potential that encapsulates the influence of background spacetime on scalar field propagation and we have used the tortoise coordinate $r_{*}$ defined by:
\begin{eqnarray}
    r_*=\int\,\frac{dr}{f(r)},\quad\quad \partial_{r_{*}}=f(r)\,\partial_r.\label{ff5}
\end{eqnarray}

The scalar perturbative potential takes the form:
\begin{eqnarray}
\mathcal{V}_\text{scalar}(r)=\left[\frac{\ell\,(\ell+1)}{r^2}+\frac{f'(r)}{r}\right]\,f(r)
=\left(\frac{\ell\,(\ell+1)}{r^2}+\frac{2\,M}{r^3}-\frac{4\,\lambda\,M^2}{r^6}-\frac{N}{r}\right)\,
\left(1-\alpha-\frac{2\,M}{r}+\frac{\lambda\,M^2}{r^4}-N\,r\right).\label{ff6}
\end{eqnarray}

The effective potential for scalar perturbation given in Eq. (\ref{ff6}) shows that various factors involved in the spacetime geometry influence this potential. These include the CS parameter $\alpha$, the quantum deformation parameter $\lambda$, and the quintessence normalization constant $N$. Additionally, the BH mass $M$, and the multipole number $\ell$ also influence on it. In the limit $\lambda=0$ and $N=0$-which corresponds to the absence of quantum deformation and the quintessential field-the expressions for perturbative potential reduce to that of the Letelier BH solution. Furthermore, setting $\alpha=0$ recovers the standard Schwarzschild BH result.

\begin{figure}[ht!]
    \centering
    \subfloat[$\lambda=0.5\,M^2,N=0.02/M$]{\centering{}\includegraphics[width=0.32\linewidth]{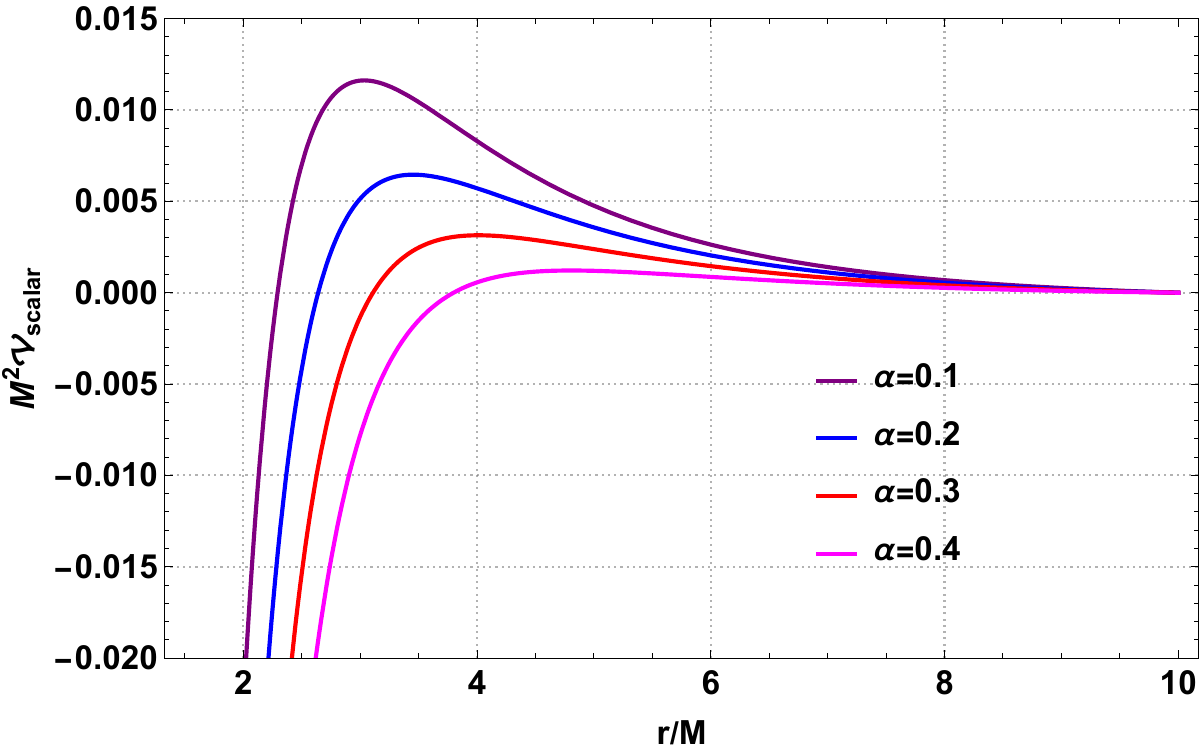}}\quad
    \subfloat[$\alpha=0.1,\lambda=0.5\,M^2$]{\centering{}\includegraphics[width=0.32\linewidth]{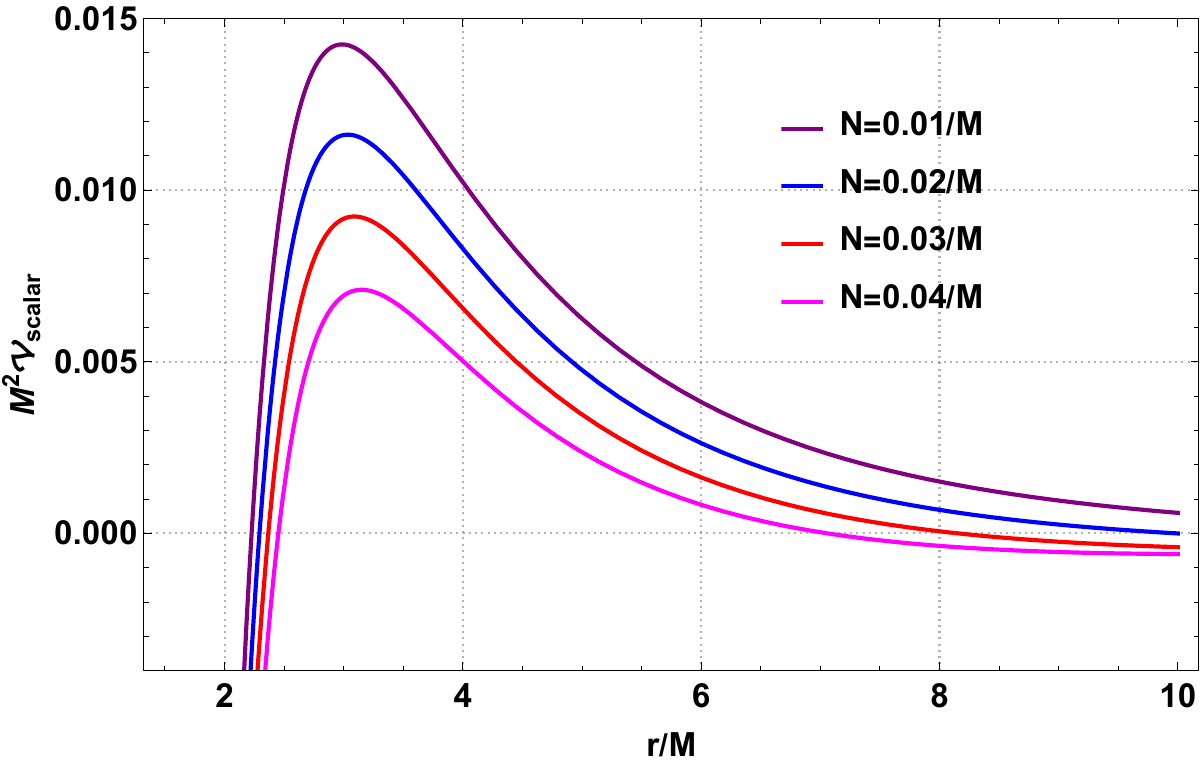}}\quad
    \subfloat[$N=0.02/M$]{\centering{}\includegraphics[width=0.32\linewidth]{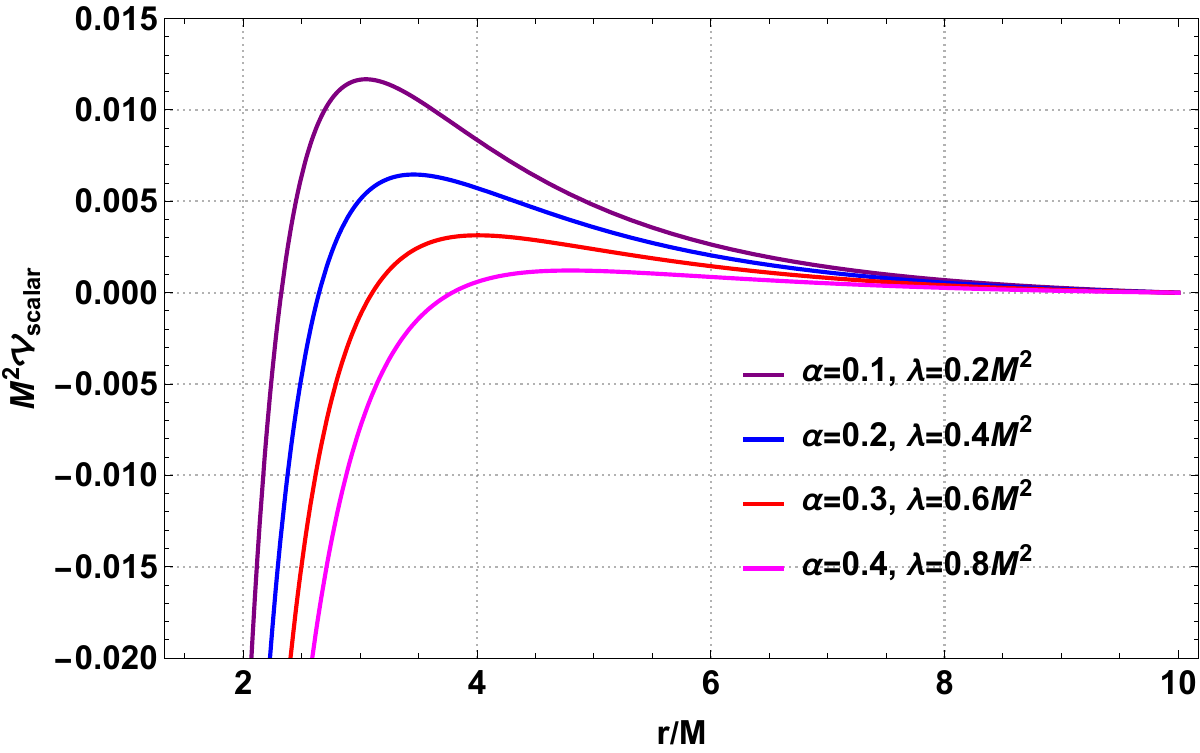}}\\
    \subfloat[$\lambda=0.5\,M^2$]{\centering{}\includegraphics[width=0.33\linewidth]{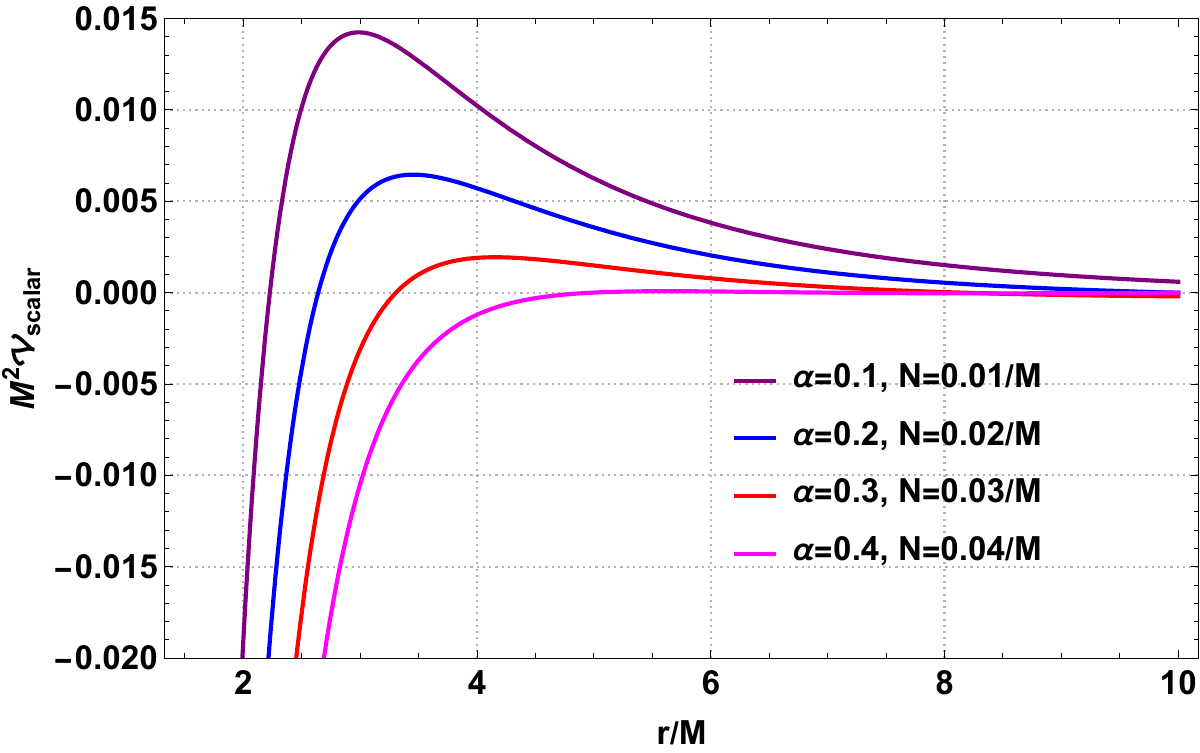}}\quad
    \subfloat[\mbox{all varies}]{\centering{}\includegraphics[width=0.33\linewidth]{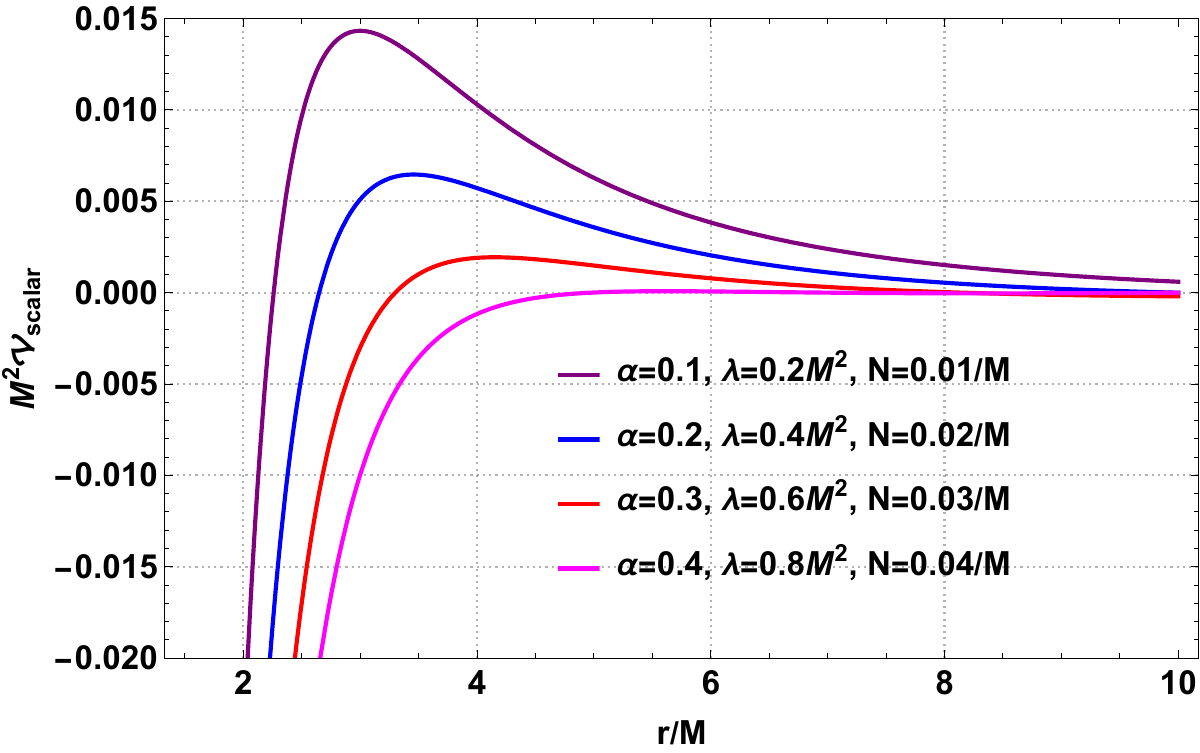}}
    \caption{Behavior of the scalar perturbative potential for different values of the string parameter $\alpha$, the deformation parameter $\lambda$, the normalization constant $N$ of field, and their combination. Here, we set $\ell=0$.}
    \label{fig:scalar}
\end{figure}

Figure~\ref{fig:scalar} displays the behavior of the mass-scaled scalar perturbative potential, $M^{2}V_{\text{scalar}}$, as a function of the dimensionless radial coordinate $r/M$ for different values of the relevant parameters. Specifically, panel~\ref{fig:scalar}(a) illustrates the effect of the CS parameter $\alpha$, panel~\ref{fig:scalar}(b) shows the influence of the QF parameter $N$, while panels~\ref{fig:scalar}(c)--\ref{fig:scalar}(e) depict their combined effects. Across all panels, a qualitatively similar structure of the effective potential is observed. In the vicinity of the event horizon ($r/M \to 2^{+}$), the potential becomes negative and exhibits a steep radial variation. This behavior reflects the dominance of strong-field gravitational effects near the BH, where scalar perturbations are significantly influenced by the spacetime curvature. At intermediate radial distances, the potential develops a well-defined positive maximum, forming an effective potential barrier for scalar wave propagation. At larger radii the potential falls off, and it vanishes again at the quintessence horizon, where $f(r_q)=0$. The correct boundary conditions are therefore those of a spacetime with two Killing horizons, purely ingoing at $r_+$ and purely outgoing at $r_q$, rather than the outgoing-wave condition at spatial infinity used for asymptotically flat backgrounds. In the tortoise coordinate the static region maps onto the whole real line, with $r_*\to-\infty$ at $r_+$ and $r_*\to+\infty$ at $r_q$, so the barrier is flanked by two flat regions exactly as in the Schwarzschild-de Sitter case. The presence of the potential peak plays a role worth spelling out. It corresponds to a region where scalar perturbations are partially reflected and partially transmitted, effectively trapping the waves for a finite duration. This temporary confinement gives rise to the characteristic oscillatory response of BHs under perturbations. In particular, the height and radial location of the peak are directly related to the QNM spectrum, governing both the oscillation frequencies and damping times of the scalar modes. A single finite peak is suggestive but proves nothing on its own, and part of the potential is negative near the horizon, which rules out the naive argument that a positive barrier forbids growing modes. Section~\ref{sec05d} settles the question by a different route. 

From Fig.~\ref{fig:scalar}, it is further evident that increasing the parameters $\alpha$, $\lambda$, and $N$, either individually or simultaneously, changes the effective potential in a definite direction. In particular: (i) the height of the potential barrier decreases, (ii) the location of the peak shifts slightly toward larger radial distances, and (iii) the overall depth of the potential well is reduced. These trends suggest that the combined effects of the CS, QF, and quantum deformation weaken the effective confining potential experienced by scalar perturbations. Physically, these additional fields and deformations alter the underlying spacetime geometry, thereby diminishing the curvature-induced trapping of scalar waves when compared to the case of smaller parameter values or the standard Schwarzschild limit.

\begin{figure}[ht!]
    \centering
    \subfloat[$\alpha=0.10$]{\centering{}\includegraphics[width=0.42\linewidth]{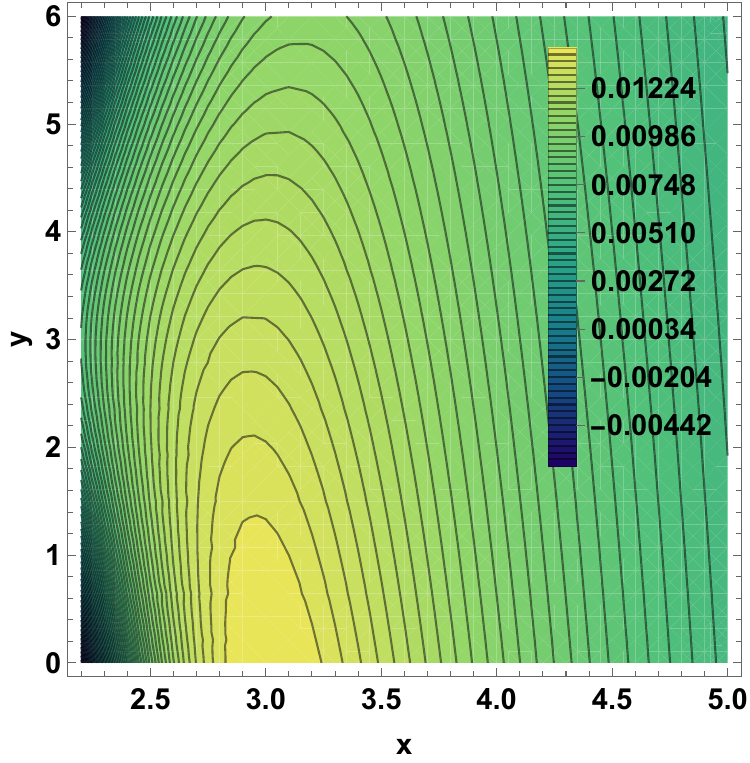}}\quad\quad\quad
    \subfloat[$\alpha=0.15$]{\centering{}\includegraphics[width=0.42\linewidth]{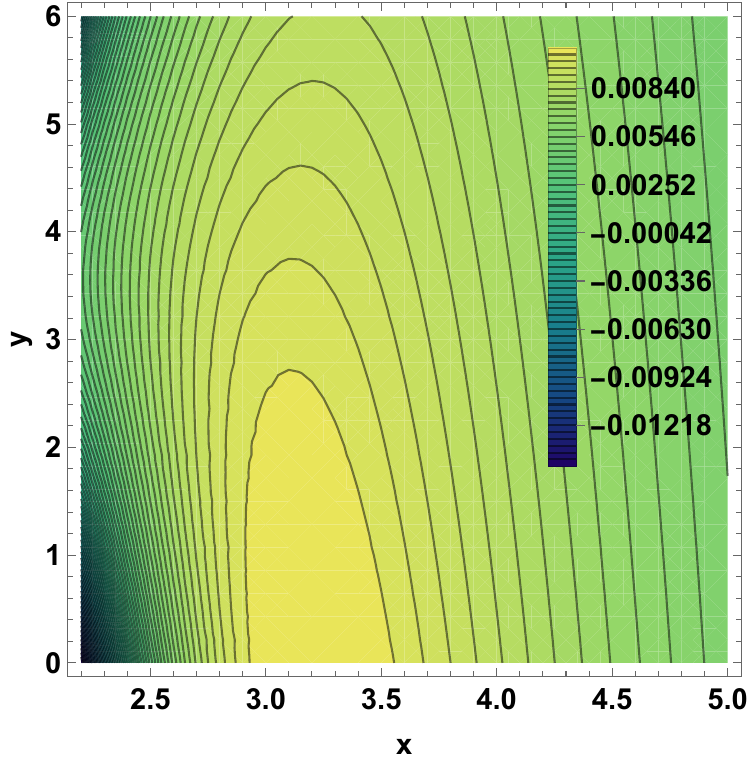}}
    \caption{The qualitative feature of zero spin scalar perturbative potential for the dominant multipole number $\ell$ = 0 is depicted in Contour plot. Blue through yellow runs from low to high. Here $N=0.01/M$.}
    \label{fig:contour}
\end{figure}

\begin{figure}[ht!]
    \centering
    \includegraphics[width=0.95\linewidth]{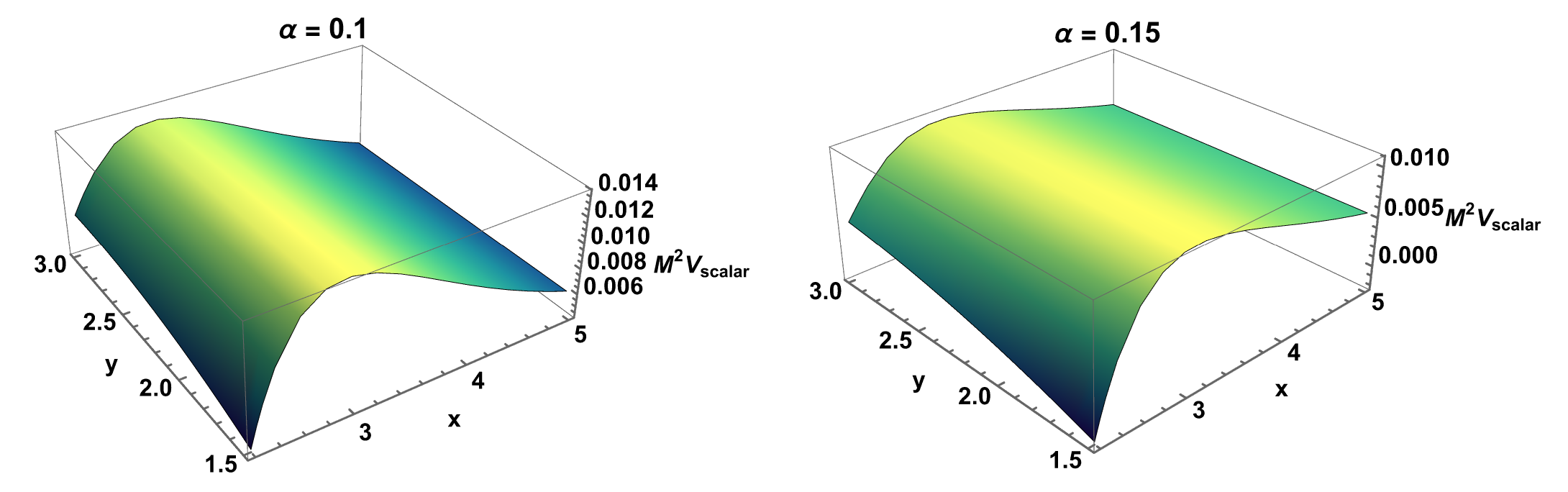}
    \caption{Three-dimensional plot of $M^2\,V_\text{scalar}$: the qualitative features of scalar perturbative potential for the dominant multipole number $\ell=0$. Here, we set $N=0.01/M$.}
    \label{fig:3dplot}
\end{figure}

To examine the qualitative features of this effective scalar potential, it is convenient to express this potential in terms of dimensionless variables. Defining the parameters $x =\frac{r}{M}$, $y = \frac{\lambda}{M^2}$, and $z=M\,N$, we find the following dimensionless quantity from Eq. (\ref{ff6}) as,
\begin{eqnarray}
M^2\,\mathcal{V}_\text{scalar}(r)&=&\left(\frac{\ell\,(\ell+1)}{x^2}+\frac{2}{x^3}-\frac{4\,y}{x^6}-\frac{z}{x}\right)\,
\left(1-\alpha-\frac{2}{x}+\frac{y}{x^4}-z\,x\right).\label{ff7}
\end{eqnarray}
For a scalar $s$-wave where the dominant multipole number $\ell=0$, we find from Eq. (\ref{ff7})
\begin{eqnarray}
M^2\,\mathcal{V}_\text{scalar}(r)&=&\left(\frac{2}{x^3}-\frac{4\,y}{x^6}-\frac{z}{x}\right)\,
\left(1-\alpha-\frac{2}{x}+\frac{y}{x^4}-z\,x\right).\label{ff8}
\end{eqnarray}

Figure~\ref{fig:contour} presents contour plots of the dimensionless scalar perturbative potential $M^2 \mathcal{V}_{\rm scalar}$ in the $(x, y) = (r/M, \lambda/M^2)$ parameter space for two representative values of the CS parameter: $\alpha = 0.10$ (left panel) and $\alpha = 0.15$ (right panel). The color gradient, ranging from blue (low values) to yellow (high values), encodes the magnitude of the effective potential. Several features merit attention. First, the diagonal ridge structure visible in both panels corresponds to the locus of potential maxima, which determines the characteristic QNM frequencies. Second, comparing the two panels reveals that increasing $\alpha$ from 0.10 to 0.15 shifts the potential peak toward larger radial distances and reduces its overall height. This trend indicates that stronger CS contributions weaken the effective scattering barrier for scalar perturbations. Third, for fixed $\alpha$, increasing the quantum deformation parameter $y = \lambda/M^2$ also diminishes the peak height, consistent with the quantum corrections softening the near-horizon geometry.
Figure~\ref{fig:3dplot} displays three-dimensional surface plots of $M^2 \mathcal{V}_{\rm scalar}$ as a function of the dimensionless radial coordinate $x = r/M$ and the deformation parameter $y = \lambda/M^2$, again for $\alpha = 0.1$ (left) and $\alpha = 0.15$ (right). These surfaces show the potential that governs scalar wave propagation. The pronounced peak near $x \approx 3$--$4$ represents the potential barrier where scalar perturbations are partially reflected and partially transmitted. The surface decays toward larger $x$ because the potential must vanish at both horizons, not because the spacetime is asymptotically flat. Comparing the two panels, the surface for $\alpha = 0.15$ exhibits a noticeably lower and broader peak compared to $\alpha = 0.10$, reinforcing the conclusion that enhanced CS density weakens the effective confinement of scalar modes near the BH.

\subsection{EM Perturbations: vector fields}

Now, we turn our attention to another perturbation of spin-1 vector field also called the EM perturbations. EM perturbations refer to small disturbances in the EM field in the curved spacetime background of a BH. EM signals originating from the vicinity of BHs can carry imprints of such perturbations. Moreover, EM perturbations are also essential for modeling the BH's response to external fields and for analyzing its stability. This perturbation has several important applications, including testing the no-hair theorem, probing the geometry of BHs, and contributing to studies involving gravitational waves and their EM counterparts (see, for examples, Refs. \cite{BW,KM} and related references therein).

For EM perturbations, the dynamics are governed by Maxwell's equations in curved spacetime:\cite{XZ}
\begin{equation}
\frac{1}{\sqrt{-g}}\partial _{\mu }\left[ F_{\alpha \beta }g^{\alpha \nu
}g^{\beta \mu }\sqrt{-g}\,\Psi\right] =0,  \label{em1}
\end{equation}
where $F_{\alpha \beta }=\partial _{\alpha }A_{\beta }-\partial _{\beta}A_{\nu}$ is the EM tensor. In the Regge-Wheeler-Zerilli formalism, one may decompose $A_{\mu}$ in terms of the scalar and vector spherical harmonics as follows:
\begin{equation}
    A_{\mu}=\sum_{\ell,m}\,e^{-i\,\omega\,t}\,\left(\left[\begin{array}{c}
         0\\
         0\\
         \psi_{em}(r)\,{\bf S}_{\ell,m}
    \end{array}\right]+\left[\begin{array}{c}
         j^{\ell,m}(r)\,Y^{m}_{\ell}\\
         h^{\ell,m}(r)\,Y^{m}_{\ell}\\
         k^{\ell,m}(r)\,{\bf Y}^{m}_{\ell}\\
    \end{array}\right] 
    \right),\label{em2}
\end{equation}
where $Y^{m}_{\ell}$ is the scalar spherical harmonics and (${\bf S}_{\ell,m}, {\bf Y}_{\ell,m}$) are the vector harmonics given by
\begin{equation}
    {\bf S}_{\ell,m}=\left(\begin{array}{c}
         \frac{1}{\sin \theta}\,\partial_{\phi}\,Y^{m}_{\ell}\\
         -\sin \theta\,\partial_{\theta}\,Y^{m}_{\ell} 
    \end{array}\right),\quad\quad {\bf Y}_{\ell,m}=\left(\begin{array}{c}
         \partial_{\theta}\,Y^{m}_{\ell}\\
         \partial_{\phi}\,Y^{m}_{\ell} 
    \end{array}\right),\label{em3}
\end{equation}
where $\omega$ is the frequency. Here $\ell$ and $m$ denote the angular momentum quantum number and the azimuthal number respectively. The first term in the right hand side of Eq. (\ref{em2}) represents the axial mode while the second term represents the polar mode. It's important to note that the axial and polar modes have parity $(-1)^{\ell+1}$ and $(-1)^{\ell}$ respectively. Moreover, the polar and axial parts have equal contributions to the final result \cite{JAW,ARF}. Thus, we focus only on the axial part. 

Now, substituting Eq. (\ref{em2}) in Eq. (\ref{em1}) and applying tortoise coordinate $r_{*}$, the radial part of Eq. (\ref{em1}) can be written in the form of Schrodinger-like wave form as:
\begin{equation}
    \frac{\partial^2 \psi_\text{em}(r_*)}{\partial r^2_{*}}+\left(\omega^2-\mathcal{V}_\text{EM}\right)\,\psi_\text{em}(r_*)=0,\label{em4}
\end{equation}
where $\mathcal{V}_\text{EM}$ represents the effective potential.

Similarly, for the EM perturbations (vector field), the effective potential is given by
\begin{equation}
\mathcal{V}_\text{EM}(r)=\frac{\ell\,(\ell+1)}{r^2}\,f(r)= \frac{\ell\,(\ell+1)}{r^2}\,\left(1-\alpha-\frac{2\,M}{r}+\frac{\lambda\,M^2}{r^4}-N\,r\right).\label{em5}
\end{equation}

 The effective potential for vector perturbations given in Eq.~(\ref{em5}) shows that various factors involved in the spacetime geometry influence the EM perturbation potential. These include the CS parameter $\alpha$, the quantum deformation parameter $\lambda$, and the QF parameter $N$. Additionally, the BH mass $M$ and the multipole number $\ell$ also influence it. In the limiting case $\lambda=0$ and $N=0$-which corresponds to the absence of quantum deformation and the QF-the expressions for the perturbative potential reduce to that of the Letelier BH case. Furthermore, setting $\alpha=0$ recovers the standard Schwarzschild BH result. Thus, it becomes evident that the combined presence of CS and QF in the modified BH spacetime alters the perturbative potential, and hence, causes the results to deviate from the standard Schwarzschild case.

\begin{figure}[ht!]
    \centering
    \subfloat[$\lambda=0.5\,M^2,N=0.02/M$]{\centering{}\includegraphics[width=0.32\linewidth]{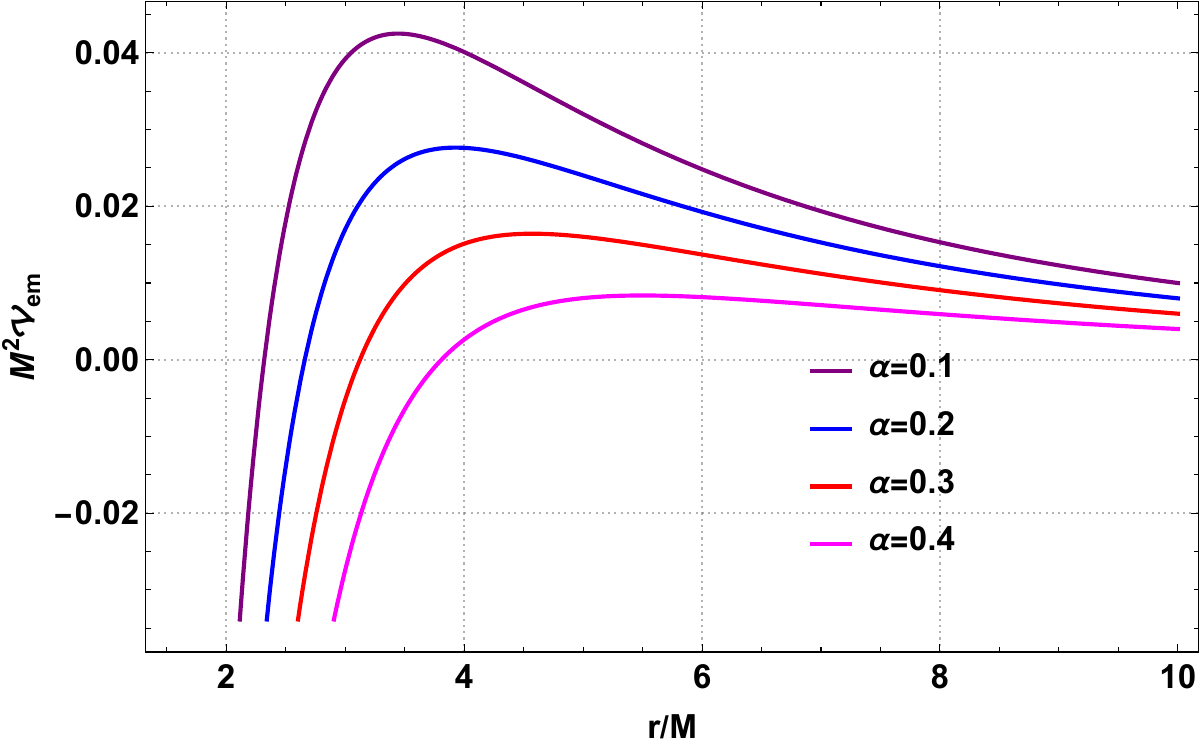}}\quad
    \subfloat[$\alpha=0.1,\lambda=0.5\,M^2$]{\centering{}\includegraphics[width=0.32\linewidth]{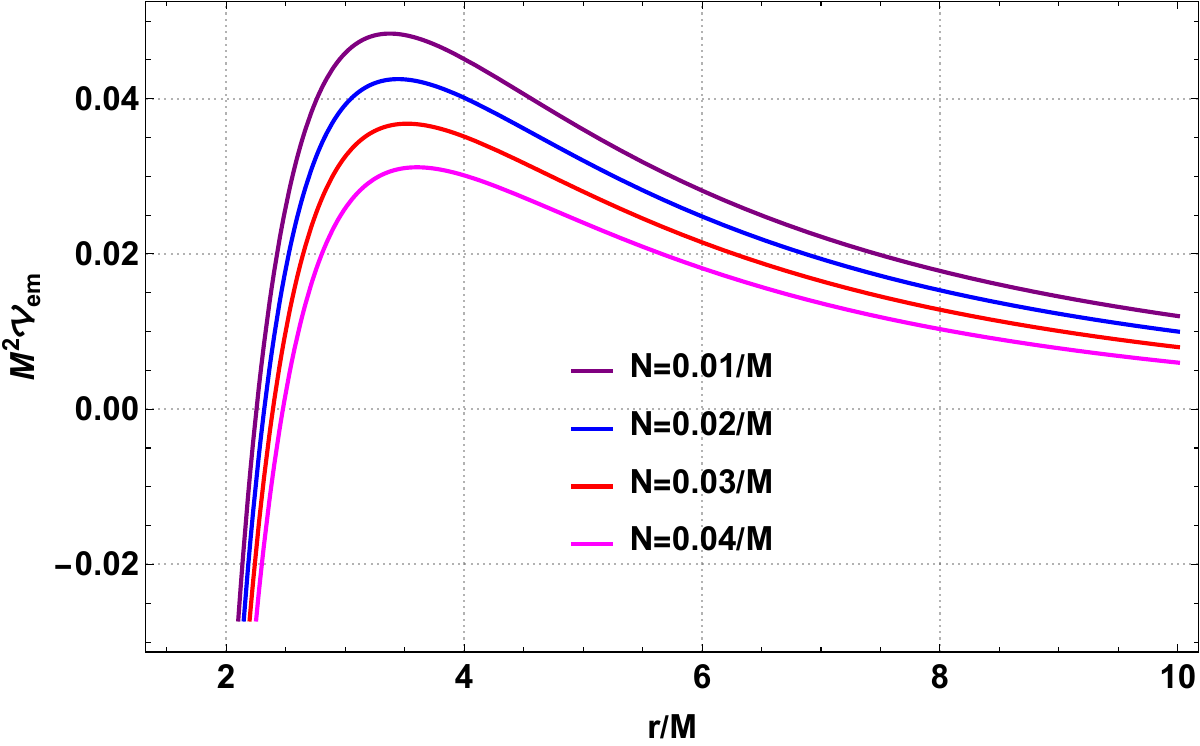}}\quad
    \subfloat[$N=0.02/M$]{\centering{}\includegraphics[width=0.32\linewidth]{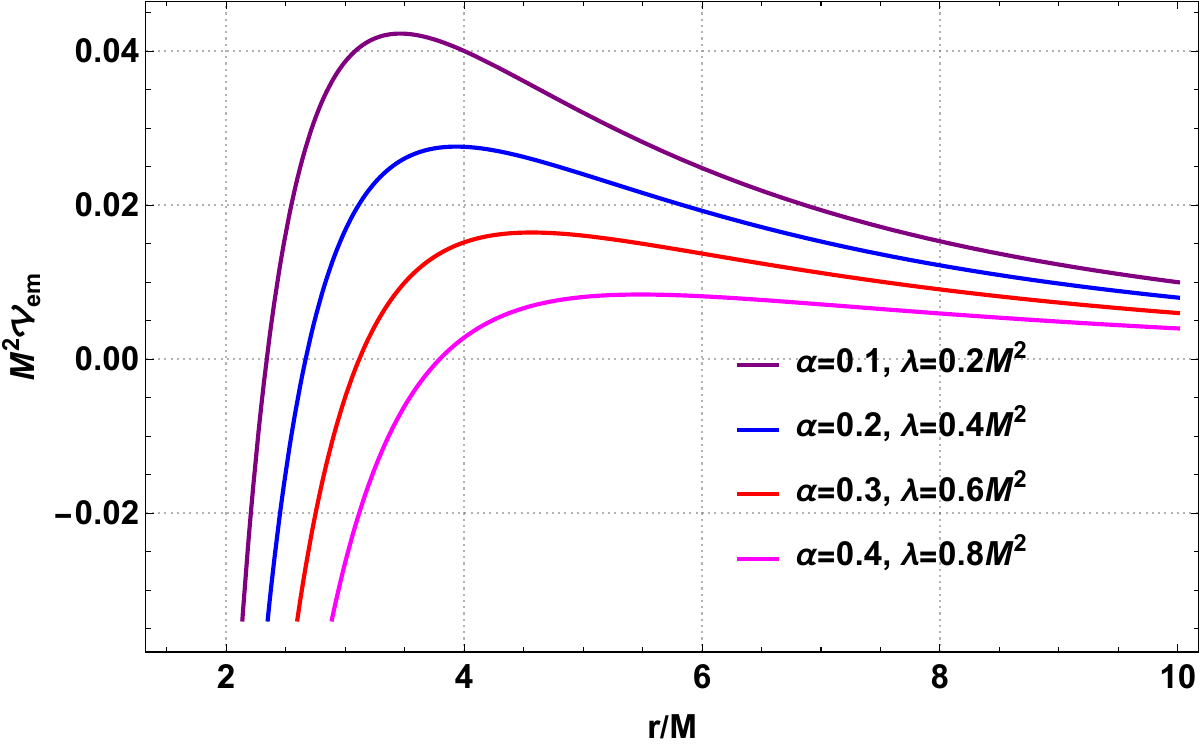}}\\
    \subfloat[$\lambda=0.05\,M^2$]{\centering{}\includegraphics[width=0.33\linewidth]{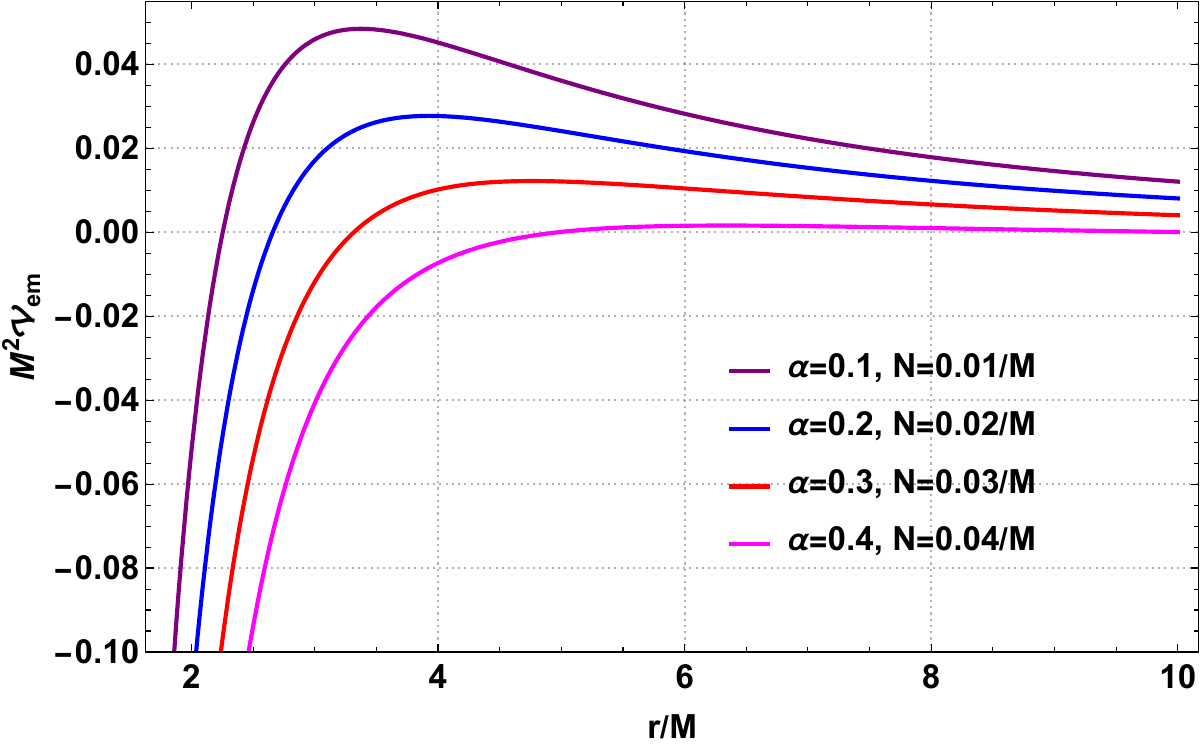}}\quad
    \subfloat[\mbox{all varies}]{\centering{}\includegraphics[width=0.33\linewidth]{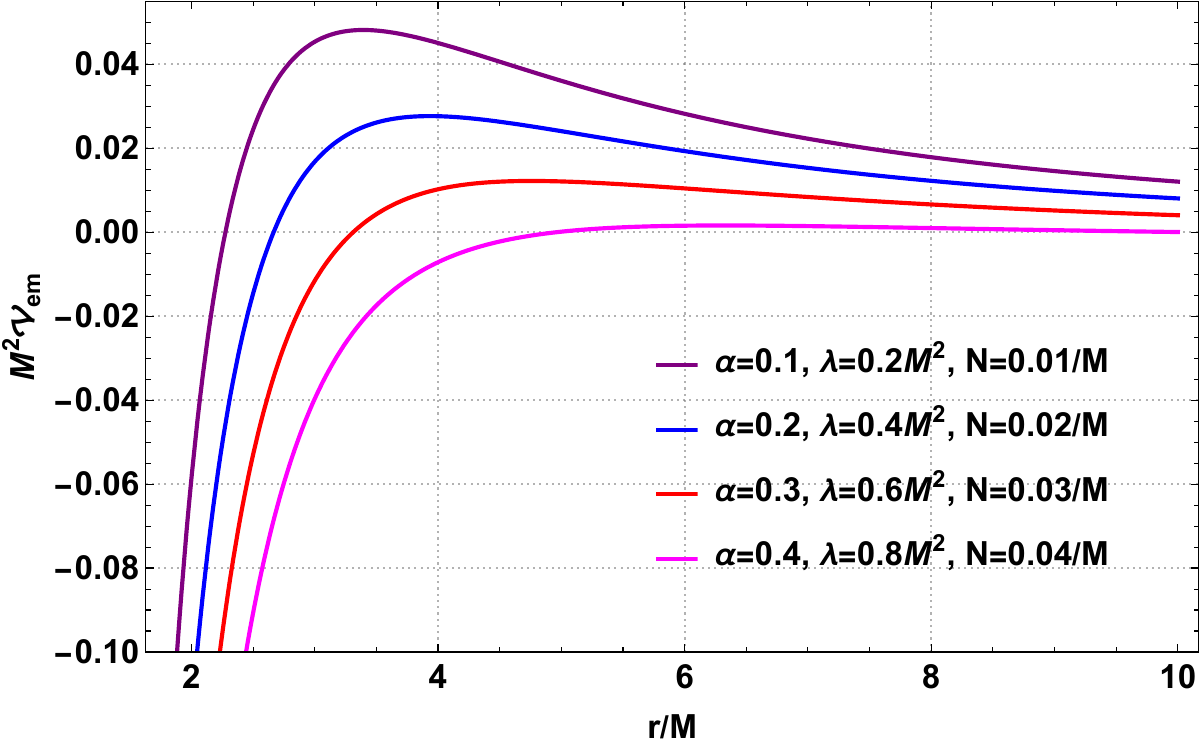}}
    \caption{Behavior of the EM perturbative potential for different values of the string parameter $\alpha$, the deformation parameter $\lambda$, the normalization constant $N$ of field, and their combination. Here, we set $\ell=1$.}
    \label{fig:electromagnetic}
\end{figure}

Figure~\ref{fig:electromagnetic} illustrates the behavior of the mass-scaled electromagnetic perturbative potential, $M^{2}V_{\text{em}}$, as a function of the dimensionless radial coordinate $r/M$ for different values of the relevant parameters. Specifically, panel~\ref{fig:electromagnetic}(a) illustrates the effect of the CS parameter $\alpha$, panel~\ref{fig:electromagnetic}(b) shows the influence of the QF parameter $N$, while panels~\ref{fig:electromagnetic}(c)--\ref{fig:electromagnetic}(e) depict their combined effects.
As in the scalar case, the electromagnetic potential exhibits a characteristic radial structure. In the vicinity of the event horizon ($r/M \to 2^{+}$), the potential is negative and varies rapidly, reflecting the dominance of strong gravitational effects close to the BH. At intermediate radial distances, the potential develops a pronounced positive maximum, forming an effective potential barrier that governs the propagation of electromagnetic waves. At larger radii the potential decreases and reaches zero at the quintessence horizon, so the EM barrier, like the scalar one, sits between two flat regions of the tortoise coordinate.
The peak of the electromagnetic perturbative potential plays a central physical role. It represents a scattering barrier where electromagnetic waves are partially reflected and partially transmitted, leading to the temporary trapping of radiation in the vicinity of the BH. This mechanism is responsible for the characteristic quasinormal ringing observed in the electromagnetic sector. Consequently, both the height and the radial position of the potential peak directly influence the QNM spectrum, determining the oscillation frequencies and decay rates of electromagnetic perturbations.
From Fig.~\ref{fig:electromagnetic}, it is evident that increasing the CS parameter $\alpha$, QF parameter $N$, and the quantum deformation parameter $\lambda$ modifies the electromagnetic potential in a definite direction. For instance, as $\alpha$ increases: (i) the height of the potential barrier decreases, (ii) the location of the peak shifts slightly toward larger radial distances, and (iii) the overall potential profile becomes progressively shallower. These trends indicate that the presence of a CS weakens the effective confining potential experienced by electromagnetic perturbations. Similarly, for the other parameters as well as their combination, the potential shows the same behavior.
Physically, the CS, QF, and quantum deformation modify the background geometry by altering the spacetime curvature, thereby reducing the efficiency with which electromagnetic waves are trapped near the BH. EM perturbations are therefore confined less tightly than in Schwarzschild, and Sec.~\ref{sec05d} converts that statement into numbers for the QNM frequencies. Whether the resulting shifts are observable is a separate question, which the present calculation does not address.

\subsection{Fermionic Perturbations: Spin-1/2 fields}

In this section, we present a preliminary analysis of spin-1/2 field perturbations through the Dirac equation in the background of the quantum Oppenheimer-Snyder BH solution, which includes a CS and quintessence. Spin-1/2 field perturbations have been studied in various BH spacetimes, such as the Schwarzschild background, and are governed by the Dirac equation (for both massive and massless cases) \cite{DRB}, given by:
\begin{equation}
\left[\gamma^{a}\,e^{\mu}_{a}\,(\partial_{\mu}+\Gamma_{\mu})+\mu\right]\,\Psi=0,\label{fermi1}
\end{equation}
Where $\mu$ is the mass of the Dirac field, and $e^{\mu}_{a}$ is the inverse of the tetrad $e^{a}_{\mu}$ defined by the metric tensor $g_{\mu\nu}$, as $g_{\mu\nu}=\eta_{ab}\, e^{a}_{\mu}\, e^{b}_{\nu}$ with $\eta_{ab}$ being the Minkowski metric, $\gamma^{a}$ are the Dirac matrices, and $\Gamma_{\mu}$ is the spin connection.

For the spherically symmetric BH background given by Eq. (\ref{bb1}), the equation of motion for massless spin-1/2 particles can be reduced to a one-dimensional Schrödinger-like wave equation \cite{DRB}:
\begin{equation}
\frac{d^{2}\psi_\text{Fermi}}{dr_{\ast }^{2}}+\left(\omega^{2}-\mathcal{V}_{\pm1/2}\right)\,\psi_\text{Fermi}=0,  \label{fermi2}
\end{equation}
where the tortoise coordinate $r_{\ast }$ is defined earlier and the effective potential is given by: {\small
\begin{eqnarray}
\mathcal{V}_{\pm 1/2}&=&\frac{\left( \frac{1}{2}+\ell\right) }{r^{2}}\,\left[\ell+\frac{1}{2} \pm \frac{r\,f^{\prime }(r)}{2\sqrt{f(r)}}\mp \sqrt{f(r)}\right]\,f(r)\nonumber\\
&=&\frac{\left(\frac{1}{2}+\ell\right) }{r^{2}}\,\left[\ell+\frac{1}{2} \pm \frac{\frac{M}{r}-\frac{2\,\lambda\,M^2}{r^4}-\frac{N}{2}\,r}{\sqrt{1-\alpha-\frac{2\,M}{r}+\frac{\lambda\,M^2}{r^4}-N\,r}}\mp \sqrt{1-\alpha-\frac{2\,M}{r}+\frac{\lambda\,M^2}{r^4}-N\,r}\right]\left[1-\alpha-\frac{2\,M}{r}+\frac{\lambda\,M^2}{r^4}-N\,r\right],\label{fermi3}
\end{eqnarray}
where $\ell$ is the standard spherical harmonics indices.} 

The effective potential for spin-1/2 fields perturbations given in Eq. (\ref{fermi3}) shows that various factors involved in the spacetime geometry influence this potential. These include the CS parameter $\alpha$, the quantum deformation parameter $\lambda$, and the quintessence normalization constant $N$, together with the BH mass $M$ and the multipole number $\ell$. No cosmological constant enters, contrary to a statement in the previous version; the model has none. In the limit $\lambda=0$ and $N=0$ the potential reduces to the Letelier case, and setting $\alpha=0$ recovers Schwarzschild.

\subsection{Mode stability and quasinormal frequencies}\label{sec05d}

The three potentials above are plotted, not solved, and a plot cannot establish stability. Two of the potentials also dip below zero over part of the static region, so the usual shortcut, a single positive barrier implies no growing modes, is unavailable. We therefore argue stability directly and then compute the spectrum.

Consider first the scalar sector. For a master equation $-\psi''(r_*)+V\psi=\omega^2\psi$ the relevant quadratic form is $\int(|\psi'|^2+V|\psi|^2)\,dr_*$, and for any smooth $S(r_*)$ it can be rewritten as
\begin{equation}
\int\left(\left|\frac{d\psi}{dr_*}+S\,\psi\right|^2+\tilde{V}\,|\psi|^2\right)dr_*,\qquad
\tilde{V}=V+\frac{dS}{dr_*}-S^2 ,\label{sdeform}
\end{equation}
after an integration by parts that produces no boundary term for perturbations decaying at both horizons \cite{Sdef}. Choosing $S=-f/r$ and using $dS/dr_*=f\,dS/dr$, the scalar potential (\ref{ff6}) gives
\begin{equation}
\tilde{V}_{\rm scalar}=\frac{\ell\,(\ell+1)}{r^2}\,f(r)\ \ge\ 0 \qquad (r_+<r<r_q),\label{sdeform2}
\end{equation}
since $f>0$ inside the static region. The $f'/r$ term, which is what made $V_{\rm scalar}$ negative near the horizon, cancels identically. The quadratic form is therefore non-negative for every $\ell$, including the $s$-wave, so the operator has no negative eigenvalue and no exponentially growing mode exists. The EM potential (\ref{em5}) is already of the form (\ref{sdeform2}) and needs no deformation.

The Dirac case follows from the structure of Eq.~(\ref{fermi3}). Setting
\begin{equation}
W(r)=\left(\ell+\tfrac{1}{2}\right)\frac{\sqrt{f(r)}}{r},
\end{equation}
a short calculation gives $\mathcal{V}_{\pm1/2}=W^2\pm dW/dr_*$ exactly. The two potentials are supersymmetric partners, the corresponding operators factorize as $A^{\dagger}A$ and $A A^{\dagger}$ with $A=d/dr_*+W$, and both are non-negative. The two sectors are isospectral, so it is enough to compute one of them.

Mode stability in hand, we turn to the frequencies. We use the third-order WKB method of Schutz, Will, Iyer, and Konoplya \cite{WKB1,WKB2,WKB3}, in which
\begin{equation}
\omega^2=V_0+\sqrt{-2\,V_0''}\;\Lambda_2-i\left(n+\tfrac{1}{2}\right)\sqrt{-2\,V_0''}\,\left(1+\Lambda_3\right),\label{wkbformula}
\end{equation}
with all derivatives taken with respect to $r_*$ at the peak of the potential, and $\Lambda_2$, $\Lambda_3$ the standard correction terms. As a check on the implementation, the Schwarzschild limit of our code returns $M\omega=0.483211-0.096805\,i$ for the scalar $\ell=2$, $n=0$ mode and $M\omega=0.457131-0.095065\,i$ for the EM $\ell=2$, $n=0$ mode, against the continued-fraction values $0.483644-0.096759\,i$ and $0.457593-0.095004\,i$ \cite{Leaver}. The agreement to three decimal places is what third-order WKB delivers for $\ell>n$, and it sets the accuracy we claim below. We quote fundamental modes only, since WKB degrades for overtones with $n\ge\ell$.

Table~\ref{tab:qnm} lists the fundamental frequencies. Every imaginary part is negative, consistent with the stability argument above. The trends are simple to read off. Raising $\alpha$ from $0$ to $0.2$ at fixed $\lambda=0.2M^2$ and $N=0.01/M$ lowers the scalar $\ell=2$ frequency from $0.461838-0.090369\,i$ to $0.319015-0.055911\,i$, a drop in both the oscillation frequency and the damping rate. This follows from Eq.~(\ref{sdeform2}) together with the shape of $f$: the CS term reduces $f$ everywhere by the constant $\alpha$, which lowers the barrier and lengthens the ringdown. Increasing $N$ from $0$ to $0.02/M$ acts in the same direction but more weakly, since the QF term is linear in $r$ and reaches the barrier region only through the combination $N r_{\rm ph}$. The quantum deformation moves the frequencies least of all. Turning $\lambda$ from $0$ to $0.5M^2$ at $\alpha=0.1$ and $N=0.01/M$ changes the scalar $\ell=2$ real part from $0.387295$ to $0.390044$, a shift in the third digit, which matches the $r^{-4}$ suppression of that term at the barrier.

\begin{table}[!htbp]
\centering
\setlength{\tabcolsep}{4pt}
\renewcommand{\arraystretch}{1.2}
\begin{adjustbox}{max width=\textwidth}
\begin{tabular}{|c|c|c|c|c|c|c|}
\hline
$\boldsymbol{\alpha}$ & $\boldsymbol{\lambda/M^2}$ & $\boldsymbol{N M}$ & \textbf{scalar} $\boldsymbol{\ell=1}$ & \textbf{scalar} $\boldsymbol{\ell=2}$ & \textbf{EM} $\boldsymbol{\ell=1}$ & \textbf{EM} $\boldsymbol{\ell=2}$ \\
\hline
0.0 & 0.0 & 0.00  & $0.291114-0.098001\,i$ & $0.483211-0.096805\,i$ & $0.245870-0.093106\,i$ & $0.457131-0.095065\,i$ \\
0.1 & 0.0 & 0.00  & $0.247085-0.079234\,i$ & $0.411605-0.078364\,i$ & $0.212355-0.075628\,i$ & $0.391559-0.077093\,i$ \\
0.1 & 0.2 & 0.00  & $0.247750-0.078692\,i$ & $0.412741-0.077895\,i$ & $0.213167-0.075096\,i$ & $0.392766-0.076626\,i$ \\
0.0 & 0.2 & 0.01  & $0.277205-0.091580\,i$ & $0.461838-0.090369\,i$ & $0.236732-0.086746\,i$ & $0.438537-0.088651\,i$ \\
0.1 & 0.0 & 0.01  & $0.231519-0.073622\,i$ & $0.387295-0.072627\,i$ & $0.201103-0.070054\,i$ & $0.369785-0.071369\,i$ \\
0.1 & 0.2 & 0.01  & $0.232161-0.073157\,i$ & $0.388375-0.072225\,i$ & $0.201859-0.069606\,i$ & $0.370918-0.070972\,i$ \\
0.1 & 0.5 & 0.01  & $0.233130-0.072394\,i$ & $0.390044-0.071566\,i$ & $0.203017-0.068857\,i$ & $0.372674-0.070317\,i$ \\
0.1 & 0.2 & 0.02  & $0.216098-0.067466\,i$ & $0.363104-0.066435\,i$ & $0.189983-0.064013\,i$ & $0.348100-0.065219\,i$ \\
0.2 & 0.2 & 0.01  & $0.189867-0.056600\,i$ & $0.319015-0.055911\,i$ & $0.168132-0.054093\,i$ & $0.306487-0.055032\,i$ \\
\hline
\end{tabular}
\end{adjustbox}
\caption{Fundamental ($n=0$) QNM frequencies $M\omega$ for scalar and EM test fields, from third-order WKB applied to Eqs.~(\ref{ff6}) and (\ref{em5}) with $M=1$ and $w=-2/3$. The first row is the Schwarzschild benchmark. Accuracy is limited by the WKB order and is at the level of the third decimal place, as the comparison with Ref.~\cite{Leaver} in the text indicates.}
\label{tab:qnm}
\end{table}

The Dirac frequencies follow the same pattern. With the convention of Eq.~(\ref{fermi3}), in which the spinor index enters as $\ell+1/2$, we obtain $M\omega=0.279242-0.097143\,i$ for $\ell=1$ in the Schwarzschild limit, in agreement with Ref.~\cite{Cho} once the index conventions are matched, and $0.226838-0.072209\,i$ at $\alpha=0.1$, $\lambda=0.2M^2$, $N=0.01/M$. The reduction in damping relative to Schwarzschild is again driven mostly by $\alpha$.

Three limitations should be recorded. The calculation treats scalar, EM, and Dirac fields as test fields on a fixed background, so it says nothing about spin-2 gravitational perturbations, which would require the perturbed field equations of the effective theory and are not obtained by any of the master equations above. The frequencies are WKB estimates rather than converged eigenvalues. And a spacetime with two horizons carries, in addition to the photon-sphere modes tabulated here, a family of purely damped modes associated with the outer horizon, which the WKB treatment of a single barrier does not capture.

\section{Thermal Properties of QOS BH} \label{sec06}

In this section, we examine the thermodynamic properties and thermal stability of the QOS BH. Our analysis focuses on key thermodynamic quantities, including the Hawking temperature, specific heat capacity, and Gibbs free energy. We investigate how these quantities are influenced by various physical parameters. In particular, we explore the roles of the CS, LQG deformations, and the presence of quintessential dark energy in modifying the thermal behavior of the BH. These parameters not only affect the location of the event horizon but also introduce significant corrections to the thermodynamic profiles, potentially altering the phase transition structures and stability conditions. By studying the interplay between these geometric and field-theoretic contributions, we aim to reveal how nontrivial matter fields and quantum-gravity-induced deformations shape the thermodynamic phase space of BH solutions.

Before proceeding with the computation of the Hawking temperature, we first obtain an expression for the ADM mass $M$ by solving the condition that defines the event horizon. This occurs at the radial coordinate $r = r_{+}$, where the metric function $f(r)$ vanishes. By applying this condition to the metric function given in Eq.~(\ref{bb2}) and whence imposing the horizon condition $f(r_+) = 0$, one gets
\begin{equation}
1 - \alpha - \frac{2M}{r_+} + \frac{\lambda M^2}{r_+^4} - N r_+ = 0. \label{horizon_condition}
\end{equation}
Multiplying both sides by $r_+^4$ to eliminate denominators yields a quadratic equation in $M$:
\begin{equation}
\lambda M^2 - 2 M r_+^3 + (1 - \alpha - N r_+) r_+^4 = 0. \label{mass_quadratic}
\end{equation}
Solving Eq.~(\ref{mass_quadratic}) for $M$ using the quadratic formula, we find
\begin{equation}
M = \frac{r_+^3}{\lambda} \left[ 1 \pm \sqrt{1 - \lambda \cdot \frac{1 - \alpha - N r_+}{r_+^{2}}} \right]. \label{mass_gen}
\end{equation}
The discriminant of Eq.~(\ref{mass_quadratic}) is $4r_+^6-4\lambda(1-\alpha-Nr_+)r_+^4$, whose square root is $2r_+^3\sqrt{1-\lambda(1-\alpha-Nr_+)/r_+^2}$. The argument therefore carries $r_+^2$, not $r_+$ as printed in the previous version; with the earlier expression the horizon condition $f(r_+)=0$ is not satisfied, as one checks by substitution at any $r_+$.
The physical branch of the solution is selected by requiring that the ADM mass $M$ approaches the Schwarzschild limit as $\lambda \rightarrow 0$, which corresponds to choosing the negative root. Hence, the final expression for the ADM mass is given by
\begin{equation}
M = \frac{r_+^3}{\lambda} \left[ 1 - \sqrt{1 - \lambda \cdot \frac{1 - \alpha - N r_+}{r_+^{2}}} \right]. \label{mass1}
\end{equation}

It is important to emphasize that the square root in Eq.~(\ref{mass1}) must remain real to ensure that the solution describes a physically meaningful BH with a well-defined event horizon. This requirement leads to the condition
\begin{equation}
\lambda \leq \frac{r_+^{2}}{1 - \alpha - N r_+}, \label{lambda_constraint}
\end{equation}
which is valid only when the denominator is positive, i.e.,
\begin{equation}
1 - \alpha - N r_+ > 0. \label{reality_condition}
\end{equation}
This inequality serves two purposes. First, it guarantees that the argument of the square root is non-negative, thereby ensuring that the ADM mass remains real. Second, when combined with the positivity of the deformation parameter $\lambda > 0$-as required by quantum gravity scenarios such as LQG-it ensures that the ADM mass itself remains positive, $M > 0$. A negative $\lambda$ would render the mass negative or even divergent in the classical limit $\lambda \to 0^-$, which would contradict physical expectations and violate energy conditions. Thus, the constraints in Eqs.~(\ref{lambda_constraint}) and~(\ref{reality_condition}) collectively ensure the existence of a BH horizon and preserve the physical viability of the solution in the presence of quantum gravitational corrections.

\subsection{Hawking Temperature}

Before writing down a temperature we have to say which horizon it belongs to and how the Killing vector is normalized, because both questions have non-trivial answers here. The spacetime carries two Killing horizons in the physical range, at $r_+$ and at $r_q$, and $\partial_t$ generates both. The surface gravity is $\kappa=|f'(r_h)|/2$ at either of them, and the temperature is
\begin{equation}
T=\frac{\kappa}{2\pi}=\frac{|f'(r_h)|}{4\pi},\label{temp_kappa}
\end{equation}
where the absolute value matters: $f'>0$ at the BH horizon and $f'<0$ at the quintessence horizon, and dropping the modulus at the second one produces a spurious negative temperature. This is the standard situation for spacetimes with a cosmological-type horizon \cite{GibbonsHawking}.

A second issue is normalization. In an asymptotically flat spacetime $\partial_t$ is fixed by $f\to1$ at infinity, which is what makes $T_H$ the temperature measured by a distant observer. Here $f\to-\infty$, so no such normalization exists, and $t$ is a coordinate whose scale has to be tied to something physical. Two options are standard. One may leave the normalization as it stands, in which case Eq.~(\ref{temp_kappa}) is a bookkeeping temperature conjugate to the entropy in the first law, which is how we use it below. Alternatively one may normalize at the radius where $f$ is maximal between the two horizons, the prescription of Bousso and Hawking \cite{BoussoHawking}, or refer everything to a static observer at $r_O$ through the Tolman relation $T_{\rm loc}=T/\sqrt{f(r_O)}$ \cite{Tolman}. The three differ by a positive factor and none of them changes any sign, so the qualitative statements below are unaffected. We use the first throughout and quote $T$ at the BH horizon unless stated otherwise.

For the BH horizon, then,
\begin{equation}
T_H = \frac{f'(r)}{4\pi} \bigg|_{r = r_+}. \label{temp2start}
\end{equation}

Using the explicit form of $f(r)$, we compute:
\begin{equation}
f'(r) = \frac{2M}{r^2} - \frac{4\lambda M^2}{r^5} - N.
\end{equation}
Substituting this into Eq.~(\ref{temp2start}) and expressing $M$ and $M^2$ in terms of $r_+$ from Eq.~(\ref{mass1}), we obtain:
\begin{equation}
T_H = \frac{r_+}{4\pi \lambda} \left[2(1 - \chi) - 4(1 - \chi)^2 \right]- \frac{N}{4\pi}, \label{temp2}
\end{equation}
where
\begin{equation}
\chi = \sqrt{1 - \lambda \cdot \frac{1 - \alpha - N r_+}{r_+^{2}}}. \label{chi_def}
\end{equation}

This compact expression highlights the dependence of the Hawking temperature $T_H$ on the horizon radius $r_+$, the CS parameter $\alpha$, the QF parameter $N$, and the quantum correction parameter $\lambda$.

In the classical limit $\lambda \to 0$, the temperature reduces to
\begin{equation}
T_H = \frac{1 - \alpha - 2 N r_+}{4\pi r_+}, \label{temp2a}
\end{equation}
which matches the expected Schwarzschild-like result with CS and QF corrections.

\begin{figure}[ht!]
    \centering
    \includegraphics[width=0.72\linewidth]{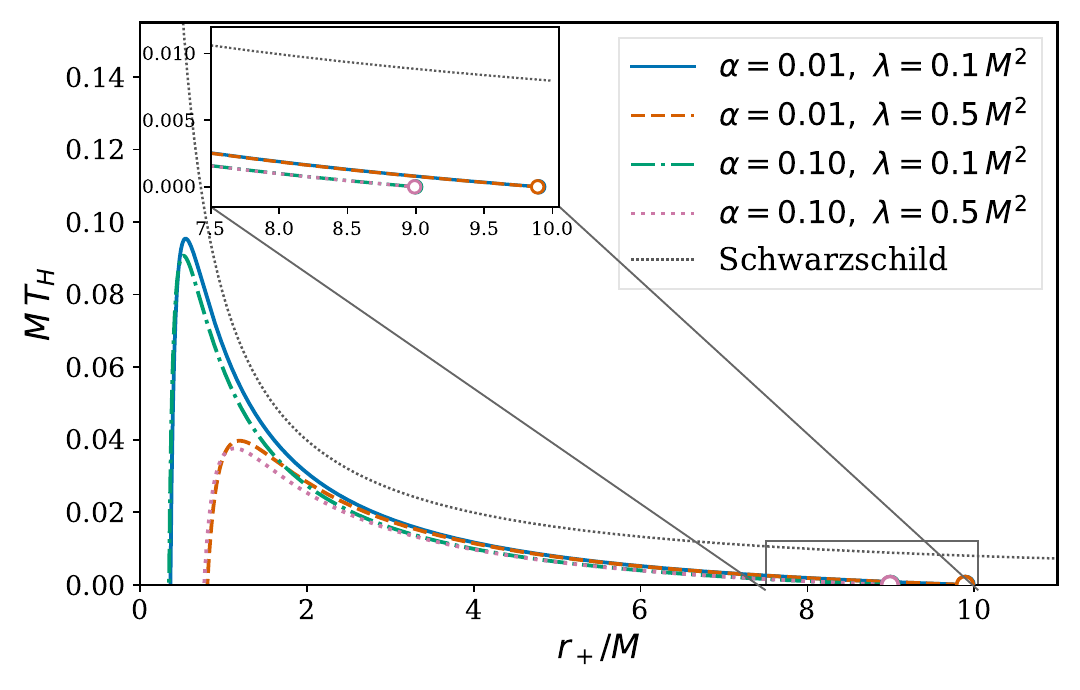}
\caption{Hawking temperature at the BH horizon against $r_+/M$ for $N=0.05/M$, from Eq.~(\ref{temp2}) with $\chi$ as given in Eq.~(\ref{chi_def}). Each curve is restricted to the interval on which $r_+$ is the middle root of Eq.~(\ref{bb4a}), bounded below by the reality condition (\ref{lambda_constraint}) and above by the degenerate configuration, marked by a filled circle, at which $r_+$ and $r_q$ coincide. The inset magnifies the approach to that endpoint. The dotted curve is Schwarzschild.}
    \label{fig:hawking_temp}
\end{figure}

Figure~\ref{fig:hawking_temp} shows $T_H$ against $r_+$ at $N=0.05/M$ for four combinations of $\alpha$ and $\lambda$. Each curve is plotted only over the range in which $r_+$ is genuinely the BH horizon, that is, the middle root of Eq.~(\ref{bb4a}). The inner end of that range is set by the reality condition (\ref{lambda_constraint}), which for $\alpha=0.01$ and $\lambda=0.1M^2$ gives $r_+\ge0.31215M$. The outer end is the degenerate configuration described below.

The previous version of this work reported negative Hawking temperatures in part of the parameter space and interpreted them as exotic thermodynamic phases. That interpretation does not survive a check of the horizon ordering, and we withdraw it. The sign change occurs exactly where $r_+$ ceases to be the BH horizon. Tracking the roots of Eq.~(\ref{bb4a}) as $r_+$ is varied at fixed $(\alpha,\lambda,N)$ makes this explicit. At $\alpha=0.01$, $\lambda=0.1M^2$, and $N=0.05/M$ the three roots are $(0.47266,\,5.00000,\,14.79462)M$ when $r_+=5M$, and $(0.51442,\,9.50000,\,10.29624)M$ when $r_+=9.5M$: the BH horizon and the quintessence horizon approach each other. They merge at $r_+=9.89812M$, where the corresponding mass is $M=2.45056$ and $T_H$ vanishes. Beyond that value the ordering has swapped, so that the radius being fed into Eq.~(\ref{temp2}) is the outer root: at $r_+=10.5M$ the roots are $(0.51409,\,9.29623,\,10.50000)M$, and the quantity computed is $f'(r_q)/4\pi<0$. What looked like a negative-temperature BH is the quintessence horizon with the modulus in Eq.~(\ref{temp_kappa}) omitted.

The vanishing point is a degenerate horizon of Nariai type \cite{Nariai}. In the classical limit it is available in closed form: setting $\lambda=0$ in Eq.~(\ref{temp2a}) gives $T_H=0$ at $r_+=(1-\alpha)/(2N)$, which for $\alpha=0.01$ and $N=0.05/M$ is $9.9M$ and for $\alpha=0.1$ is $9.0M$. The numerical roots quoted above, $9.89812M$ and $8.99812M$ at $\lambda=0.1M^2$, sit just below these values, so the quantum deformation shifts the degenerate radius inward by a small amount. On the admissible branch $T_H\ge0$ everywhere, with equality only at the endpoint. No negative-temperature phase exists in this model.

The quintessence horizon carries a temperature of its own, $T_q=|f'(r_q)|/4\pi$, which is small because $r_q$ is large. For the parameters of Table~\ref{istab_new} with $\alpha=0.1$, $\lambda=0.2M^2$, and $N=0.01/M$ one finds $M T_q=7.751\times10^{-4}$ against $M T_+=2.929\times10^{-2}$ at the BH horizon. The two differ by more than an order of magnitude, so the spacetime is not in thermal equilibrium and there is no single temperature for the whole static region. The equilibrium thermodynamics used in the rest of this section applies to the BH horizon separately, which is the usual practice for Schwarzschild-de Sitter and is an approximation whose accuracy improves as $r_q/r_+$ grows.

The temperature maximum visible in Fig.~\ref{fig:hawking_temp} deserves a comment, since it has no Schwarzschild counterpart. For Schwarzschild, $T_H=1/(4\pi r_+)$ falls monotonically. Here each curve rises from the small-$r_+$ end, peaks, and then decreases to zero at the degenerate radius. The rise comes from the quantum term, which dominates $f'$ when $r_+$ is small, and the fall comes from the QF term, which dominates when $r_+$ approaches $r_q$. At $\alpha=0.01$ and $\lambda=0.1M^2$ the peak reaches $M T_H=0.0954$ near $r_+=0.55M$, while at $\lambda=0.5M^2$ it drops to $0.0397$ near $r_+=1.2M$, so a larger deformation flattens the profile. The presence of a maximum, rather than its height, is what matters thermodynamically, because $\partial T_H/\partial r_+=0$ is where the specific heat diverges.

\subsection{Specific Heat Capacity and Thermal Stability}

The specific heat capacity is a central quantity in BH thermodynamics, providing direct information about thermal stability and the existence of phase transitions~\cite{Davies1977,Hut1977}. For a thermodynamic system, the heat capacity measures the energy required to produce a unit change in temperature. In the context of BH physics, where the Bekenstein-Hawking entropy is given by $S = A/4 = \pi r_+^2$, the specific heat capacity can be expressed as

\begin{equation}
    C_p=T_H\,\left(\frac{\partial S}{\partial T_H}\right)=2\pi r_{+} \frac{T_H}{\left(\frac{\partial T_H}{\partial r_{+}}\right)}.\label{temp3}
\end{equation}

The second equality follows from applying the chain rule and using $\partial S/\partial r_+ = 2\pi r_+$. The sign of $C_p$ carries important physical meaning: positive heat capacity ($C_p > 0$) indicates a locally stable BH that can achieve thermal equilibrium with a heat reservoir, while negative heat capacity ($C_p < 0$) signals thermodynamic instability where the BH heats up as it loses energy. Furthermore, divergences in $C_p$ mark critical points corresponding to second-order phase transitions, where the BH undergoes qualitative changes in its thermodynamic behavior~\cite{Davies1977}.

Using the ADM mass $M$ given in Eq.~(\ref{mass1}) and the Hawking temperature $T_H$ in Eq.~(\ref{temp2}), the general expression for the specific heat capacity becomes
\begin{equation}
    C_p=2\,\pi\,r_{+}\,\frac{\frac{r_+}{4\pi \lambda} \left[ 2(1 - \chi) - 4(1 - \chi)^2 \right] - \frac{N}{4\pi}}{\frac{\partial}{\partial r_{+}}\left[\frac{r_+}{4\pi \lambda} \left[ 2(1 - \chi) - 4(1 - \chi)^2 \right] - \frac{N}{4\pi}\right]},\label{temp3a}
\end{equation}
Though this expression is quite lengthy in the general case, it captures the complete thermodynamic behavior including quantum corrections.

However, in the limit $\lambda=0$, corresponding to the absence of quantum deformation effects in the BH solution, we find the specific heat capacity as
\begin{equation}
    C_p=-\pi\,r^2_{+}\,\frac{(1-\alpha-N\,r_{+})}{(1-\alpha-3\,N\,r_{+})}.\label{temp5}
\end{equation}

\begin{figure}[ht!]
    \centering
    \includegraphics[width=0.72\linewidth]{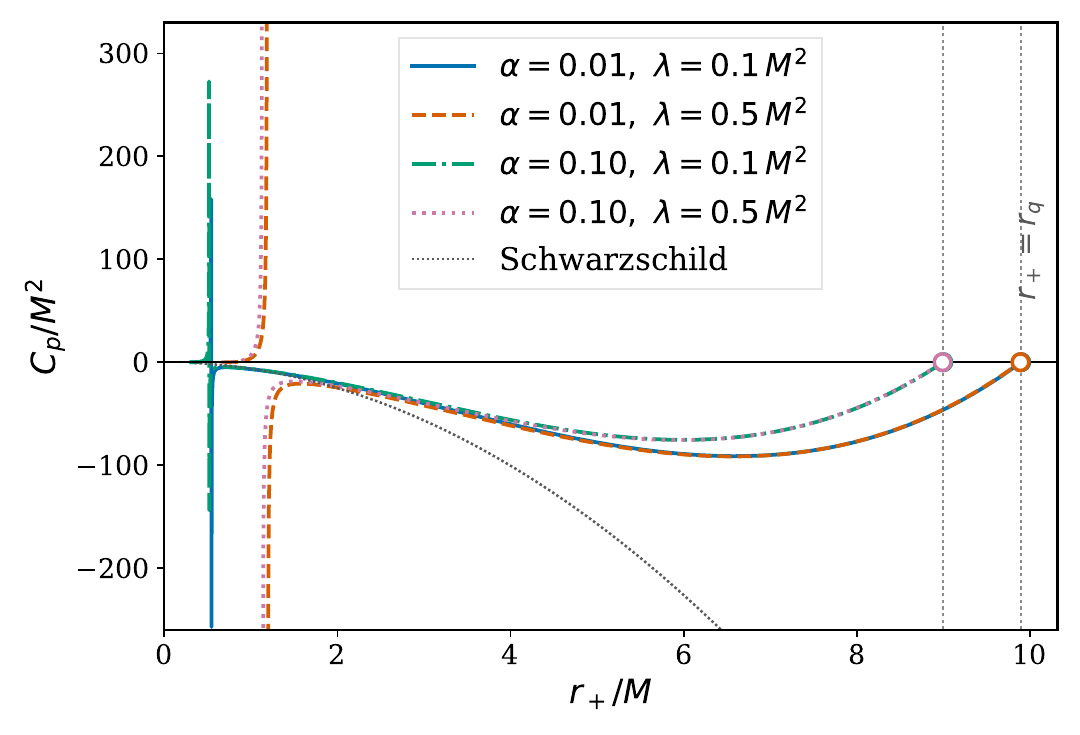}
 \caption{Specific heat against $r_+/M$ at $N=0.05/M$, from Eq.~(\ref{temp3a}) with the corrected $\chi$, over the same admissible interval as Fig.~\ref{fig:hawking_temp}. The divergence in each curve sits at the temperature maximum, where $\partial T_H/\partial r_+=0$, and separates a small stable branch from the large unstable branch. Each curve ends at its own degenerate radius, marked by an open circle and by a vertical dotted guide, where the BH horizon merges with the quintessence horizon so that $T_H=0$ and hence $C_p\to0$ at finite entropy. The two endpoints lie at $r_+=8.998M$ for $\alpha=0.10$ and $r_+=9.898M$ for $\alpha=0.01$, the pairs differing only in the fourth digit with $\lambda$, so the markers of each pair overlap. The dotted curve is Schwarzschild.}
    \label{fig:specific_heat}
\end{figure}

Figure~\ref{fig:specific_heat} shows the specific heat capacity $C_p$ as a function of the horizon radius $r_+$ for various combinations of the CS parameter $\alpha$ and quantum deformation parameter $\lambda$, with the QF parameter fixed at $N = 0.05$. The analysis is restricted to the physically admissible parameter space satisfying the reality conditions $1 - \alpha - Nr_+ > 0$ and $\lambda \leq r_+/(1 - \alpha - Nr_+)$.
The sign of $C_p$ determines local thermodynamic stability. When $C_p > 0$, the BH can reach thermal equilibrium with a surrounding heat reservoir: absorbing energy raises the temperature, which enhances radiation until balance is achieved. When $C_p < 0$, the system is thermodynamically unstable---absorbing energy causes cooling, which promotes further absorption and leads to runaway behavior. From Figure~\ref{fig:specific_heat}, the branch to the left of the divergence has $C_p>0$ and is locally stable, while the whole branch to its right has $C_p<0$, as in Schwarzschild. The stable branch is narrow. For $\alpha=0.01$ and $\lambda=0.1M^2$ it spans roughly $0.31\lesssim r_+/M\lesssim0.55$, and it is a quantum-deformation artifact in the sense that it disappears as $\lambda\to0$, where the divergence migrates to the origin.
The divergences visible in Figure~\ref{fig:specific_heat} mark second-order phase transitions where the thermal response changes abruptly. These critical points occur at $\partial T_H/\partial r_+ = 0$, i.e., at extrema of the Hawking temperature. For the Schwarzschild BH (dashed curve), $C_p = -2\pi r_+^2$ remains negative and monotonic, so no phase transition exists. Introducing nonzero $\alpha$ and $\lambda$ generates critical points that separate distinct thermodynamic branches.
The parameter dependence runs as follows. Raising $\lambda$ from $0.1M^2$ to $0.5M^2$ moves the divergence outward, from $r_+\simeq0.55M$ to $r_+\simeq1.2M$, because it shifts the temperature maximum outward. Raising $\alpha$ from $0.01$ to $0.10$ moves it inward by a smaller amount and lowers the temperature at the peak. The resulting phase structure has one Davies-type critical point rather than none, which is the whole of the difference from Schwarzschild in this sector.
These features carry direct physical consequences. A BH in an unstable phase ($C_p < 0$) evaporates continuously via Hawking radiation without reaching equilibrium. If the evaporation trajectory crosses a divergence into a stable branch ($C_p > 0$), the BH may settle into a quasi-equilibrium state. The modified phase structure also suggests that remnant formation in this model could differ from standard predictions, with the ultimate fate governed by the interplay of $\alpha$, $\lambda$, and $N$.

\subsection{Gibbs Free Energy}

The Gibbs free energy provides a criterion for global thermodynamic stability and governs the spontaneous direction of phase transitions at fixed temperature~\cite{HawkingPage1983,Kubiznak2012}. In BH thermodynamics, the Gibbs free energy is defined as
\begin{equation}
G = M - T_H S,
\end{equation}
where $M$ represents the ADM mass of the BH, $T_H$ is the Hawking temperature, and $S = \pi r_+^2$ is the Bekenstein-Hawking entropy. This expression mirrors the standard thermodynamic definition $G = H - TS$ with the mass $M$ playing the role of enthalpy. For systems at constant temperature, the configuration with lower Gibbs free energy is thermodynamically preferred. Consequently, sign changes and extrema in $G$ signal phase transitions between distinct BH states or between BH and non-BH (e.g., thermal radiation) phases. The celebrated Hawking-Page transition~\cite{HawkingPage1983} between Schwarzschild-AdS BHs and thermal AdS space provides the prototype example of such behavior.

Using the general expressions for the ADM mass $M$ given in Eq.~(\ref{mass1}) and the Hawking temperature $T_H$ in Eq.~(\ref{temp2}), the Gibbs free energy becomes
\begin{equation}
   G=\frac{r_+^3}{\lambda} \left[ 1 - \sqrt{1 - \lambda \cdot \frac{1 - \alpha - N r_+}{r_+}} \right] - \left[\frac{r_+}{4\pi \lambda} \left[ 2(1 - \chi) - 4(1 - \chi)^2 \right] - \frac{N}{4\pi}\right]\pi r_+^2,\label{temp5a}
\end{equation}
where $\chi$ is defined in Eq.~(\ref{chi_def}).

In the simplified case where $\lambda=0$, the free energy becomes
\begin{equation}
   G=\frac{r_{+}}{2}\,\left(1-\alpha-N\,r_{+}\right)-\frac{\pi r_{+}^2}{4}\,\frac{(1-\alpha-2\,N\,r_{+})}{4\pi r_{+}}.\label{temp6}
\end{equation}
Simplifying this expression, we obtain
\begin{equation}
   G=\frac{r_{+}}{2}\,\left(1-\alpha-N\,r_{+}\right)-\frac{r_{+}}{4}\,(1-\alpha-2\,N\,r_{+}).\label{temp6a}
\end{equation}

\begin{figure}[ht!]
    \centering
    \includegraphics[width=0.72\linewidth]{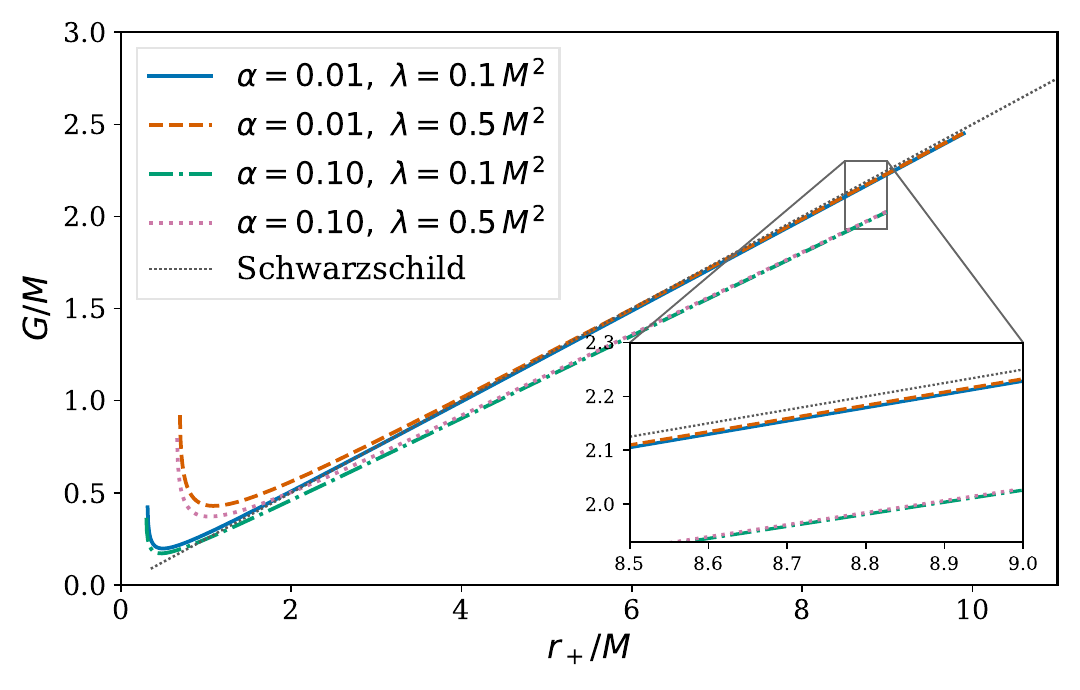}
    \caption{Gibbs free energy against $r_+/M$ at $N=0.05/M$, from Eq.~(\ref{temp5a}), over the admissible interval. The dotted curve is the Schwarzschild result $G=r_+/4$. On this branch $G$ stays positive and rises monotonically toward the degenerate endpoint. The inset magnifies the window $8.5\le r_+/M\le9.0$, where the two pairs of curves separate by less than one line width on the main panel; the guide lines mark the magnified region. Inside the inset the $\alpha=0.10$ pair terminates at its degenerate radius, $r_+=8.99M$.}
    \label{fig:gibbs_free_energy}
\end{figure}

While the specific heat capacity determines local thermodynamic stability, the Gibbs free energy governs global stability by identifying which phase is thermodynamically preferred at a given temperature. Among all configurations accessible to the system, the one with the lowest $G$ will be realized in equilibrium. Figure~\ref{fig:gibbs_free_energy} displays the Gibbs free energy as a function of the horizon radius $r_+$ for several combinations of the CS parameter $\alpha$ and quantum deformation parameter $\lambda$, with the QF parameter fixed at $N = 0.05$.
The sign of $G$ carries direct physical meaning. Regions where $G > 0$ indicate that the BH phase is thermodynamically disfavored relative to a reference background (such as flat spacetime or thermal radiation at the same temperature). Regions where $G < 0$ indicate that the BH configuration is globally preferred and represents the equilibrium state of the system. The zeros of $G$ mark phase equilibrium points where the BH and reference phases have equal free energy; small perturbations can drive the system toward either phase.
Figure~\ref{fig:gibbs_free_energy} shows that $G$ stays positive across the whole admissible interval and increases with $r_+$, approaching $M$ itself at the degenerate endpoint where $T_H S\to0$. There is no zero crossing and hence no Hawking-Page-like transition in this sector, which differs from the statement made in the earlier version of this work. That statement rested on curves extended past the degenerate radius, into the region where $r_+$ is no longer the BH horizon.
The parameter dependence is as follows. Increasing the CS parameter $\alpha$ raises the Gibbs free energy across all horizon radii, making BH configurations less favorable compared to the Schwarzschild case. This behavior reflects the additional energy cost associated with the conical deficit created by the CS. The quantum deformation parameter $\lambda$ has a more subtle effect: it modifies the curvature of the $G(r_+)$ profile near small radii without drastically changing the large-$r_+$ behavior. The interplay between $\alpha$ and $\lambda$ determines whether stable BH configurations exist and, if so, their characteristic sizes.
A Hawking-Page transition \cite{HawkingPage1983} needs a reference background at the same temperature whose free energy can cross that of the BH, and in an asymptotically anti-de Sitter setting thermal radiation supplies one. This geometry supplies no such background, since it is neither flat nor anti-de Sitter at large radius, and the region outside $r_q$ is not static. What the figure does show is the Davies-type second-order transition already identified in the specific heat, at the temperature maximum. We make no claim beyond that.

Three results survive from this section. The Hawking temperature is non-negative on the entire admissible branch and vanishes only where the BH and quintessence horizons merge, so the negative-temperature phases reported earlier were an artifact of extending the formulas past that point. The temperature acquires a maximum, absent in Schwarzschild, whose position is set by $\lambda$ and whose height is reduced by both $\alpha$ and $\lambda$; the maximum carries a Davies-type second-order transition, separating a narrow stable branch at small $r_+$ from the unstable branch that occupies the rest of the interval. The Gibbs free energy stays positive throughout, with no first-order transition of Hawking-Page type. All of this refers to equilibrium at the BH horizon alone. Since $T_q$ and $T_+$ differ by more than an order of magnitude for every parameter set considered, the two-horizon system is not in equilibrium as a whole, and a full treatment would have to handle the horizons jointly.

\section{Summary and Conclusions } \label{sec07}

We have re-examined a QOS BH spacetime carrying both a QF and a CS, with attention to the global structure that the earlier version of this work left unresolved. The metric function (\ref{bb2}) is a solution of the effective field equations with an additive source, and Sec.~\ref{sec02a} shows why the addition is legitimate: the Einstein tensor of a mass-function metric is linear in $m(r)$, so the CS, quantum, and QF contributions enter separately and no cross terms are generated. What that argument does not supply, and what we do not claim, is a derivation of the quantum deformation with the exotic fields switched on.

The global structure turned out to matter for nearly every result that follows. For $w=-2/3$ the metric function goes to $-Nr$ at large radius, so the spacetime is not asymptotically flat, and even at $N=0$ the CS leaves a solid-angle deficit. The horizon condition is a quintic, Eq.~(\ref{bb4a}), whose real positive roots are an inner horizon generated by the quantum term, the BH horizon, and an outer quintessence horizon; Table~\ref{istab_new} lists all three. The outer root is large, from $45.6M$ at $\alpha=0.5$ and $N=0.01/M$ up to nearly $10^3M$ at $N=0.001/M$, but it is finite, and it bounds the static region on which every quantity in this paper is defined.

Geodesic motion was analyzed inside that region. Circular photon orbits remain unstable for all parameters tested, with the Lyapunov exponent reduced as $\alpha$, $\lambda$, and $N$ grow. The comparison of orbital speeds with Schwarzschild reverses sign at a finite radius, and the earlier claim that the speed is uniformly enhanced holds only outside that radius, near $13.9M$ for $\lambda=0.5M^2$ and $N=0.01/M$. The geodetic precession frequency exceeds the Schwarzschild value over $3\lesssim r/M\lesssim10$ for the parameters plotted, and we make no statement outside that window.

The shadow had to be recomputed. With no asymptotically flat region available, the observer sits at finite radius, and we quote $R_s$ for $r_O=10M$. The dependence on $\alpha$ is strong and monotonic, taking $R_s/M$ from $4.64758$ in the Schwarzschild case to $8.14091$ at $\alpha=0.5$. The dependence on $\lambda$ is weak. The dependence on $N$ is not monotonic, since the outward shift of the photon sphere competes with the reduction of the redshift factor at the observer, and the two exchange dominance near $\alpha\simeq0.4$. A measured shadow therefore does not fix the sign of $N$ without independent information about $\alpha$.

On perturbations, the earlier version showed potentials and inferred stability from their shape. That inference was not available, since two of the potentials go negative near the horizon. We replaced it with an argument: an S-deformation with $S=-f/r$ turns the scalar potential into $\ell(\ell+1)f/r^2$, which is non-negative on the static region, and the Dirac potentials are supersymmetric partners built from a superpotential, so both associated operators factorize and are non-negative. Mode stability follows for all three test fields. The quasinormal frequencies were then computed at third-order WKB order and benchmarked against continued-fraction values for Schwarzschild, agreeing to three decimal places. Damping falls as $\alpha$ and $N$ increase and is almost insensitive to $\lambda$. These are test-field results on a fixed background; spin-2 perturbations were not treated, and the WKB estimates do not capture the purely damped family associated with the outer horizon.

The thermodynamic section required the largest revision. An algebraic error in the ADM mass, where the deformation entered as $\lambda/r_+$ rather than $\lambda/r_+^2$, propagated into the temperature and the reality condition and has been corrected. With the correction in place, and with the surface gravity taken as $|f'(r_h)|/2$ at whichever horizon is under discussion, the Hawking temperature is non-negative over the whole admissible branch and vanishes at the degenerate configuration where the BH and quintessence horizons merge. The negative temperatures reported earlier are withdrawn: they arise from evaluating the temperature formula at radii beyond that degenerate point, where the root being used is the quintessence horizon rather than the BH horizon. A temperature maximum appears that has no Schwarzschild counterpart, and it carries a Davies-type second-order transition separating a narrow stable branch from the unstable one. The Gibbs free energy remains positive throughout, so no Hawking-Page-type transition occurs. Because the two horizon temperatures differ by more than an order of magnitude, the system is not in global thermal equilibrium and the analysis applies to the BH horizon alone.

Several problems follow naturally. The most pressing is a proper two-horizon thermodynamics, in which the BH and quintessence horizons are treated jointly rather than separately, along the lines developed for Schwarzschild-de Sitter \cite{GibbonsHawking,BoussoHawking}; the effective-temperature and entropy-sum approaches would both be worth applying here. Next, the QNM spectrum deserves a converged treatment by continued fractions or time-domain integration \cite{Leaver,isz27}, which would test the WKB estimates and expose the purely damped modes we could not reach. Third, the assumption that the CS and QF do not feed back on the quantum deformation is uncontrolled, and quantizing the collapsing dust with those components included would either justify it or replace $\lambda$ with a parameter that depends on $\alpha$ and $N$. Finally, extending the geometry to include rotation would bring frame dragging and the ergoregion into the picture and would make contact with the accretion-flow observables \cite{isz30,isz31} that the static model cannot address. A further direction is the corrected-entropy analysis of multi-horizon configurations of this kind \cite{iszinfo}.

\section*{Acknowledgments}
{\footnotesize F.A. expresses gratitude to the Inter University Centre for Astronomy and Astrophysics (IUCAA), Pune, India, for a visiting associateship. \.{I}.~S. acknowledges academic and/or financial support from EMU, T\"{U}B\.{I}TAK, ANKOS, and SCOAP3, along with networking support from COST Actions CA18108, CA22113, CA21106, CA23130, CA23115, and CA21136.}

\section*{Data Availability Statement}

This study is purely theoretical and does not involve the generation or analysis of any datasets. 
Therefore, data sharing is not applicable to this article as no new data were created or analyzed in this study.

\section*{Funding}
This research received no external funding.

\end{document}